# Computational Approaches for Disease Gene Identification

**YANG PENG**

School of Computer Engineering

A Thesis Submitted to

Nanyang Technological University

In Fulfillment of the Requirement

For the Degree of Doctor of Philosophy

May 2013

# Acknowledgements

First and foremost, my special gratitude goes to my supervisors, Professor Kwoh Chee-Keong, Professor Li Xiaoli and Professor Ng See-Kiong for their immense patience and invaluable advice they provide me during the essential part of my life. The work that I have done here would not have been possible without them. During working with them, I have gained both the scientific knowledge and the way to do original research work. I am benefit from their thoroughness and diligence in revising every word and sentence in my research write-ups and strict attitude toward scientific research work.

I have to thank Assist Professor Zheng Jie, from whom I am impressed by his sharp eyes in finding the key underlying problems and his critical thinking skills to let me see many issues with my own eyes.

I have to thank Associate Professor Lin Feng, Assist Professor Manoranjan Dash for their effective comments and support during my Ph.D qualifying examination.

Besides professors, I would warmly thank my colleague and friend, Dr. Mei Jianping for her friendly help and kindly sharing with me her MATLAB codes of many machine learning algorithms used in this study. I also extend my special appreciation to Dr. Li Yongjin, from whom I was inspired to learn a lot of knowledge and approaches related with disease gene prediction. I also enjoyed the time that I spent with my other colleagues in Bioinformatics Research Center including Dr. Liu qian, Dr. Wu min, Dr. Zhao Liang, Dr. Li Zhenhua, Dr. Piyushkumar and with my lab mates including Ouyang Xuchang, Su Tran To Chinh, Liu Wenting, Chen Haifen, Zhang fan, who worked closed with me to discover new problems in Bioinformatics.

Especially, I would like to give my special thanks to my brother, girlfriend and parents. Their encouragement and understanding are my motivation to finish my Ph.D study.

Finally, I express my sincere gratitude to everyone who has contributed to this thesis.

# Contents

# Abstract

Identifying disease genes from human genome is an important and fundamental problem in biomedical research. Despite many publications of machine learning methods applied to discover new disease genes, it still remains a challenge because of the pleiotropy of genes, the limited number of confirmed disease genes among whole genome and the genetic heterogeneity of diseases. Recent approaches have applied the concept of 'guilty by association' to investigate the association between a disease phenotype and its causative genes, which means that candidate genes with similar characteristics as known disease genes are more likely to be associated with diseases. However, due to the imbalance issues (few genes are experimentally confirmed as disease related genes within human genome) in disease gene identification, semi-supervised approaches, like label propagation approaches and positive-unlabeled learning, are used to identify candidate disease genes via making use of unknown genes for training – typically in the scenario of a small amount of confirmed disease genes (labeled data) with a large amount of unknown genome (unlabeled data). The performance of Disease gene prediction models are limited by potential bias of single learning models and incompleteness and noise of single biological data sources, therefore ensemble learning models are applied via combining multiple diverse biological sources and learning models to obtain better predictive performance. In this thesis, we propose three computational models for identifying candidate disease genes.

1. In this Ph.D thesis, we first propose a computational algorithm Random Walk on Protein Complex Network (RWPCN) to prioritize disease genes. Different from traditional two-layer phenotype-gene heterogeneous network, the basis of RWPCN is a novel three-layer heterogeneous network, where links from disease phenotypes to human protein complexes are interacted based on confirmed phenotype-gene associations, and links from protein complexes to individual proteins are interacted when individual proteins are members of protein complexes. To evaluate the performance of the proposed RWPCN, we conducted experiments to compare RWPCN with other state-of-art techniques and the results show that our RWPCN significantly outperforms existing disease gene prediction approaches. In addition, the RWPCN is applied to investigate candidate disease protein complexes and predict novel disease genes associated with two representative diseases, namely, breast cancer and diabetes.

2. Disease gene identification is a positive-unlabeled problem. Currently, only a few disease genes have been identified from large number of unlabeled human genome. In the second part of the thesis, we propose a novel disease gene classification model, Positive-Unlabeled learning for Disease gene Identification (PUDI). Unlike traditional learning models that use unknown genes as a negative training set $N$, PUDI treats unknown genes as an unlabeled set $U$. Since unknown genes may contain unconfirmed disease genes, it is inappropriate to label all unknown genes as negative class. Therefore, we design a positive-unlabeled (PU) learning method to partition unlabeled training genes into multiple sets and then apply a weighted support vector machine (SVM) to build a disease gene classifier. We find that PUDI could model the classification problem for disease gene prediction more effectively as it achieves significantly better results than the state-of-the-art methods.

3. Due to inherent complex characteristics of phenotype-gene associations, such as pleiotropy of genes and genetic heterogeneity of diseases, disease gene identification requires various and sufficient biological data to represent genes and reliable computational approaches to build robust classifiers. To further improve the performance of disease gene identification, we focus on an ensemble positive unlabeled learning model, namely Ensemble Positive-Unlabeled learning for disease gene identification (EPU), to combine network-based approaches and positive-unlabeled learning in chapter 3. The random walk with restart (RWR) algorithm, a network propagation approach, is applied to three gene networks to assign combined confidence scores to unlabeled genes. Using weighted unlabeled genes and initial labeled genes, we build three individual PU learning classifiers to predict 'soft' classes for test genes. Finally, an ensemble strategy EPU is applied to build an ensemble model where individual classifiers' predictions are linear combined together to make final decisions for the test gene class. The experimental results show that our proposed EPU is able to produce favorable performance compared to state-of-the-art techniques over six disease groups, indicating our ensemble framework could make use of multiple data sources and multiple learning models to build an accurate classifier.

# Important Abbreviations Used

| | |
|---|---|
| BP | Biological Process |
| CC | Cellular Component |
| CORUM | Comprehensive Resource of Mammalian protein complexes |
| CP | Candidate Positive |
| EPU | Ensemble-based PU learning for Disease gene Identification |
| DNA | Deoxyribonucleic acid |
| GO | Gene Ontology |
| GWAS | Genome-wide Association Study |
| HPRD | Human Protein Reference Database |
| LP | Likely Positive |
| LN | Likely Negative |
| LOO-CV | Leave-One-Out Cross Validation |
| MeSH | Medical Subject Headings vocabulary |
| MF | Molecular Function |
| NB | Naïve Bayes |
| NN | Neuron Network |
| OMIM | Online Mendelian Inheritance in Man |
| OPHID | Online Predicted Human Interaction Database |
| OSMED | Oto-Spondylo-Mega-Epophyseal Dysplasia |
| ROC | Receiver Operating Characteristic |
| RWPCN | Random Walk on Protein Complex Network |

RNA             Ribonucleic acid

PPI             Protein-Protein Interaction

PPMCC           Person Product-Moment Correlation Coefficient

PU              Positive-Unlabeled Learning

PUDI            Positive-Unlabeled Learning for Disease Identification

RN              Reliable Negative

RWR             Random Walk with Restart

SVM             Support Vector Machine

WN              Weak Negative

WSVM            Weighted Support Vector Machine

# List of Figures

# List of Tables

# Chapter 1.

# Introduction

This chapter begins with a brief introduction of the problem of disease gene identification, and then provides the motivation of this research. This is followed by a summary of current research contributions and an outline of the PhD thesis.

## 1.1   Background

### 1.1.1   Motivation and Objective of Disease Gene Identification

There are more than 2000 monogenic syndromes (the syndromes found associated with a single causative gene) in human beings [1]. Each syndrome has a specific combination of phenotypic features, which are the biological implementations of their underlying genes, and each differs from other syndromes by only one or a few of those features [1]. Therefore, discovering phenotype-gene association is a fundamental and critical biomedical task, which assists biologists and physicians to discover pathogenic mechanism of syndromes. Knowledge of which genes cause which disorders will simplify diagnosis of patients and provide insights into the functional characteristics of the mutation.

Disease gene identification is a process by which scientists identify the mutant genotypes responsible for an inherited genetic disorder. Mutations in these genes can include single nucleotide substitutions, single nucleotide additions/deletions, deletion of the entire gene, and other genetic abnormalities.

Traditionally, disease gene identification composes of two main steps: genetic linkage analysis and positional clone, followed by mutation analysis. Firstly, linkage analysis is performed in human pedigrees to detect the susceptible chromosome interval that is the approximate location of candidate genes associated with diseases [2]. Secondly, the technique 'positional cloning' is proposed to sequence a set of the candidate genes in the region [3]. This process includes a physical mapping and a transcript mapping. From the candidate disease gene set, one or more genes can be the real disease gene. Since 1980, positional cloning techniques have been used to prioritize candidate genes and these techniques have successfully identified disease genes for a number of diseases, such as Duchene muscular dystrophy, Huntington's disease and cystic fibrosis, etc. However, both the positional cloning and mutation analysis experiments are labor intensive, tedious and time consuming, requiring computational methods to select highly suspicious genes for experiment validation. The human genome contains in the range of 20 to 25 thousand genes. It is a challenging task to identify real disease genes from hundreds of candidate genes on the whole genome. To address the challenge, it is necessary to prioritize candidate genes from hundreds of experimentally suspicious genes using computational techniques, which would greatly reduce the numbers of genes for wet-lab experimental analysis. In other words, computational approaches help to prioritize the genes, which are likely to associate with disease of interest, for further web-lab experiment validation. The common strategy is to rank these candidate genes according to their functional similarity to known disease genes. In this thesis, this is called disease gene prioritization.

## 1.1.2  Challenges of Disease Gene Identification

Human genetic disease is a genetic disorder caused by abnormalities in genes or chromosomes, especially a condition that is present before birth. Genetic diseases are generally divided into two types: single gene disorder and complex disorder.

A single-gene (monogenic) disorder occurs as a direct result of a single mutation in the structure of the DNA, leading to a single basic defect with pathologic consequences. Such disorders are passed on to subsequent generations in simple patterns according to Mendel's Laws. As such, these kinds of disorders are often called Mendelian disorders [4]. The Mendelian Inheritance in Man (MIM) is a comprehensive knowledge base of human genes and genetic disorders. Its Online Version, Online Mendelian Inheritance in Man (OMIM) database, currently provides information on more than 5000 genetic disorders.

Genetic disorders may also be multi-factorial, namely complex diseases, which reflect the pathologic consequences associated with the effects of a combination of genetic mutations, lifestyle and environmental factors, and genetic factors represent only part of the phenotypes associated with the disorders [5]. These kinds of diseases are called multi-factorial and polygenic (complex) diseases. Out of 5080 disease phenotypes in OMIM, there are a total of 168 phenotypes associated with multiple causative genes in our experiment data which do not obey the standard

Mendelian patterns of inheritance [6]. The complex diseases include Alzheimer's disease, asthma, Parkinson's disease, connective tissue disease, kidney disease, and many more. For example, diabetes mellitus type 2, a metabolic disease, is found to be associated with multiple gene mutations, and multiple susceptible chromosome loci, including chromosomes 2, 3, 4, 5, 6, 7, 13, 15, 17 and 19 [7].

In addition, disease gene identification remains a daunting problem due to the pleiotropy of genes, the genetic heterogeneity of disease as well as other environment factors. Pleiotropy describes the genetic effect of a single gene on multiple phenotypic traits. Pleiotropy occurs when one gene influences multiple, seemingly unrelated phenotypic traits, such as Phenylketonuria that is a human genetic disease that affects multiple systems but is caused by one gene defect. Consequently, a mutation in a pleiotropic gene may have an effect on some or all traits simultaneously. Contrast to pleiotropy where a single gene may cause multiple phenotypic expression or disorders, genetic heterogeneity is a phenomenon in which a single phenotype or genetic disorder may be caused by any one of a multiple number of alleles or non-allele (locus) mutations. Genetic heterogeneity can be classified as either "allelic" or "locus". Allelic heterogeneity means that different mutations within a single gene locus cause the same phenotype expression. For example, there are over 1,000 known mutant alleles of the CFTR gene that cause cystic fibrosis. Locus heterogeneity means that variations in completely unrelated gene loci cause a single disorder. For example, retinitis pigmentosa has three origins from autosomal dominant, autosomal recessive and X-linked.

Recent approaches are strive to discover the patters of pleiotropy of genes as well as the multi factors of genetic disorders using the gene sequences, gene expression and PPI network based on the assumption that similar disease phenotypes are caused by similar genes in terms of sequence, expression profile similarity and network topology. However, above methods merely focus on views of single genes. As genes cannot function alone, they are likely to form "functional modules", where genes are likely to attached together to perform a biological function or process. This kind of "module" can be protein complex, pathway or metabolic network [8] [9] [10] [11] [12] [13].

### 1.1.3  Modular Nature of Genetic Diseases

It is shown that similar phenotypes are caused by functionally related genes. Evidence from many sources suggests that genetically heterogeneous diseases [14] [15] (such as Fanconi anemia [16], breast cancer [5] [17] and diabetes [18]) are caused by many genes which work together in a single biological module. Such module can be a multi-protein complex, or a pathway. For example, it is now becoming clear that protein interactions play a key role in the mechanisms of cellular functions at the molecular level and determine the outcomes of biological processes [19], such as signal transduction, enzyme-mediated metabolism, DNA replication and transcription [20]. From the analysis above, the modular nature of human genetic diseases could indicate or reflect the modularity in true biological interaction networks.

Typically, multiple syndromes can be caused by mutations in the same genes, and a single disorder can be caused by mutations in different genes. Different alleles of genes in different individuals integrate differently with each other to create individual final phenotypes [10]. This is the basis of human phenotypic diversity and an important factor contributing to the fact that no two individuals are identical.

For instance, genes involved in the same protein complex or biochemical pathway work together to perform specific biological functions. While there is a tendency for similar disease phenotypes to be caused by functionally related genes, mutations that affect different functions of a pleiotropic gene can result in different phenotypic manifestations. As shown in Fig. 1.1, the similar phenotypes of Stickler, Marshall and oto-spondylo-mega-epophyseal dysplasia (OSMED) syndromes are caused by mutations in the functionally closely related genes COL2A1, COL11A1 and COL11A2. The phenotypically distinct Pallister-Hall syndrome is caused by mutations in the functionally unrelated or only weakly related GLI3 gene. Several genes can underlie one phenotype, as in the case of Stickler syndrome, which can be caused by mutations in each of the three collagen genes. Conversely, one gene can lead to different phenotypes as in the case of COL11A1 (Stickler and Marshall phenotypes) and COL11A2 (Stickler and OSMED phenotypes). The thickness of black lines linking genes indicates the (hypothetical) degree of functional relatedness between them.

Although complex disorders are likely to form functional modules, they do not have a clear-cut pattern of inheritance (such as the pleiotropy of genes), which makes it more difficult to determine their characteristics than single-gene (Mendelian) disorders. In addition, the functional modules, such as protein complexes, have temporal dynamics characteristics – structure of functional module may vary with "time" in terms of cell cycle phase. Finally, there are no literatures to tell how to establish association of functional modules to disease phenotypes.

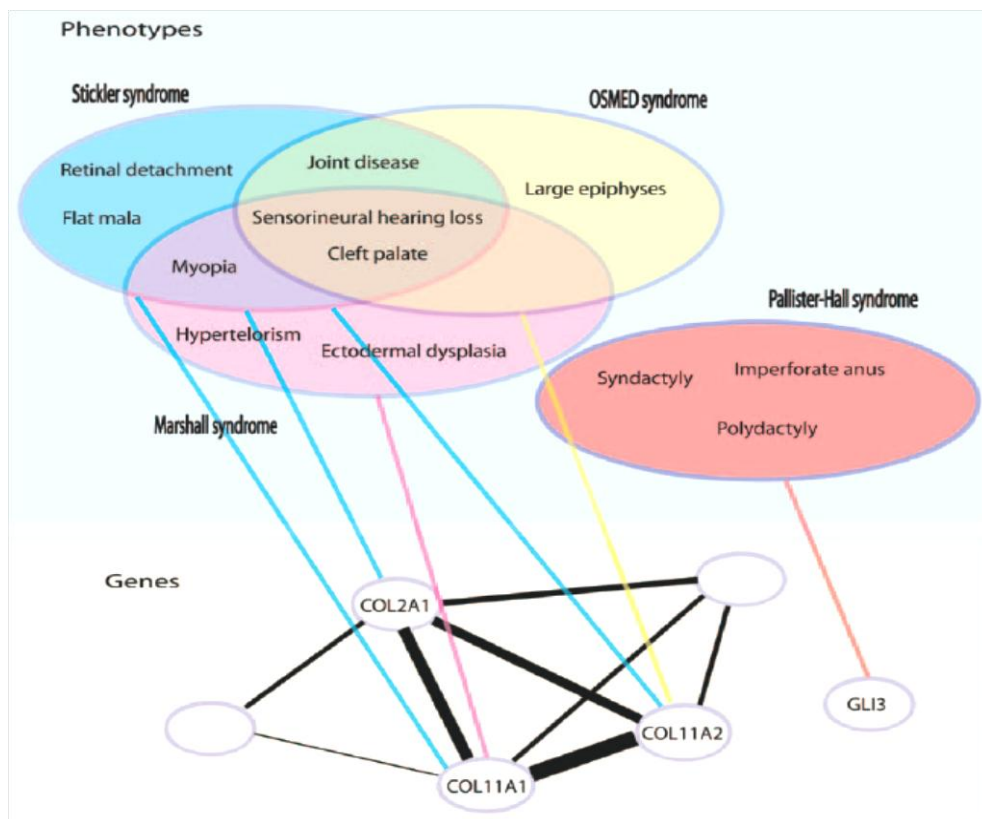

Figure 1.1: Possible relationships between genes and phenotypes, taken from [10].

## 1.2 Related Prior Works

Uncovering the associations between genetic diseases and their causative genes is one of the fundamental objectives of human genetics [21] due to its significant impact in healthcare.

Gene association could represent that genes have physically interaction in the protein-protein interaction network or have similar/relevant biological functions in terms of gene ontology. The recent approaches have applied protein-protein interaction (PPI) to detect the association between disease and candidate genes in the PPI network [22] [23] [24] [25]. The underlying assumption in these studies is that the interacting partners of a disease-causing gene (or more precisely, its gene product protein) in the PPI network are likely to cause either the same or similar diseases [26]. In order to find similar diseases, some of the existing methods compute the similarities between the phenotypes, generating a phenotypically similarity network where two phenotypes are linked if they are phenotypic similar. These methods first construct a protein interaction network and subsequently compute the closeness between candidate genes and known disease genes based on network topological information. In particular, Wu et al. [27] built a regression model measuring the correlation between phenotype similarity and gene closeness in the PPI network for prioritizing candidate disease genes. Vanunu et al. [28] and Li et al. [22] designed a global network-based method by formulating constraints on the genes' score function that smooth over the whole network. However, these methods only focus on protein-level function prediction. Moreover, proteins cannot function isolate; they are more likely to be attached together as functional modules (such as protein complexes and pathways) to perform biological functions. Recently, Lage et al. assigned a candidate gene to protein complexes and then applied a Bayesian model to rank the candidate protein complexes with confidence scores to disease phenotypes [29]. However, they simply assembled those neighboring proteins as complexes. Given that protein complexes are molecular groups of proteins that work together as 'protein machines' for common biological functions, only those tight-knit substructures in PPI networks correspond to actual protein complexes [30]. As such, the protein complexes constructed were not accurate. Additionally, they have not considered the associations among individual protein complexes even though many complexes share common/pleiotropic proteins and the proteins from different complexes do interact with each other. Biologically, genes associated with similar disorders demonstrate a higher probability of physical interactions between their gene products [31] [1]. As such, if two protein complexes share common proteins or have physical interactions between them, then the mutations of certain genes in a protein complex could lead to identical or similar phenotypes of its connected protein complexes [1] since the mutations could potentially disrupt these complexes' functions. In this thesis, we believe that constructing a protein complex network where nodes are individual complexes and the interactions between two complexes are measured by the connection strength between them would be a proper model for disease gene prioritization.

It should be noted that the above methods only provide a gene rank list and a threshold is needed to decide whether a specific gene is disease related or not. A more biologically meaningful approach would build a binary classification model that can automatically identify a gene as disease or not, according to various features of biological datasets, such as protein sequence, PPI and gene expression.

To address this problem, Lopez Bigas and Ouzounis [32] investigated the distinguishing features of protein sequences between disease and non-disease genes. Adie et al. [33] further improved on this method by employing a decision tree algorithm based on a variety of genomic and evolutionary features (such as coding sequence length, evolutionary conservation and presence). In particular, Xu et al. [34] employed the K-nearest neighbor (KNN) classifier to predict disease genes based on the topological features in PPI networks, including protein degree and the percentage of disease genes in the protein neighborhood (the proteins directly linking to disease-related proteins). Smalter et al. [35] applied the support vector machines (SVMs) classifier using PPI topological features, sequence-derived features, evolutionary age features, etc. The above works employ machine learning methods to build a binary classifier by using the confirmed disease genes as positive training set $P$ and some unknown genes as the negative training set $N$. However, the negative set $N$ will contain unconfirmed disease genes (false negatives), which confuses the machine learning techniques for building accurate classifiers. As such, the classifiers built based on the positive set $P$ and noisy negative set $N$ do not perform as well as they should in identifying new disease genes. To address this issue, Mordelet et al. proposed a bagging method ProDiGe for disease gene prediction. This method iteratively chooses random subsets ($RS$) from $U$ and trains multiple classifiers using bias SVM to discriminate $P$ from each subset $RS$. It then aggregates all the classifiers to generate the final classifier [36]. However, as the random subsets $RS$ from $U$ could still contain unknown disease genes, selecting candidate disease genes and reliable non-disease genes could be helpful for building an accurate classification model. In this thesis, we design a novel PU learning algorithm to build a more accurate classifier based on $P$ and $U$ [37] [38] [39].

The above recent approaches only use a single learning model (such as SVM in Smalter et al. [35], KNN in Xu et al. [34]) or single biological datasets (PPI network in Xu et al. [34] and protein sequence in Lopez Bigas and Ouzounis [32]) to prioritize candidate disease genes from the unlabeled gene set. A classification model built on single datasets may be limited by incompleteness and noise of single biological datasets. For a classification model built on a single hypothesis, it is not easy to achieve competitive performance on all disease groups. To address the above issue, an ensemble-based approach is applied to build a combined classifier by integrating multiple learning models that could be obtained from any of the constituent models [40] [41], such as boosting and bootstrap aggregation (bagging). In this thesis, we investigate how to integrate multiple biological datasets and learning models to identify disease genes based on the disease gene set and unlabeled gene set.

## 1.3  Major Contributions and Organization

As mentioned earlier, the ultimate goal of this research is to prioritize candidate disease genes from hundreds of candidate genes on the whole genome. In this thesis, this is done through introducing reliable biological modules, especially protein complex, applying PU learning algorithm on imbalance genetic dataset, and combining reliable PU learning approaches and various biological datasets. We describe a set of novel disease gene prediction models, where efficiencies are verified by different disease groups. Our computational tools are following the 'Guilt-by-association' rules that exploit the underlying modularity of disease phenome, protein interactome and genome. Through providing an effective framework for disease gene prediction, we can integrate additional biological and computational resources in the future. Figure 1.2 presents a whole schema of our three contributions: RWPCN is a network-based model that applied flow propagation algorithm on a novel protein complex interaction network, while PUDI exploits PU learning techniques for disease-gene identification based on gene biological features. Finally, EPU is an ensemble framework that combines the RWPCN and PUDI.

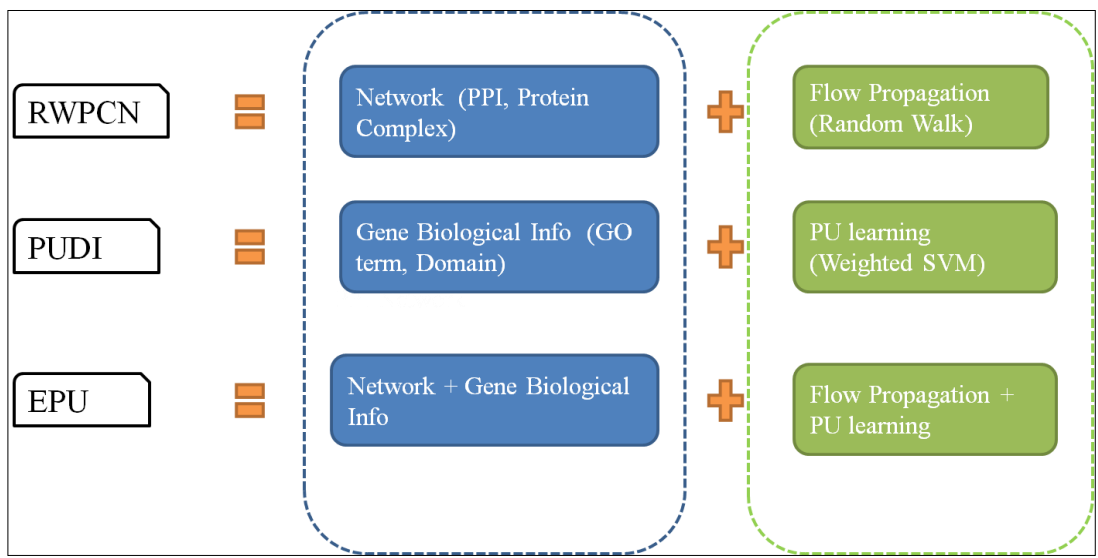

Figure 1.2 Overall Schema for Three Contributions

- A novel complex-based disease gene prediction algorithm, namely RWPCN, is different from the existing methods as our network propagation algorithm is operated at the complex level instead of the protein level. We use reliable human protein complexes from CORUM, the comprehensive resource of mammalian protein complexes, since these protein complexes are curated from the biological literatures. To the best of our knowledge, this is the first attempt to exploit the biological modularity of the protein complexes, and exploit effect protein complexes to disease phenotypes to detect disease genes in an explicit way. We construct a novel Protein Complex Network as in our proposed methods. Our experimental results show that such an effort is indeed worth the while, for the proposed algorithm is able to discover gene-phenotype associations more effectively compared to existing state-of-the-art methods. This suggests that the protein complex network can reflect the underlying modularity in the biological interaction networks better than simple protein interaction networks. Our method is also applied to uncover novel candidate genes on specific complex genetic diseases.

- A novel PU learning approach, PUDI, is applied to build a multi-level classifier for disease gene prediction. A new feature selection method is introduced to identify the discriminating features. A random walk algorithm is applied on the gene similarity network to perform a further partitioning of the unlabeled set U into multiple training sets for a more refined treatment of U to build the final classifier. We found that PUDI could better model the classification problem for disease gene prediction as it achieved significantly better results than the state-of-the-art methods. In addition, we investigated the efficiency and time complexity of PUDI and other state-of-the-art techniques for disease gene prediction and showed the time spent by each method. Given that many machine learning problems in biomedical research involve positive and unlabeled data instead of negative data, we believe that the performance of machine learning methods for these problems can potentially be improved further by adopting a PU learning approach, as we have done here for disease gene identification. For future work, we will consider integrating more biological resources, such as gene expression data. In addition, we may explore more complicated machine learning methods to better model the positive and unlabeled data distributions.

- A novel ensemble positive-unlabeled learning approach EPU is applied to identify disease genes. Firstly we perform the random walk with restart algorithm on three networks (protein interaction network, gene expression similarity network and GO similarity network) to extract multiple positive and negative samples from the unlabeled set $U$. Then we exploit these extracted positive and negative samples as training data to build three independent PU learning models. Finally, we design a novel ensemble strategy EPU by minimizing the overall error rate and giving different weights to different PU learning models. We have compared EPU with various state-of-the-art techniques. The experimental results show that EPU outperforms the existing methods significantly in identifying the disease genes on six disease groups.

## 1.4 Outline

The PhD thesis is organized as follows:

Chapter 1 begins with a brief introduction of the basics of human genetic disease and the problem of candidate gene prioritization, followed by the motivation of the research, with an outline of our work and what it has achieved.

Chapter 2 provides a survey on the literatures relevant to this topic, including prioritization algorithms on single data sources and integration algorithms on multiple sources.

Chapter 3 introduces the protein complex network model. We propose a random walk method on this model that integrates the protein-protein interaction network and protein complex information for disease gene prioritization.

Chapter 4 focuses on the positive unlabeled learning algorithm for disease gene identification. We conduct the experiments to compare our PUDI with several state-of-the-art techniques in general disease genes and specific disease classes, and investigate the capacity of PUDI to identify novel disease genes.

Chapter 5 proposes a novel ensemble framework, EPU, to identify disease genes in six disease classes. The predicted novel candidate genes for metabolic and cancer diseases are shown in this part.

Finally, Chapter 6 concludes this thesis and presents possible directions to extend the current scope of this entire PhD research.

# Chapter 2.

# Literature Review

Disease gene identification is a process to detect the mutant genotypes that cause a corresponding genetic disorder. Mutations in these genes are mainly divided into three conditions: 1. Single nucleotide substitution/additions/deletions; 2. Deletion of the entire gene; 3. Other genetic abnormalities. Disease gene identification follows two procedures: first DNA is collected from several patients who suffer same genetic disease; then DNA samples are screened to determine regions where the mutations could reside [2]. Genes in this region (more than 10 Mb) are called candidate genes, one or more of which might be the real disease gene. Identification of the most probable of these candidate disease genes for further wet-lab experimental analysis is a significant challenge because the number of genes in the region is in the range of dozens, or even hundreds. Identification of all the genes in the region is time-consuming and expensive. The common strategy is to rank these candidate genes according to their functional similarity to known disease genes, and then to prioritize top ranked candidate genes as novel disease genes. A number of computational methods have been developed to address this problem. In this chapter, we provide a comprehensive survey on disease gene discovery methods using various biological data sources and computational strategies that combine multiple data sources and learning methods.

## 2.1 Prioritization of Candidate Genes Based on Biological Data Source

In this section, we introduce disease gene discovery methods based on different biological data sources, including sequence-based methods, gene expression based methods, ontology-based methods, and Protein-Protein Interaction (PPI) network based methods.

### 2.1.1 Sequence-Based Methods

A protein is involved in a genetic disease when its corresponding gene is mutated, impairing its function or expression strongly enough to produce one or several abnormal phenotypes (called disease phenotypes). With the completion of the human genome sequence project, there is an opportunity to investigate disease susceptibility loci on a large scale (genome size of Homo sapiens: 3.2 Gb), in terms of the likely or known functions of the annotated genes present within them. Several research groups predicted disease genes through sequence-based features, because they found that human genes involved in hereditary disease share some common distinct sequence characteristics which render them more susceptible to mutations causing genetic disorders [42] [43] [32] [33].

Lopez-Bigas et al. [32] investigated some features of protein sequences between disease genes and non-disease genes. They found that proteins involved in hereditary diseases tend to be long, with highly conserved amino acid sequences, wide phylogenetic extent, and without similar paralogues. Adie, Adams et al. [33]

found 24 sequence-based features, which were significant differences between disease genes and unknown genes. For example, disease genes have a significantly larger number of cDNA length and encoded larger proteins, significantly longer 3' UTR, and longer distance to the nearest neighboring genes. Based on these features, they created an automatic classifier using the decision tree algorithm which typically produces a tree that is predictive, concise and easy to understand. It is called PROSPECTR, which ranks genes in the order of likelihood of involvement in diseases.

### 2.1.2 Gene Expression Based Methods

Gene expression measurements on a genome-scale, representing the transcriptome, have been accomplished through the technological advancement of microarrays. Since genes involved in identical functions tend to show very similar expression profiles, co-expression analysis could be a powerful approach for inferring functional relationships which may correlate with similar disease phenotypes [44] [45]. Genes that are co-expressed tend to be involved in the same biological processes. Co-expression between genes is usually calculated based on the microarray data. However the microarray data is noisy, therefore co-expression does not strongly suggest a functional relation. Conserved co-expression could be a much stronger criterion than single species co-expression to the genes relevant to similar disease phenotypes, because significant co-expression existing in more than one species (more orthologous) indicates the significance and stability of the relationship in evolutionary history [46]. Ala et al. [47] showed that reliable disease-relevant relationships may be identified from massive microarray datasets by concentrating only on genes sharing similar expression profiles in both humans and mice. Integration of human-mouse conserved expression with a phenotype similarity map systematically allows the efficient identification of disease genes in large genomic regions. Combining evolutionarily distant species to calculate evolutionary co-expression will further increase the reliability. Oti et al. [48] used co-expression data from yeast (S. cerevisiae), nematode worms (C. elegans), fruit flies (D. melanogaster), mice (Mus musculus) and humans (homo species), and the co-expression predictive value could be improved using evolutionary conservation. Figure 2.1 illustrates the method of calculating conserved co-expression between humans and flies involving KOG0011 and KOG3438, which are defined by the eukaryotic clusters of the Orthologous Groups (KOG) database. KOG0011 contains two genes in humans (RAD23A and RAD23B) and flies (FBgn0026777 and FBgn0039147) while KOG3438 contains one in humans (CKS1B) and two in flies (FBgn0010314 and FBgn0037613). For one species, the KOG0011-KIG3438 co-expression score is computed as the mean of all gene-gene correlations between the two KOGs. The final conserved co-expression correlation between humans and flies is calculated by taking the mean of two single species KOG0011-KIG3438 co-expression values. They found that the evolutionary conservation of co-expression between species improved the predictive power of co-expression data and was more reliable than co-expression vectors in single species. Microarray signal values were measured by the Spearman Rank Correlation Coefficients (that is defined as the Pearson correlation coefficient between the rank variables) as shown in the following equation:

$$\rho = \sum_i (x_i - \bar{x})(y_i - \bar{y}) \Big/ \sqrt{\sum_i (x_i - \bar{x})^2 \sum_i (y_i - \bar{y})^2} \tag{1}$$

where $x_i$ and $y_i$ are the $i^{th}$ elements of two gene expressions of vector $x$ and vector $y$, $\bar{x}$ and $\bar{y}$ are the means of values of the gene expression vectors $x$ and $y$. The authors took the mean of species-specific KOG-based co-expression scores over all species considered.

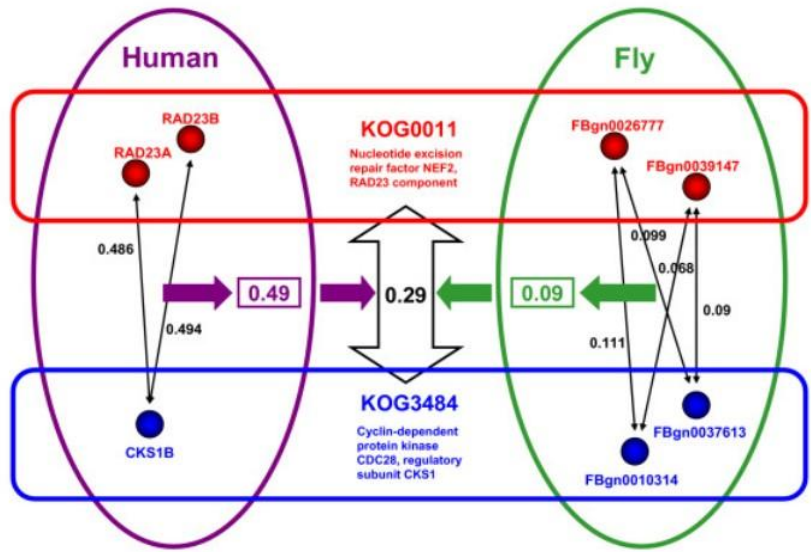

**Figure 2.1**: Procedure for calculating conserved co-expression scores

### 2.1.3   Ontology-Based Methods

Ontologies define concepts/terms and their relationships within a specific subject area. It is a formal way of representing knowledge in which concepts are described by their meanings and their relationships to each other [49]. There are several ontologies in the field of biomedical research, such as Gene Ontology (GO) [50], eVOC anatomical ontology [51], mammalian phenotype ontology (MP) [52] and human phenotype ontology [53]. Gene Ontology (GO) describes the biological process, molecular function and cellular location of action of a protein in a generic cell. The eVOC ontologies provide simple sets of controlled terms describing human anatomical systems, cell types, diseases and developmental stages.

Several candidate-gene identification systems that rely on grouping GO terms have been reported [54] [55] [56]. Turner et al. [54] proposed an approach called POCUS (prioritization of candidate genes using statistics) that prioritizes candidate genes across multiple susceptibility loci that share GO terms. Perez-Iratxeta et al. [55] developed a methodology based on biomedical literatures that associate pathological conditions with particular Gene Ontology (GO) terms, which then allow candidate disease genes to be ranked according to the number of these terms they share. Freudenberg and Propping [56] produced clusters of known disease genes based on a measure of phenotypic similarities that are computed according to their phenotypic appearances, using the indices 'periodicity', 'etiology', 'tissue', 'age of onset' and 'mode of inheritance'. Candidate genes were then scored according to the GO terms shared with known disease genes in the clusters.

Tiffin et al. [57] used another kind of ontology, the eVOC anatomical system ontology [51], to predict disease genes. In Figure 2.2, they first identified the association between eVOC anatomy terms and disease names according to their co-occurrence in the PubMed literature, and then ranked each term based on the frequency of association. Finally, candidate disease genes identified by linkage analysis were prioritized based on their corresponding annotation with the selected disease-related eVOC terms.

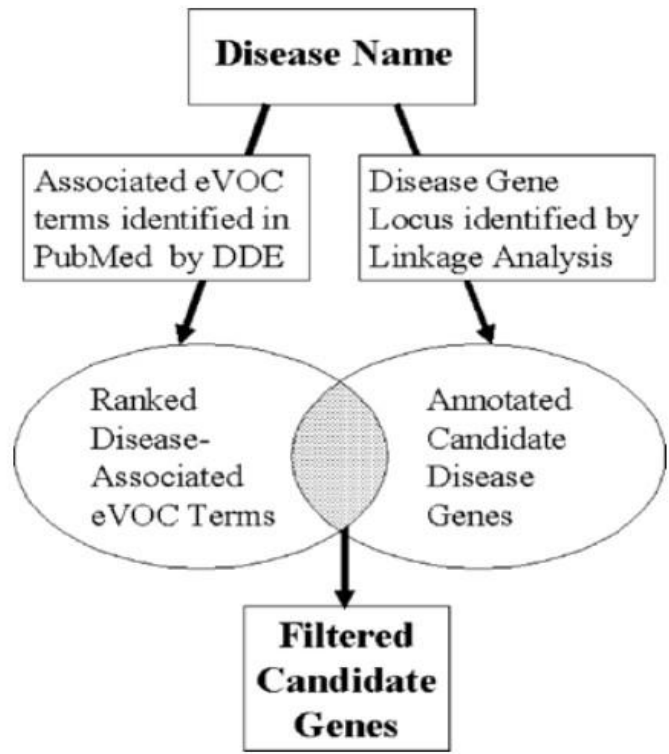

**Figure 2.2**: Identifying candidate genes using eVOC terms

### 2.1.4   PPI Network Based Methods

Physically interacting proteins tend to be involved in the same cellular process [58]. Hence, proteins encoded by genes mutated in inherited genetic disorders are likely to interact with proteins known to cause similar disorders, suggesting the existence of disease sub-networks [31]. Oti et al. [24] predicted interacting partners of disease genes in the PPI network. Xu et al. [34] found that disease genes share some distinct topological features in the PPI network compared to unknown genes. Based on these distinguishing features, they employed the K-nearest neighbor classifier to predict novel disease genes that are functionally similar to known disease genes. Köhler et al. [25] used the random walk with restart (RWR) algorithm [59] to prioritize candidate genes. In the susceptible region detected by QTL, there are **n** genes called candidate genes. Firstly, candidate genes and known disease genes are mapped to the PPI network. Then, the disease genes are used as source nodes to run the RWR algorithm, and the candidate genes are scored by their proximity to known disease genes. Finally, the candidate genes are prioritized based on the proximity scores.

In Figure 2.3, one genetic disease is represented as five phenotypes in the database of Online Mendelian Inheritance of Man (OMIM) [60]. The corresponding known disease genes are mapped to the PPI network, represented as yellow nodes without an integer number. Candidate genes are represented as yellow nodes with a number, say, 1, 2 and 3. As shown in Figure 2.3, the third candidate gene is well connected to known disease genes; therefore it is very likely to be a real disease gene. After running the RWR algorithm, this gene is given the highest proximity score and ranked at the top.

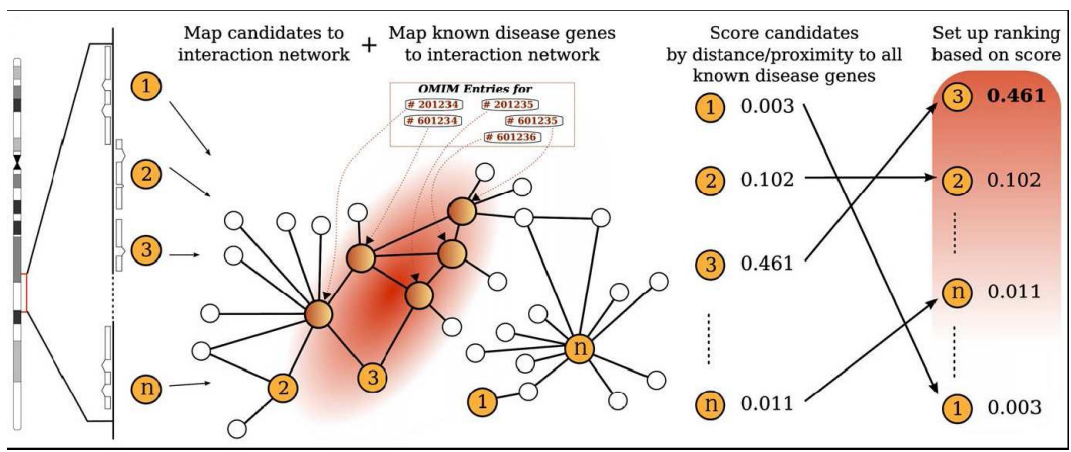

**Figure 2.3**: Candidate genes prioritization using random walk with restart (RWR) algorithm, taken from [25].

## 2.2 Integration Methodologies on Candidate Genes Prioritization

As more genes are being sequenced and annotated, and gene/protein interaction data are accumulating [61] [62], an ever-increasing wealth of biological data is now available in public databases. Each data source covers part of the human genome, therefore these data sources are complementary to each other. For example, a gene that has been extensively studied for a long time will have a large amount of associated literature and have a better chance of being annotated in GO [50]. Genes that have well-characterized protein products are more likely to be found in the PPI network. In this section, we introduce several representative integration algorithms on disease gene identification.

**SUSPECTS**: Adie et al. [63] proposed a tool named SUSPECTS for the prioritization of candidate genes. In this tool, gene annotation, gene expression and protein sequence data are used to prioritize disease candidate genes. Given a set of confirmed disease genes associated with a particular disease as the training set, each test gene is scored based on four lines of evidence: first by Prospectr [33] on the basis of its sequence features; second by the extent of co-expression with the training set based on expression data [64]; third by the number of rare (found in <5% of all proteins) protein domains, obtained from the Interpro [65] database, shared with the training set; and finally by the functional semantic similarity [66] to genes in the training set. The four scores are then combined to final scores that are weighted depending on the amount of information available for each line of evidence.

This approach relies on good quality functional annotation for each candidate gene. Genes that are lacking in the GO, expression and protein domain may limit the prediction performance of the method.

**CAESAR**: Gaulton et al. [67] integrated data from several ontologies to discover disease genes associated with disease phenotypes that are of interest to users. Their integration algorithm is called CAESAR. CAESAR requires a user-defined body of text (referred to as a corpus) to represent a disease of interest. This corpus is ideally an authoritative and comprehensive source of biological knowledge about the diseases of interest. It can be the clinical symptom of a disease or an OMIM identifier [60]. CAESAR uses the OMIM record as the corpus when an OMIM identifier is available.

Four ontologies are used in CAESAR, namely the Biological Process (BP) and Molecular Function (MF) ontologies (from GO) [50], the mammalian phenotype ontology (MP) [52], and the eVOC anatomical ontology [51]. For each ontology, CAESAR uses text mining techniques to extract the ontology term and description that comprise a document. Each ontology document and the corpus are represented as vectors in word space, $(c_1, c_2, \ldots, c_m)$, where elements are weighted counts of the words within the document. The similarity score of each ontology to the corpus is calculated as the Cosine similarity between the ontology vector and corpus vector. A score $\varepsilon_{ij}$ of gene $i$ for source $j$ is then calculated as either the maximum, sum or mean of the disease similarity scores of the matched ontology terms annotating gene $i$. Then the normalized score for each gene is defined as

$$Z_{ij} = \left(\varepsilon_{ij} - v_j\right)\big/ S_j \qquad (3)$$

where $v_j$ is the mean and $S_j$ is the standard deviation of the scores from data source $j$. Finally, the combined score for one candidate gene is obtained by taking the maximum, sum or mean of the modified scores $Z_{ij}$ from different data sources.

Finally, CAESAR may be ineffective for those genes and traits with insufficient annotations and text description. To overcome this issue, this method could include other data sources, such as functional gene interaction and other species systems.

**PRIORITIZER**: Franke et al. [68] proposed a method named Prioritizer to discover disease genes from the susceptibility loci. They first compiled a functional human gene network that comprises known interactions derived from different databases, i.e., the Kyoto Encyclopedia of Genes and Genomes (KEGG) [69], the Biomolecular Interaction Network Database (BIND) [70], Reactome [71], and the Human Protein Reference Database (HPRD) [72] using a Bayesian classifier. Then they predicted some other functional relationships from GO and microarray data using the Bayesian framework, and constructed a human gene functional network. Finally, candidate genes were prioritized based on the shortest path distance to known disease genes in the functional network.

This method can be further improved by both the quality of the data sets making up gene networks and the efficiency of statistical methods incorporating the networks.

**TOM**: Rossi et al. [73] proposed a web-based system called Transcriptomics of OMIM (TOM) to predict novel disease genes by integrating gene expression data and GO data [50]. They used confirmed disease genes from OMIM [60] as seeds, and defined an expression neighborhood (a set of candidate genes with similar expression to the seeds). Next, they further filtered candidate genes based on their functional annotation in GO, with which they were able to extract the genes involved in the same or similar biological processes as the seeds and filter the other genes with different biological process. This statistically validated filtering allows the targeted extraction of a shortlist of candidate genes, thus saving resources for the following costly and time-consuming genetic analysis. However, this method may be ineffective for disease genes that are poorly characterized by GO and gene expression.

**Prediction based on Disease Protein Complex Method**: Genes sharing mutant phenotype are highly correlated in their biological functions [74]. The first work to integrate phenotype similarity was proposed by Lage et al. in 2007 [29]. For a disease phenotype, they found the most probable candidate genes by finding the most probable corresponding protein complexes, which comprised of the candidate genes and disease genes associated with similar disease phenotypes. The procedures are described in Figure 2.4, using Leber Congenital Amauros (LCA) as an example. The first step is to find protein complexes including the candidate genes. The protein complexes are named as the candidate complexes. In the second step, proteins known to be involved in similar disorders are identified in the candidate complexes. In this case, proteins that are involved in different disorders comparable to LCA are scored according to the phenotype similarities. The next step involves scoring and ranking the candidate complexes using the Bayesian Disease Gene Predictor as shown in Equation 4. Finally, the ranking of candidate genes are obtained according to their ranking within corresponding candidate complexes.

$$P(p_i = dis|DATA) = \frac{P(DATA|p_i=dis) \times P(p_i=dis)}{\sum_{j=1}^{N} P(DATA|p_j=dis) \times P(p_j=dis)} \tag{4}$$

where $P(dis = i|DATA)$ is the posterior probability that the protein $p_i$ is the disease-related protein after evaluating all the data. The $P(DATA|p_i = dis)$ is the probability of obtaining the data if the protein $p_i$ is disease related.

However, this method does not use actual protein complexes but simply assembles neighboring proteins as complexes (consisting of a protein and all their direct interaction partners).

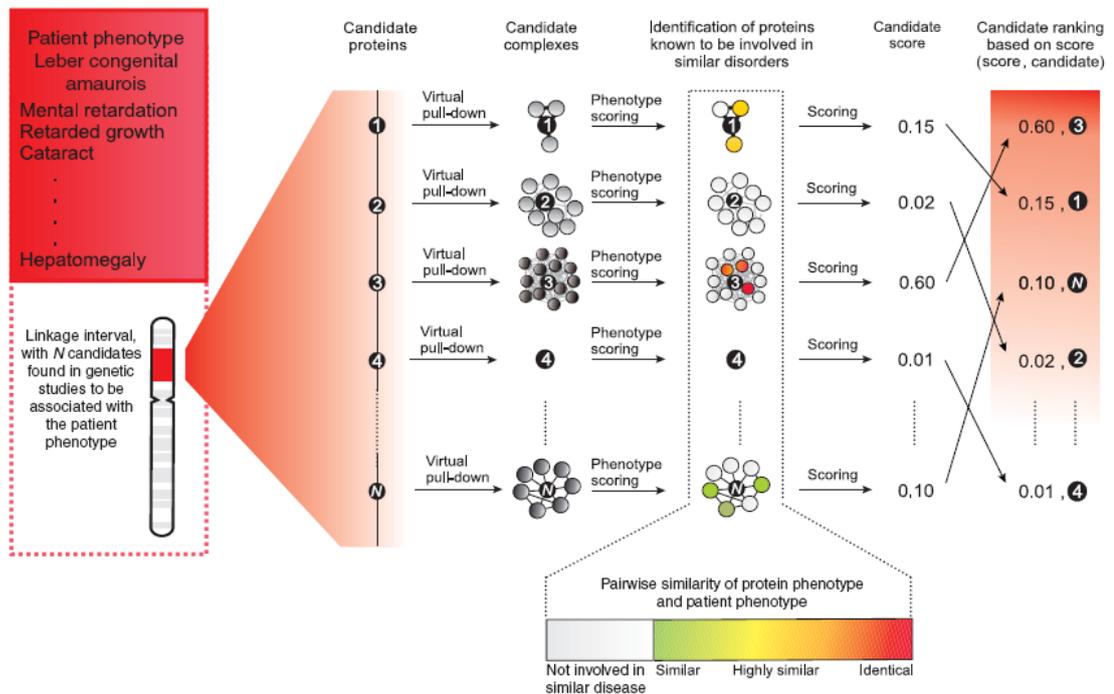

**Figure 2.4**: Steps in scoring each candidate in the linkage interval, taken from [29]: First, *N* candidate genes associated with the target disease phenotype (Leber congenital amaurois) are found within the linkage interval identified by linkage studies. Then, protein complexes including the *N* candidate genes are virtually pulled down. Third, all disease-related proteins in protein complexes are identified and are colored based on their phonotypical similarity to the target disease, Leber congenital amaurois. Finally, each candidate gene is scored according to phenotypes associated with the proteins in the candidate protein complex by the Bayesian predictor.

**CIPHER**: Wu et al. [27] proposed an integration method called CIPHER, which integrates the PPI network and the phenotype information obtained from OMIM.

Given a query phenotype and a set of candidate genes, CIPHER calculates a *concordance* score between the query phenotype and the candidate gene as shown in Figure 2.5. In the first step, it calculates a similarity profile of the query phenotype. The similarity profile is a numerical vector, consisting of the similarity scores between the query phenotype and all phenotypes. The similarity score between two phenotypes is calculated based on the topological distance between two sets of associated disease genes in the PPI network as shown in Equation 5:

$$S_{pp'} = C_p + \sum_{g \in G(p)} \left( \beta_{pg} \sum_{g' \in G(p')} e^{-L_{gg'}^2} \right) \tag{5}$$

where Gaussian kernel $e^{-L_{gg'}^2}$ is used to transfer gene-gene distance to gene-gene closeness. $C_p$ is a constant, and $\beta_{pg}$ is the coefficient of this regression model, respectively. In the second step, the closeness profile is calculated from a candidate gene to all the phenotypes. The closeness profile is a numerical vector, an element of which denotes the proximity from the candidate gene to a phenotype. It is calculated based on the topological distance between the candidate gene and the set of disease genes associated with the phenotype in the PPI network, as shown in Equation 6:

$$\phi_{gp'} = \sum_{g' \in G(p')} e^{-L_{gg'}^2} \tag{6}$$

where $L_{gg'}^2$ is the topological distance of two genes. Finally, a concordance score is calculated as the correlation between the similarity profile and closeness profile. The candidate genes are ranked based on corresponding concordance scores as

shown in Equation 7:

$$CS_{pg} = COV(S_p, \phi_g) / (\sigma(S_p)\sigma(\phi_g))$$  (7)

where $cov$ and $\sigma$ denote the covariance and standard deviation, respectively.

However, Wu's method is limited by the consideration of only small localized regions in both the protein interaction network and phenotype network. The global network analysis may provide ways to interpret the relationships between different diseases.

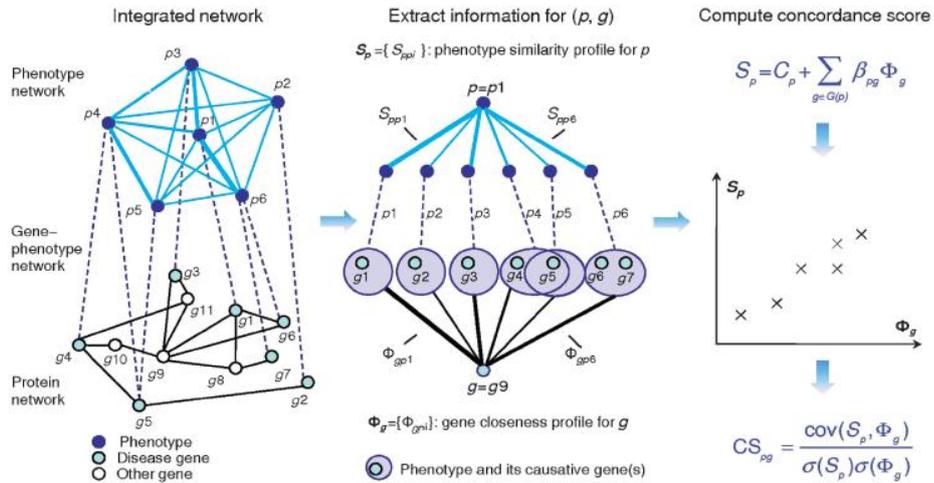

**Figure 2.5**: Scoring scheme of CIPHER, taken from [27]

**Multiple Kernel Learning**: One challenge in computational biology is to integrate heterogeneous biological datasets that are derived from various types of experimental data. To address this issue, Kernel-based methods are applied to represent each data source by means of kernel function, which defines similarities between pairs of genes, proteins and so on. Such similarities can be the relationships that capture the patterns of the underlying biological characteristics.

Therefore, the kernel-based method, Multiple Kernels Learning (MKL), has been successfully used to integrate heterogeneous data sources [75].

De Bie et al. [76] extended the MKL method to one-class classification problems. Similar to Lanckriet et al. [75], they represented each data source (including gene expressions, protein sequences and protein interactions) by means of a specific kernel matrix, which was called a gene functional similarity matrix $K_i$. Different kernel functions corresponded to different data sources and interpreted different notions of similarity. Then they combined multiple kernels into one in terms of linear combination $K = \sum_{i=1}^{m} \mu_i K_i$. To find the optimal linear discriminant, the SVM ranking model was trained based on the combined kernel, using semi-definite programming (SDP) [77].

**RWRH**: Li et al. [22] proposed a RWRH (Random Walk with Restart on Heterogeneous Network) algorithm to infer the gene-phenotype relationships as shown in Figure 2.6. It connects the protein interaction network and the phenotype network by gene-phenotype relationships to construct a heterogeneous network. It then extends the random walk with restart (RWR) algorithm to the heterogeneous network.

The RWRH algorithm is inspired by the co-ranking framework and ranks phenotypes and genes at the same time. It uses this algorithm to disclose the relationship between diseases. For one given disease, the disease phenotypes and disease genes are used as seed nodes to run the RWRH algorithm. Other phenotypes are ranked based on their relevance to the disease. On the other hand, the top ranked genes are used to identify disease associations. In the RWRH algorithm, two data sources are complementary to each other and reinforce each other.

The RWRH algorithm only relies on the protein interaction network and does not consider protein complexes, which are viewed as molecular machines that integrate multiple gene products to perform biological functions.

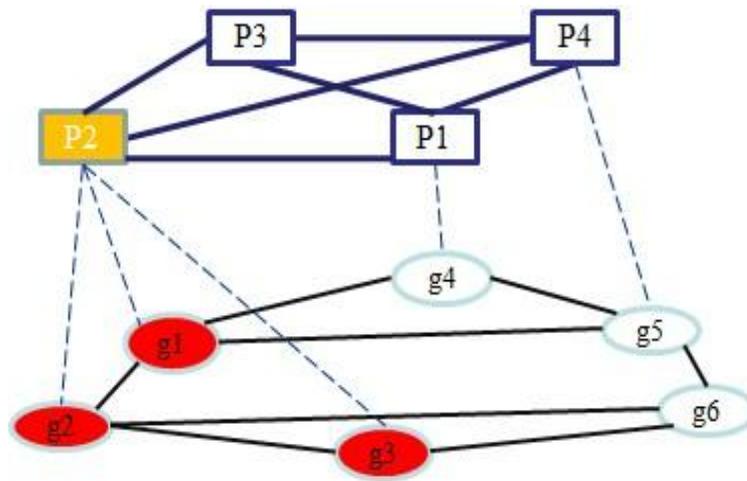

**Figure 2.6**: Illustration of Heterogeneous Network in RWRH, taken from [22]:

## 2.3   Summary

This chapter provides a literature review of the representative computational methods on disease gene prioritization, including prioritization algorithms using various biological data sources and ensemble strategies to integrate data sources and learning models.

Network propagation methods are exploited to prioritize disease genes based on the PPI network and phenotype similarity network [25] [22] [27]. However, these methods only focus on protein-level associations instead of complex-level associations. As proteins cannot function isolation, they are more likely to be attached together as functional modules (such as protein complexes and pathways) to perform biological functions. In Chapter Three, we build protein complex network model where nodes are individual complexes and the interactions between two complexes are measured by the connection strength between them. The experimental results show that the protein complex network model is able to discover gene-phenotype associations more effectively than protein-level models.

It should be noted that the above methods prioritize candidate disease genes based on gene rank scores and a threshold is needed to identify whether a specific gene is disease related or not. A more biologically meaningful approach would be to build a binary classification model that can automatically identify a gene as disease-related or not, according to various features of biological datasets, such as protein sequence [32] and PPI topological features. However, these machine learning methods typically treat the unknown genes as the negative set for building disease gene classifiers. Such a kind of classifiers do not perform very well because unknown genes may include unconfirmed disease genes. To address this issue, we have designed a novel PU learning algorithm in which we treat unknown genes as the unlabeled set instead of negative examples, as described in Chapter Four [34].

In addition, it should be pointed out that the above recent approaches only use a single learning model [34] [35] or a single biological dataset [32] [34] to identify candidate disease genes. However, ensemble-based approaches are more robust and reliable compared to the classification methods built on single biological datasets and learning models. In Chapter Five, we propose a novel ensemble positive unlabeled learning model for disease gene identification. The experimental results demonstrated that our proposed ensemble method significantly outperforms the three component learning models.

# Chapter 3.

# Predicting Disease Gene via Protein Complex Network Propagation

Recent studies have revealed that proteins associated with similar disease phenotypes have high probability of physical interactions between their products. And proteins cannot function alone, they are likely to be attached together to perform biological functions. In this chapter, we construct a novel human protein complex network by integrating human PPI network and CORUM protein complexes. We conduct a genome-wide disease gene prioritization for multi-factorial diseases using such a human protein complex network. Using our approach, the top ranked candidate disease genes that are found to be closely associating with protein complex can potentially be used to guide the prediction of disease-related protein complexes.

## 3.1   Introduction

Phenotypically similar diseases are found to be caused by functionally related genes, suggesting a modular organization of the genetic landscape of human diseases that mirrors the modularity observed in biological interaction networks. Protein complexes, as molecular machines that integrate multiple products to perform biological functions, express the underlying modular organization of protein-protein interaction network. As such, protein complexes can be useful for interrogating the networks of phenome and interactome to elucidate gene-phenotype associations of diseases.

We propose a technique called RWPCN (Random Walker on Protein Complex Network) for predicting and prioritizing disease genes in this chapter. The basis of RWPCN is a protein complex network constructed using existing human protein complexes and protein interaction network. To prioritize candidate disease genes for the query disease phenotypes, the associations between the protein complexes and query phenotypes are computed in their respective protein complex and phenotype networks. RWPCN is evaluated on predicting gene-phenotype associations and the method is observed to outperform existing approaches. We also apply RWPCN to predict novel disease genes for two representative diseases, namely, Breast Cancer and Diabetes.

Our proposed method is different from the existing methods as our network propagation algorithm is operated at the complex level instead of the protein level. We use reliable human protein complexes from the Comprehensive Resource of Mammalian protein complexes [78] since the protein complexes are curated from the biological literatures. To the best of our knowledge, this is the first attempt to present and exploit the biological modularity of the protein complexes and their relationships in an explicit way.

Guilt-by-association prediction and prioritization of disease genes can be enhanced by fully exploiting the underlying modular organizations of both the disease phenome and the protein interactome. As the protein complex network can reflect the underlying modularity in the biological interaction networks better than simple protein interaction networks, RWPCN is found to be able to detect and prioritize disease genes better than traditional approaches that used only protein-phenotype associations.

## 3.2 Method

In this section, we will first introduce the overall network structure for RWPCN algorithm, which includes the phenotype network, protein complex network, protein interaction network, as well as gene-phenotype associations. Then, we describe the construction of the phenotype network and protein complex network. With these, we then present the RWPCN algorithm to prioritize disease-related genes.

### 3.2.1 Overall Network Structure in RWPCN

Figure 3.1 depicts the overall network structure used in RWPCN. It consists of three levels of networks, namely, the phenotype network (top), protein complex network (middle), and the protein interaction network (bottom). In the phenotype network at the top level, we connect phenotypes using K-NN model (K-Nearest Neighbor). In Figure 3.1, the links are marked with blue lines, where the thicker lines denote higher phenotypic similarities.

The protein complex network in the middle layer is where phenotypically related protein complexes are connected. Within the protein complex networks, the links are marked with gray lines, with the thicker lines indicating stronger linkage strengths between the two corresponding protein complexes. We will describe how to compute the protein complexes' linkage strengths later. The links between the phenotypes and complexes capture the known gene-phenotype associations, denoted by dashed blue lines.

At the bottom level is the PPI network. Two proteins are connected if they are reported to be interacting to each other. Across the networks, each protein complex in the middle level links with all its component proteins (yellow nodes) in the PPI network.

Given a query disease phenotype (a query node in the top level), our objective is to predict disease genes for this phenotype in the bottom level PPI network, guided by the protein complex relationships in the middle level. Our proposed RWPCN algorithm will traverse between the three networks and exploit the structural relationships accordingly.

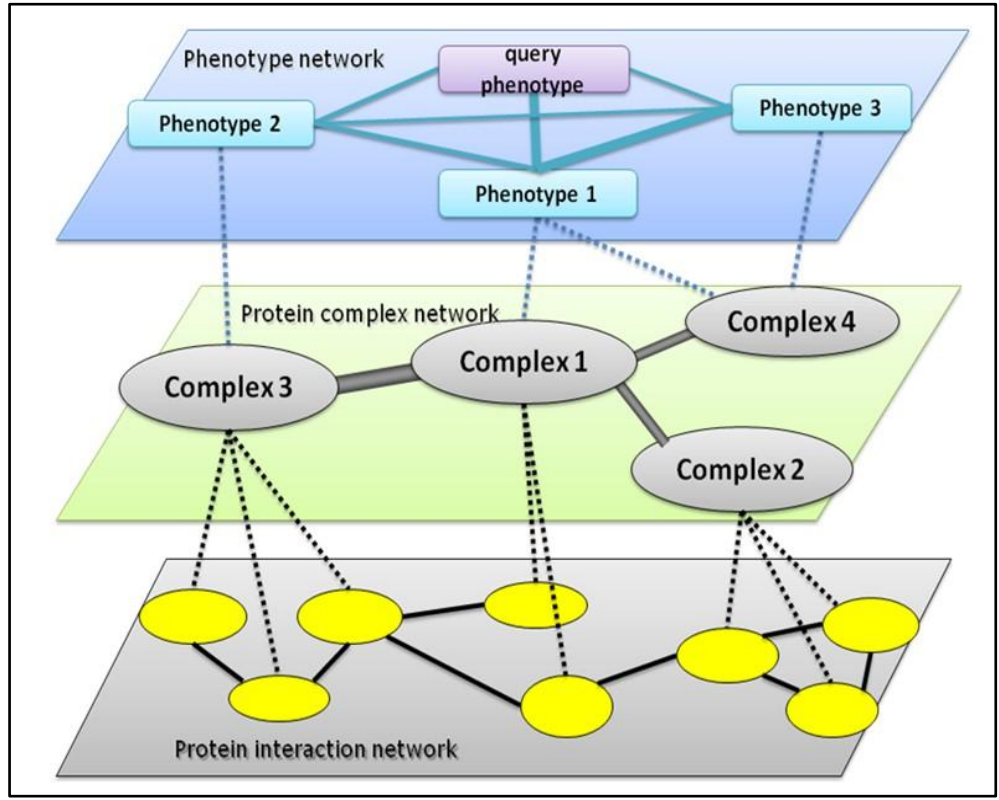

**Figure 3.1** Illustration of the overall network structure in RWPCN

## 3.2.2 Constructing Phenotype Network

Biologists already have a detailed knowledge of the phenotypes that are associated with each other. These phenotype associations have been used to prioritize candidate disease genes as well as to discover functional relations between genes and proteins [74].

Phenotype network is constructed using k-NN model (k-Nearest Neighbor). That is, for each phenotype $p_{ti}$, we compute its top $k$ most similar phenotypic neighbors (i.e. having the $k$ highest phenotypic similarities with $p_{ti}$) to link to it. We experimentally test the effects of different values of $k$ on the performance of the proposed algorithm, and set $k$ = 10 as the default value.

As recommended in [74], similarity values in the range [0, 0.3] are believed to be uninformative and noisy while those in [0.6, 1] are considered to be reliable. Therefore, we re-compute the phenotypic similarity between $pt_i$ and $pt_j$ using a logistic function $L(pt_i, pt_j) = 1/(1 + e^{C*sim(pt_i, pt_j)+d})$ used in [79]. We used the default values recommended in [79] for the parameters $c$ and $d$, namely $c$ = -15 and $d$ = log(9999) respectively.

### 3.2.3  Constructing Protein Complex Network

A PPI network (in the bottom level) is an undirected graph $G_{PPI}$ = ($V_{PPI}$, $E_{PPI}$), where $V_{PPI}$ is the set of nodes (proteins) and $E$ ={$(u,v)$| $u,v \in V_{PPI}$} is the set of edges (protein interactions). To construct protein complex network in the middle level, we need to collect known protein complex data or use some computational methods to predict protein complexes. For the former, we use the Comprehensive Resource of Mammalian protein complexes (CORUM) database [78], which is a collection of high quality experimentally verified mammalian protein complexes. However, the CORUM complex database is still far from complete and they are built from 2400 different genes, covering 12% of protein-coding genes in human [78]. As such, the protein complex set COM consists of a set of multi-protein complexes from CORUM (set $C_M$) as well as a set of *individual complexes* (set $C_I$) — namely those individual proteins that are not involved in any of the current CORUM complexes. As such, we have the following:

$$COM = C_M \cup C_I,$$ (10)

$$C_M = \{c_A | \ c_A \in \text{CORUM}, \ c_A \text{ is a complex}\} \tag{11}$$

$$C_I = \{\{p\} | \ \forall \ c_A \in \text{CORUM}, \ p \notin c_A, \ p \in V_{PPI}\} \tag{12}$$

Given the protein complex set *COM*, the protein complex network is defined as a directed super graph $G_{COM} = (V_{COM}, E_{COM})$, where the super node set $V_{COM} = COM$ denotes a set of protein complexes and $E_{COM} = \{(c_A, c_B) | \ c_A, c_B \in V_{COM}\}$ represents the set of links between protein complexes. Note that a link $(c_A, c_B) \in E_{COM}$ can be categorized into one of three types depending on the nature of complexes $c_A$ and $c_B$, namely, $E_{C2C}$ (*C2C* links between two multi-protein complexes), $E_{I2I}$ (*I2I* links between two individual complexes), and $E_{I2C}$ (*I2C* links between an individual complex and a multi-protein complex). Next, we describe how to assign weight for these three types of links.

Note that each complex $c_A \in C_M$ is a super node that can be represented as a graph $c_A = (V_{cA}, E_{cA})$ where the set $V_{cA}$ represents all the proteins in the complex $c_A$, and the set $E_{cA}$ represents the protein-protein interactions among the proteins in $V_{cA}$. Given two complexes $c_A = (V_{cA}, E_{cA})$ and $c_B = (V_{cB}, E_{cB})$, $c_A, c_B \in C_M$ , a *C2C* link between $c_A$ and $c_B$, $E_{C2C}(c_A, c_B)$ can be quantified as follows:

$$E_{C2C}(c_A, c_B) = \frac{\sum_{P_A \in V_{cA}, \ P_B \in V_{cB}, \ P_A, P_B \notin V_{cA} \bigcap V_{cB}} I(P_A, P_B)}{(|V_{cA}| - |V_{cA} \bigcap V_{cB}|) * (|V_{cB}| - |V_{cA} \bigcap V_{cB}|)} \tag{13}$$

where

$$I(P_A, P_B) = \begin{cases} 1, & \text{if } (P_A, P_B) \ \in \ E_{PPI} \\ 0, & \text{Otherwise} \end{cases}$$

$$\tag{14}$$

Basically, Equation (13) evaluates how closely the protein members from different complexes interact with each other. If there are a lot of physical interactions between the members from two complexes (non-overlapping proteins), then the two complexes are likely to be highly related as mutations of proteins in one protein complex could correspondingly disrupt the function of other complexes, thereby producing similar disease phenotypes. Note that according to equation (13), it is easy to know that $E_{C2C}(c_A, c_B) = E_{C2C}(c_B, c_A)$.

In the case that we have one multi-protein complex $c_A \in C_M$ and one *individual* protein complex $I_A \in C_I$, then the C2I link $E_{C2I}(c_A, I_A)$ and the *I2C* link $E_{I2C}(I_A, c_A)$ can be defined as follows:

$$E_{C2I}(\boldsymbol{c}_A, \boldsymbol{I}_A) = \frac{\sum_{P_A \in V_{c_A}} \boldsymbol{I}(\boldsymbol{P}_A, \boldsymbol{I}_A)}{|V_{c_A}|}, \quad E_{I2C}(I_A, c_A) = \frac{\sum_{P_A \in V_{c_A}} I(I_A, P_A)}{\deg(I_A)} \tag{15}$$

Finally, given two *individual* protein complexes $\boldsymbol{I}_A$ and $\boldsymbol{I}_B$ ($I_A, I_B \in C_I$), then the I2I link $E_{I2I}(I_A, I_B)$ and the *I2I* link $E_{I2I}(I_B, I_A)$ are computed as follows:

$$E_{I2I}(I_A, I_B) = \frac{1}{\deg(I_A)}, \quad E_{I2I}(I_B, I_A) = \frac{1}{\deg(I_B)} \tag{16}$$

where $deg(I_A)$ is the number of neighbors of vertex $I_A$.

### 3.2.4   Random walk with restart on the protein complexes network (RWPCN)

We are now ready to present our proposed algorithm. Given a query phenotype $pt_i$,

the aim is to prioritize the candidate disease genes based on the known disease genes which are associated to $pt_i$'s similar phenotypic neighbors in the phenotype network.

## Initialization of seed genes and complexes

Let $\boldsymbol{N(pt_i)}$ represents the $k$-NN phenotype neighbor set of the query phenotype $\boldsymbol{pt_i}$ where each $\boldsymbol{pt_j} \in \boldsymbol{N(pt_i)}$ is similar to $\boldsymbol{pt_i}$. Let $\boldsymbol{dis(pt_i)}$ be the set of causative genes of the phenotype $\boldsymbol{pt_i}$. We define the seed disease gene set with respect to $\boldsymbol{pt_i}$ as

$S = \bigcup\limits_{pt_j \in N(pt_i)} dis(pt_j)$. For a seed disease gene s $\in S$, we assign to it a score

$seed(s, pt_i) = \sum\limits_{s \in dis(pt_j)} L(pt_j, pt_i)$. Given a phenotype $\boldsymbol{pt_i}$ and the score for its seed

gene set $seed(s, pt_i)$, the protein complex $c_A$ can be scored as follows:

$$F(c_A, pt_i) = density(c_A) * \sum\limits_{s \in V_{c_A}} seed(s, pt_i) \qquad (17)$$

$F(c_A, pt_i)$ denotes initial score of the protein complex $c_A$ with respect to $pt_i$.

The density of a graph $G=(V_G, E_G)$, denoted as *density* ($G$), quantifies the richness of edges within $G$ and it is defined as shown in equation (18):

$$density(G) = \frac{2*|E_G|}{|V_G|*(|V_G|-1)} \qquad (18)$$

Note that $0 \leq density$ ($G$) $\leq 1$. If *density* ($G$) $= 1$, then $G$ is a complete graph, which means every pair of distinct vertices in $V_G$ is connected by an edge. As each protein complex can be viewed as a graph, we apply *density($C_A$)* to quantify the richness of protein interactions within $C_A$.

## Propagating the seeds' influence to the complexes in the whole network

The Random Walker algorithm [80] is applied to the protein complex network. First, the seed protein complexes are each assigned a score with respect to the query phenotype if they contain the genes in the seed disease gene set. Then all the protein complexes are scored in *COM* by propagation. We propose to do flow propagation for this. The disease influence flows initialized in seed complexes are distributed and pumped from seed complex vertices to their neighboring complexes in the network. These super vertices will then spread the influence flows received from previous iteration to their neighbors.

Formally, let $F_0$ be a vector of the initial probabilities of all the protein complexes in the protein complex network computed using equations (13-16). $F_{r-1}$ denotes the vector after $r$-1 iterations. The probability vector at step $r$, $F_r$, can be calculated by equation:

$$F_r = (1-\alpha)W^{'}F_{r-1} + \alpha F_0, r \geq 2, \qquad (19)$$

where $F_1 = F_0$, and $W'$ is the column normalized form, the transpose matrix of adjacency matrix $W$ which is the transition matrix of the whole protein complex interaction network. We construct matrix $W$ based on the three different links between protein complexes. Recall that the protein complex set *COM* consists of both multi-protein complexes ($C_M$) and individual complexes ($C_I$). The matrix $W$ is thus defined as:

$$W = \begin{pmatrix} A_{C2C\,(n*n)} & A_{C2I\,(n*m)} \\ A_{I2C\,(m*n)} & A_{I2I\,(m*m)} \end{pmatrix}$$

(20)

where $A_{C2C\,(n*n)}$, $A_{C2I\,(n*m)}$, $A_{I2C\,(m*n)}$ and $A_{I2I\,(m*m)}$ are the adjacency sub-matrices. In particular, $A_{C2C\,(n*n)}$ represents the sub-network links between multiple-protein complexes, $A_{C2I\,(n*m)}$ represents the sub-network links from multi-protein complexes to individual complexes, $A_{I2C\,(m*n)}$ represents the sub-network links from individual complexes to multi-protein complexes, and $A_{I2I\,(m*m)}$ represents the sub-network links between individual protein complexes respectively, where $n=|C_M|$ and $m=|C_I|$ are the numbers of multi-protein complexes and individual complexes respectively.

Note that in Equation (19), the parameter $\alpha \in (0,1)$ provides a probabilistic weighting of spreading the prior information of the seed complex vertices to other protein complexes at every step. $\alpha$ is set as 0.8 in the experiments. At the end of the iterations, the prior information held by every vertex in protein complex network will reach a steady state. This is determined by the probability difference between $F_r$ and $F_{r-1,}$ represented as Dif=$|F_r - F_{r-1}|$ (measured by $L1$ norm). When Dif=$|F_r - F_{r-1}| <= 10^{-10}$, as suggested in Li *et al.* [22], we consider that a steady stage has been reached and stop the iterative process. Note that the function $F$ is smooth over the whole protein complex-complex network, and each vertex complex is assigned a value to represent its association with the disease phenotype of interest.

## Scoring disease gene based on associations of protein complexes to diseases

Once the vector $F_r$ reaches a steady state, we obtain the final scores of protein complexes with respect to query phenotype. Recall that the final objective of our algorithm is to prioritize the candidate disease genes amongst the genes in the $G_{PPI}$. The final step is therefore to prioritize the candidate disease genes based on their associations with protein complexes. Given a candidate gene $g$, its association with query phenotype $pt_i$, denoted by $S(g, pt_i)$, is computed as

$$S(g, pt_i) = \sum_{g \in c_A} F_r(c_A, pt_i) \qquad (21)$$

where $C_A$ is the set of complexes containing the gene $g$. Especially, mutations on the genes shared by multiple protein complexes may lead to multiple similar phenotypes, so scores of these shared genes should be the accumulated score of protein complexes that contain them.

The detail of the process of RWPCN is listed in Algorithm 3.1.

Algorithm 3.1 Random Walk on Protein Complex Network (RWPCN)

Input:

  $W(A_{(m+n)} \times A_{(m+n)})$ //the adjacency matrix of the protein complex network

  $F_0$ // the initial probability vector ($A_{(m+n)} \times 1$)

  $K$: // the number of direct neighbors of one phenotype entry

Output: RS:

  1.  Calculate the density of protein complexes using Eq. 18

2. Initialize the probability vector $F_0$ using Eq. 17

3. $r = 2; F_{r-1} = F_0; \delta = 1;$

4. While $\delta \geq 10^{-10}$ do

5. $\quad F_r = (1 - \alpha)W'F_{r-1} + \alpha F_0;$ // the random walk with restart algorithm

6. $\quad \delta = \sum_{i=1}^{m+n}(|F_r(i) - F_{r-1}(i)|);$

7. $\quad r = r + 1;$

8. end while

9. $F_f = F_r;$

10. Calculate the score of candidate genes based on $F_f$;

11. $RS(g, pt_i) = \sum_{g \in A_j} F_f (A_j, pt_i)$

12. Return $RS$;

## 3.3 Experiment Results

In this section, we will first describe the experimental data used. Then the experimental settings and evaluation metrics will be introduced. Finally, we present the experimental results compared to the state-of-the-art techniques.

### 3.3.1 Experimental settings and evaluation metrics

The objective is to uncover novel gene-phenotype relationships. In order to compare different techniques, we employ standard leave-one-out cross-validation in the experiments. Each known gene-phenotype association ($g$, $p$) is employed as one test case where the phenotype $p$ is the query phenotype and the gene $g$ is the test disease gene. In each round of cross-validation test, we will first intentionally remove the association ($g$, $p$) from our data. Then the proposed algorithm is run to score the genes based on their associations with protein complexes with respect to the query phenotype $p$. If the test disease gene $g$ is ranked as top 1, we will consider it as a successful prediction; otherwise it is a failed case. We use the number of overall successful predictions to evaluate the performance of different prediction methods. Depending on the genes involved in the ranking, we further categorize our evaluation metrics into the following two classes, namely, *whole genome* evaluation and *ab initio* evaluation [27]. *Whole genome* evaluation basically ranks all the genes to scan for disease genes, e.g. we can consider all HPRD genes which do not link to the query phenotype (exactly identical setting as RWRH [22]) and check how many known test disease genes are still ranked as top 1 in the cross-validation test. However, there are no causative genes for half of the OMIM phenotypes [81]. *Ab initio* prediction has been proposed to identify disease genes without any known disease genes for those query phenotypes [27]. For each phenotype entity, the gene-phenotype associations are removed from this phenotype $p$ to all of its known causative genes[1] and we can only use the other disease genes associated with $p$'s neighbor phenotypes as the seed disease gene set. If one of the known causative genes (assuming $p$ is associated with multiple disease genes) related to the phenotype $p$ is ranked top 1, we consider it a successful prediction. Noted in our experiments, the same experimental data and evaluation metrics have been consistently used to evaluate all the prediction techniques.

---

[1]The difference between the *whole genome* evaluation and *Ab initio* evaluation lies in that for the *whole genome* evaluation, I only remove one phenotype gene association each time, but for *Ab initio* evaluation, multiple phenotype gene associations may be removed if the phenotype is associated with more than one causative genes.

## 3.3.2   Experimental Results

In this section, we first compare our algorithm with two state-of-the-art techniques, namely, CIPHER-DN (CIPHER with the topological distance feature of Direct Neighbor) [27] and RWRH [22]. Next, the sensitivities of the parameters are tested in our proposed method. For discussion, we present a case study of predicting disease genes for two representative diseases i.e., Breast cancer and Diabetes. Finally, the scores for protein complexes are computed to discover if the protein complexes are disease related.

### Comparison with CIPHER-DN and RWRH

We compare the performance of RWPCN algorithm with current computational techniques, namely, CIPHER-DN and RWRH, using the two evaluation metrics presented above, namely, *whole genome* evaluation and *ab initio* evaluation. Table 3.1 shows the overall comparison results of different algorithms. In terms of *whole genome* evaluation (second column in Table 3.1), we observe that the proposed RWPCN is able to achieve the best result, successfully predicting 253 genes, which are 8 and 88 more genes predicted than RWRH and CIPHER-DN respectively. In terms of *ab initio* evaluation (third column in Table 3.1), the RWPCN is able to predict 226 disease genes successfully, which are 25 and 69 more than the RWRH and CIPHER-DN respectively.

Note that in the original CIPHER-DN paper [27], the authors have adopted a less strict evaluation metric for *ab initio* evaluation than mine. As long as the target gene was ranked among the top $N$ (*instead of the top 1*), it was regarded as a successful prediction where $N$ ($N>=1$) denoted the number of known disease genes for the query phenotype. Using this less stringent evaluation metric, our method predicts 240 genes successfully while CIPHER-DN could only predict 157 genes in the *ab initio* evaluation.

**Table 3.1**: Overall performance of RWRH, CIPHER-DN and RWPCN algorithm

| Algorithm | Whole genome evaluation | Ab initio evaluation |
|-----------|------------------------|---------------------|
| RWPCN     | 253                    | 226                 |
| RWRH      | 245                    | 201                 |
| CIPHER-DN | 165                    | 157                 |

In the evaluations above, we have used the standard (but old) gene-phenotype association data which were also used in [27] [22] for comparison. To further validate the predicted associations, we collect a new version of gene-phenotype association data extracted from OMIM recently [82]. It contains 1614 gene-phenotype associations, which includes 274 novel gene-phenotype associations where the disease genes were unknown in the previous version (other 1340 associations are shared by both versions). Table 3.2 shows that using the new gene-phenotype association data, RWPCN successfully ranks the 273 (a sensitivity of 0.169) genes as top 1 in terms of *whole genome* evaluation, and 247 (a sensitivity of 0.153) in terms of *ab initio* evaluation, which are only slightly lower than the results of BIOMART in *whole genome* (a sensitivity of 0.177) and *ab initio* (a sensitivity of 0.158) respectively, indicating our method is certainly capable of detecting the novel knowledge which are absent in the older reference data.

We have constructed three levels of networks in our model. It is thus not strange that our method needs more computations. We may explore how to improve the efficiency of our algorithm in our future work, using better data structures and specific libraries for matrix operation, etc.

**Table 3.2**: Overall performance of BIOMART06, 09 and 06+09 phenotype-gene

| Phenotype-gene data | Whole genome evaluation | Ab initio evaluation |
|---|---|---|
| BIOMART06 | 253 | 226 |
| BIOMART09 | 273 | 247 |
| BIOMART06+09 | 285 | 253 |

## Effect of parameters $\alpha$ and $k$ in RWPCN

Recall that we have two parameters $\alpha$ and $k$ in RWPCN algorithm. The flow parameter $\alpha$ is used in our RWPCN algorithm to control the proportion of information that flows back into the seed nodes/protein complexes at each iteration of the algorithm. A larger $\alpha$ represents that information flows are likely to return to the seed nodes, therefore those protein complexes near to seed nodes are more likely to be ranked forward. On the contrary, a smaller $\alpha$ represents that information flows are likely to flow out of the seed nodes, therefore those protein complexes near to seed nodes are more likely to be ranked backward. The second phenotype parameter $k$ decides the number of related phenotypes with regard to the query phenotype. An unnecessarily large $k$ will include many related phenotypes which are not relevant while a smaller *k* will include smaller number of related phenotypes and may miss out some important relevant phenotypes as a result.

We first investigate how the flow parameter α affects the performance of the algorithm. We run our algorithm using leave-one-out cross-validation with values of $\alpha$ ranging from 0.2 to 0.9 in steps of 0.1, while keeping the phenotype *k* fixed to 10. The performance of the algorithms is measured using *whole genome* evaluation and *Ab initio* evaluation mentioned above. The results are shown in Figure 3.2.

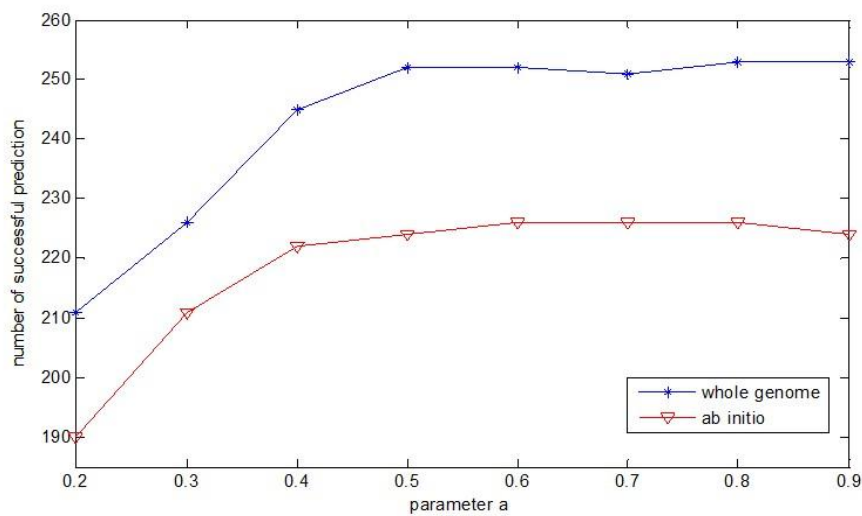

**Figure 3.2**: Effect of value  α  based on *whole genome* and *ab initio* evaluation

With increasing values of  α, we are able to obtain increased numbers of successful predictions for both *whole genome* evaluation and *ab initio* evaluation. This is expected since the seed nodes in protein complex network are more likely to hold the information flows, thus few flows will be distributed to the distant neighbors in the network. Biologically, this is reasonable since the protein complexes (and the corresponding proteins in the complexes) that directly interact with the disease complexes/proteins are more likely to be disease/phenotype related. We observe that the performance of RWPCN with $\alpha >= 0.4$ are better than the existing CIPHER-DN and RWRH algorithms. In fact, we find that the optimal values of $\alpha$ can be found within a large range of $0.5 <= \alpha < = 0.9$. As such, selecting a suitable value for $\alpha$ for good performance is not a problem.

To study the effect of the parameter $k$ that decides the number of related phenotypes, we run RWPCN with $k$ varying from 7 to 12 and $\alpha = 0.8$, based on *whole genome* and *ab initio* evaluations. Results are shown in Figure 3.3. The performance of RWPCN algorithm is improved with increased value of $k$ from 7 to 10, indicating that incorporating more related phenotypes is helpful for prioritizing target disease genes. However, if we further include more phenotypes (e.g. $k>10$) with low phenotypic similarities, noisy and un-meaningful phenotypes will be included [74] and eventually affects the performance of disease gene prediction. For example, the results in Figure 3.3 show that the performance with $k$ in the range of [11, 12] has worsened. Nevertheless, the performance of RWPCN algorithm with $k$ in the wide range [7, 12] is consistently better than that of RWRH, suggesting that RWPCN is insensitive to the specific values of $k$ as far as comparison with RWRH is considered.

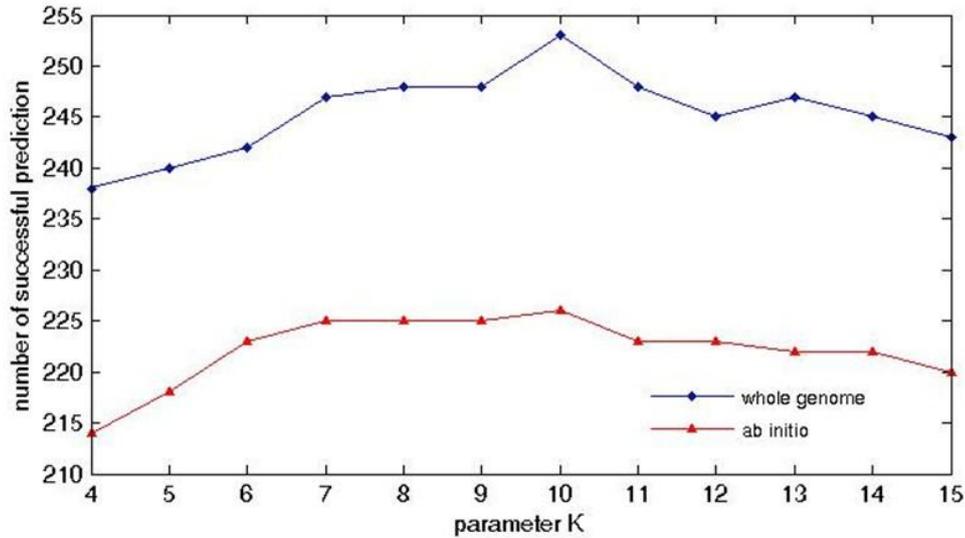

**Figure 3.3**: KNN phenotype network on *whole genome* and *ab initio* evaluation.

## Inferring novel causal genes for breast cancer and diabetes

We also apply our method for uncovering novel candidate genes on specific complex genetic diseases. We have chosen Breast Cancer (MIM: 114480) and Diabetes Mellitus type 2 (MIM: 125853) for our case study here.

We run the RWPCN algorithm (with $k$=10 and $\alpha = 0.8$) for Breast Cancer and Diabetes Mellitus type 2. Note that no gene-phenotype associations are removed since the aim is to predict the disease genes instead of cross-validation. We rank the resulting candidate genes over the whole genome and selected the top 20 ranked genes associated with target phenotypes (Breast Cancer and Diabetes Mellitus type 2). The experimental results are listed in Tables 3.3 and 3.4 for Breast Cancer and Diabetes Mellitus type 2 respectively.

Genes marked with * are known disease genes associated with target phenotype, genes marked with √ are genes associated with target phenotype either extracted from literature or from database, genes marked with ～ are un-related to target phenotype.

Table 3.3: Breast cancer genes prediction

| Rank | Score | HGNC Gene symbol | Mark |
|------|-------|------------------|------|
| 1 | 3.61665 | BRCA1 | * |
| 2 | 2.64458 | RBBP8 | ✓ |
| 3 | 1.04115 | HDAC1 | ✓ |
| 4 | 1.02108 | HDAC2 | ✓ |
| 5 | 1.00632 | CTBP1 | ✓ |
| 6 | 0.983392 | LMO4 | ✓ |
| 7 | 0.814445 | RAD51 | * |
| 8 | 0.812762 | BRCA2 | * |
| 9 | 0.807072 | NBN | * |
| 10 | 0.806886 | BRIP1 | * |
| 11 | 0.801356 | PIK3CA | * |
| 12 | 0.671104 | ZNF350 | ✓ |
| 13 | 0.142519 | SMAD3 | ✓ |
| 14 | 0.141945 | ELAC2 | ✓ |
| 15 | 0.141729 | RNASEL | ✓ |
| 16 | 0.140748 | PTEN | ✓ |
| 17 | 0.0947266 | TP53 | ～ |
| 18 | 0.0849672 | SMAD4 | ～ |
| 19 | 0.0831955 | EP300 | ～ |
| 20 | 0.0721527 | CREBBP | ～ |

Table 3.3 shows six highly ranked genes that are also known to associate with the Breast Cancer. However, we are more interested in investigating whether the predicted *novel* susceptible genes are also associated with the disease phenotypes. We search for additional gene-phenotype associations from GENECARDS database [83] and also perform literature search from Pubmed on the other susceptible genes predicted by the algorithm to be associated with the disease phenotypes of breast cancer. Eight additional genes are found, namely RBBP8, HDAC1, HDAC2, CTBP1, LMO4, ZNF350, SMAD3, ELAC2, RNASEL and PTEN that are also reported to be related to the Breast Cancer. For CtIP (also known as retinoblastoma binding protein 8, RBBP8, ranked at top 2), the expression of this gene has been shown to be a novel mechanism for tamoxifen resistance development in breast cancer [84]. HDAC1 and HDAC2 (ranked at top 3 and 4), among class I HDACs, are reported to regulate the changes in histone acetylation and are associated with HDAC inhibitors that are expected to reverse hypoacetylation levels observed even at the early stages of breast cancer progression [85]. CtBP1 (ranked at top 5) is confirmed to be associated with breast cancer and its activation has a potential impact in breast cancer development [86]. LMO4 (ranked at top 6) is a novel cell cycle regulator with a key role in mediator of ErbB2/HER2/HER2/Neu-induced breast cancer cell cycle progression [87]. Genetic variants and haplotype analyses of the ZNF350 (ranked at 12) gene suggest that it is associated with high-risk non BRCA1/2 French Canadian breast and ovarian cancer families [88]. Germ line mutation in RNASEL (ranked top 15) predicts increased risk of breast cancer [89]. SMAD3 (ranked at top 13) has critical roles in stimulation of breast cancer growth and metastasis [90]. Finally, Tsou HC et al. [91] reported three novel MMAC1/PTEN (ranked at 16) mutations in CS (Cowden syndrome) are associated with breast cancer. All these show that our prediction method can discover novel disease genes for breast cancer beyond the original disease gene set.

In the table list of candidate genes unmatched with breast cancer, TP53 is direct neighbor of confirmed disease gene BRCA1, two suspicious disease genes, HDAC1 and HDAC2. SMAD4 is directly linked to disease gene BRCA1, and share identical neighbor EP300 with confirmed disease gene NBN. CREBBP has protein interactions to confirmed disease genes BRCA1 and suspicious disease gene HDAC1.

**Table 3.4**: Diabetes genes prediction

| Rank | Score | HGNC Gene symbol | Mark |
|------|-------|------------------|------|
| 1 | 1.34591 | PIK3R1 | ✓ |
| 2 | 1.33691 | IRS1 | * |
| 3 | 1.33691 | INSR | * |
| 4 | 1.33691 | KHDRBS1 | ✓ |
| 5 | 0.821847 | NEUROD1 | * |
| 6 | 0.812877 | IPF1 | * |
| 7 | 0.810154 | SLC2A4 | * |
| 8 | 0.802705 | MAPK8IP1 | * |
| 9 | 0.802453 | TCF2 | * |
| 10 | 0.802404 | PPP1R3A | * |
| 11 | 0.354724 | TCF1 | ∼ |
| 12 | 0.194629 | CREBBP | ∼ |
| 13 | 0.15557 | EP300 | ✓ |
| 14 | 0.102423 | PCAF | ∼ |
| 15 | 0.0807789 | PLN | ∼ |
| 16 | 0.0806853 | RPS6KA1 | ∼ |
| 17 | 0.0652625 | CUL3 | ∼ |
| 18 | 0.0652625 | SPOP | ∼ |
| 19 | 0.0595811 | POLR2A | ∼ |
| 20 | 0.0471911 | ABCC8 | ✓ |

Table 3.4 shows our prediction results for Diabetes Mellitus type two. Out of the top 20 predicted disease genes, eight genes are known to associate with the phenotype. We find three additional genes PIK3R1, EP300 and ABCC8 to be related to the disease phenotypes. PIK3R1 (ranked as top 1) has been tested for their influence on insulin action, showing significant associations with diabetes [92]. KHDRBS1 (ranked at top 4, aliases SMA68) is reported that its RNA binding protein is a potential target to treat diabetes and obesity [93]. EP300 (ranked at top 13, aliases p300), as a transcriptional coactivator, can cause diabetes via regulating fibronectin expression via PARP and NF-kappaB activation [94]. For ABCC8, a rare mutation in ABCC8/SUR1 (ranked at top 20) has been reported to have an effect on K(ATP) channel activity and beta-cell glucose sensing, leading to diabetes in adulthood [95].

In the table list of candidate genes unmatched to diabetes mellitus type 2, TCF12 and CREBBP have protein interactions to suspicious disease gene EP300. PLN and RPS6KA1 are directly interacted to confirmed disease gene PPP1R3A at molecular level. CUL3 and SPOP are involved in same protein complex. This complex and POLR2A share identical protein interaction neighbor with suspicious disease gene EP300.

From Tables 3.3 and 3.4, we find our predicted disease genes indeed mapped significantly with disease genes that are either curated in existing database or reported in the literature. Though unmatched genes are not associated with any positive evidences in current databases or literatures, they are closely interacted with confirmed disease genes in protein-protein interaction network or involve in protein complexes containing confirmed disease genes. As such, they are good candidates for biologists and clinicians to do experiments for validation.

To investigate the significance of our top selected candidate genes, two disease gene prioritization approaches are applied to prioritize novel disease genes associated with breast cancer and diabetes mellitus type 2. One is random walk with restart (RWR) that was used in [25], who run the algorithm on PPI network without considering protein complex information. The other one is random selection, in which we randomly permutate all the genes and select five groups of top 20 candidate genes for breast cancer and diabetes mellitus type 2 respectively. Same measurement has been proposed to five groups of 20 genes to evaluate the association to target phenotypes. We compare top 20 candidate genes selected by our method with genes that are predicted by RWR and random selection, and report the result in the table 3.5. The results show that our method is better than RWR and random selection on breast cancer and diabetes gene prioritization.

**Table 3.5**: Comparison with random selected genes

| disease / method | breast cancer (114480) | | | diabetes mellitus type 2 (125853) | | |
|---|---|---|---|---|---|---|
| | confirmed | suspicious | unmatched | Confirmed | suspicious | unmatched |
| RWPCN | 6 | 10 | 4 | 8 | 4 | 8 |
| RWR | 6 | 5 | 9 | 8 | 2 | 10 |
| Random | 0 | 2.8 | 17.2 | 0 | 0.6 | 19.4 |

## Detecting disease-related protein complexes

Recall that we have assigned scores to the protein complexes to indicate the degree of association of the protein complexes to the query disease phenotypes. Protein complexes assigned high scores indicate strong associations to corresponding phenotypes. Based on the scores, we have ranked the protein complexes and studied the top two complexes here: sarcoglycan-sarcospan complex (SG-SPN) and Pex26-Pex6-Pex1 complex. For evaluation, a set of 248 disease protein complexes from Lage *et al.* [29] are used as our benchmark.

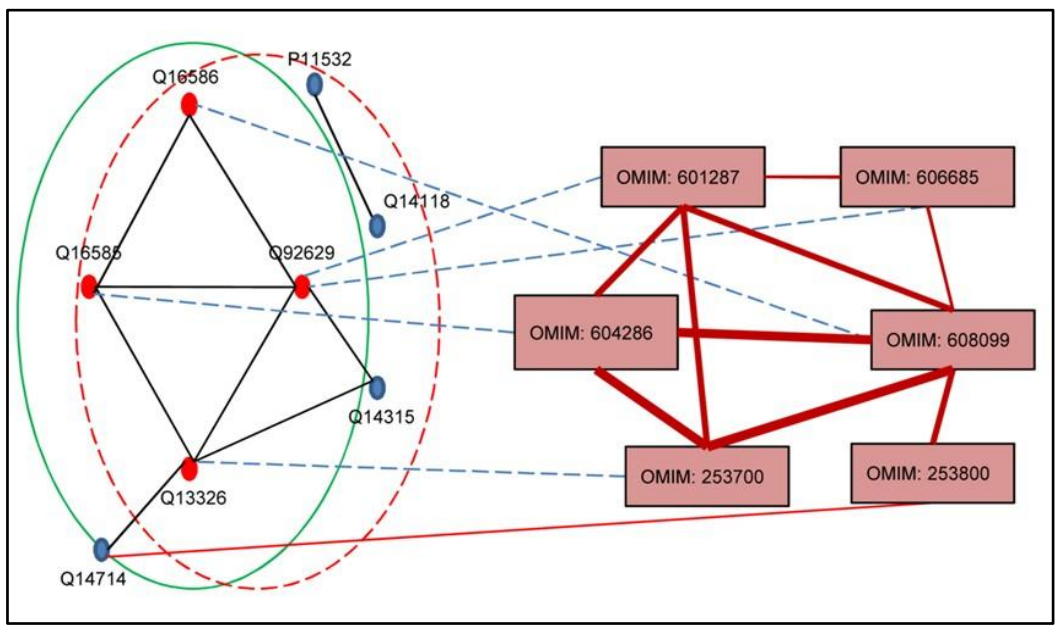

**Figure 3.4**: SG-SPN overlaps with the disease complex No. 230.

Figure 3.4 shows that the SG-SPN complex (surrounded by green line) contains five human proteins: Q16586, Q16585, Q92629, Q13326, Q14714, and it rank at top 1 protein complex for phenotype OMIM: 608099 by RWPCN algorithm. We find that this SG-SPN complex has a large overlap (shared four proteins) with the disease complex No. 230 (surrounded by red dash line) in our benchmark set. We also find that the shared four proteins are linked to disease phenotypes (blue dash links), which have high phenotypic similarity among them. Note that gene Q14714 (SSPN) in SG-SPN complex is associated with phenotype Fukuyama Congenital Muscular

Dystrophy (FCMD) (MIM: 253800) [96] which is closely related to phenotype OMIM: 608099 in our phenotype network, indicating that SG-SPN complex could indeed be a valid disease complex.

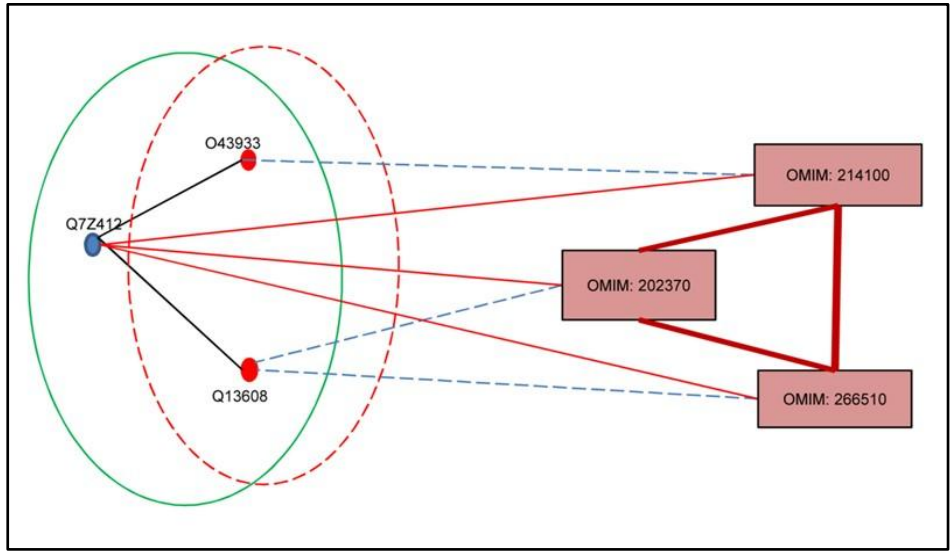

**Figure 3.5**: Pex26-Pex6-Pex1 overlaps with the disease complex No. 335.

Similarly, Figure 3.5 shows the Pex26-Pex6-Pex1 complex (surrounded by green line) which covers a benchmark disease complex (surrounded by red dash line) that consists of proteins O43933 (PEX 1) and Q13608 (PEX 6). The Pex26-Pex6-Pex1 complex is involved in peroxisome biogenesis disorders (PBDs), which includes the Zellweger syndrome spectrum (PBD-ZSD) and rhizomelic chondrodysplasia punctata type 1 (RCDP1). PBD-ZSD represents a continuum of disorders including infantile Refsum disease (MIM: 266510), neonatal adrenoleukodystrophy (MIM: 202370), and Zellweger syndrome (MIM: 214100). Note that the Q7Z412 (PEX 26) protein in our predicted disease complex is also a known disease gene associated with all the three phenotypes, suggesting that the mutations of proteins in the same CORUM protein complexes are likely to induce the same or similar phenotypes. It also shows that our highly ranked protein complexes are indeed disease related.

## Disease gene modules in PAGES database

While RWPCN is significant for complementing the weaknesses of individual molecular interaction networks, it relies on the human protein complex interaction network. Therefore the spotty coverage of the protein complex data can affect the performance of prediction. To increase the coverage, we extract "disease gene modules" in the PAGED database [97].

We utilize PAGED disease gene modules in our RWPCN algorithm to uncover novel candidate genes on Breast Cancer (MIM: 114480) and Diabetes Mellitus type 2 (MIM: 125853) for our case study here.

We renew our protein complex network with additional PAGED disease modules and then run the RWPCN algorithm for Breast Cancer and Diabetes Mellitus type 2. Following experimental setting in Table 3.3 and 3.4, the top 20 ranked genes are selected to assess our algorithm performance. The experimental results are listed in Tables 3.6 and 3.7 for Breast Cancer and Diabetes Mellitus type 2 respectively.

Genes marked with *, $\sqrt{}$ and $\sim$ are represented as confirmed disease genes, genes with literature support and un-related genes to target phenotype.

**Table 3.6**: Breast cancer genes prediction using PAGED dataset

| Rank | HGNC Gene symbol | Score | Mark |
|------|------------------|-------|------|
| 1 | BRCA1 | 4.32062 | * |
| 2 | EP300 | 4.31291 | ∼ |
| 3 | CASP8 | 4.18623 | * |
| 4 | FANCC | 3.54481 | ✓ |
| 5 | FANCG | 3.54481 | ✓ |
| 6 | FANCF | 3.54481 | ✓ |
| 7 | FANCA | 3.54481 | ✓ |
| 8 | ERCC1 | 3.38491 | ∼ |
| 9 | ERCC4 | 3.38491 | ✓ |
| 10 | ESR1 | 3.36958 | * |
| 11 | NCOA3 | 3.31292 | ✓ |
| 12 | RBBP8 | 3.1943 | ✓ |
| 13 | FADD | 3.073 | ✓ |
| 14 | CREBBP | 3.05405 | ∼ |
| 15 | SMAD3 | 2.93813 | ✓ |
| 16 | KAT2B | 2.82028 | ✓ |
| 17 | PARD3 | 2.54331 | ∼ |
| 18 | RPA1 | 2.54269 | ✓ |
| 19 | RPA2 | 2.54269 | ✓ |
| 20 | RPA3 | 2.54269 | ∼ |

Table 3.6 shows three highly ranked genes that are also known to associate with the Breast Cancer. We search for additional gene-phenotype associations from GENECARDS database and also perform literature search from Pubmed on the other susceptible genes predicted by the algorithm to be associated with the disease phenotypes of breast cancer. 12 genes are associated with breast cancer in our literature research, which are better than predicted results on CORUM in Table 3.3.

**Table 3.7**: Diabetes genes prediction using PAGED dataset

| Rank | HGNC Gene symbol | Score | Mark |
|------|------------------|-------|------|
| 1 | PIK3R1 | 1.75811 | ✓ |
| 2 | IRS1 | 1.56845 | * |
| 3 | INSR | 1.56845 | ✓ |
| 4 | KHDRBS1 | 1.56845 | ✓ |
| 5 | HNF1A | 1.15679 | ✓ |
| 6 | ADIPOQ | 1 | ✓ |
| 7 | ADIPOR2 | 0.9 | ✓ |
| 8 | KCNJ11 | 0.895163 | * |
| 9 | ABCC8 | 0.882255 | * |
| 10 | ACP1 | 0.843614 | ✓ |
| 11 | CYP3A4 | 0.833333 | ✓ |
| 12 | FBXO38 | 0.833333 | ∼ |
| 13 | CCR5 | 0.831326 | ✓ |
| 14 | ADRB2 | 0.829597 | ✓ |
| 15 | CREB1 | 0.82954 | ✓ |
| 16 | NEUROD1 | 0.824896 | ✓ |
| 17 | SPINK1 | 0.819611 | ✓ |
| 18 | CLU | 0.819265 | ∼ |
| 19 | EXT2 | 0.817123 | ✓ |
| 20 | AGT | 0.816536 | ✓ |

Table 3.7 shows our prediction results for Diabetes Mellitus type 2 under PAGED disease modules and there are three known disease genes and 15 suspicious genes in top20 ranking list. Although merely three known disease genes have high rank, we are more interested in novel disease gene prediction. In the table, 15 genes are associated with diabetes mellitus type 2, which are supported by GENECARDS database as well as PubMed literatures. The significant prediction result may be attributed to confirmed gene modules from PAGED, which collected gene/protein modules associated with particular diseases. Therefore, genes involved in PAGED gene modules are more likely to be prioritized in renewed protein complex network.

## Algorithm convergence rate and computational complexity

It is known that the network matrix, defined as $M$, has $|V|$ eigenvalues $\lambda_1$, $\lambda_2$, ..., $\lambda_{|V|}$ such that $1 = \lambda_1 > |\lambda_2| \geq \cdots \geq |\lambda_{|V|}|$. The eigengap of $M$ is defined as $\Delta_M = 1 - |\lambda_2|$, which provides a bound of the convergence time. A larger eigengap means shorter convergence time [98]. Therefore, the computational complexity of random walk is related to network structure. The proposed protein complex network (PCN) is induced from the human PPI network without increasing nodes and edges, therefore the complexity of random walk on PCN equals to that on PPI network. Compared to RWRH [22] and the work [25], proposed RWPCN has comparative efficiency in computational complexity and convergence rate.

## Discussion on RWPCN advantages

Many specific examples show that genes causing similar phenotype tend to be linked at biological levels as components of a multi-protein complex. Protein complexes, as molecular machines that integrate multiple gene products to perform biological functions, are direct manifestations of biological modules. In the other side, protein complexes sharing common proteins, the mutations of genes in one protein complex could lead to same or similar phenotypes of the other protein complex. Therefore, a novel protein complex network is constructed where nodes are individual complexes and the interactions between two complexes are measured by connection strengths. The proposed protein complex network can be a useful basis for interrogating the networks of phenome and interactome to elucidate gene-phenotype associations of diseases.

## 3.4   Summary

While great progress has been made in genomics and proteomics, discovering the associations between genes and phenotypes have remained as challenges. In this chapter, we construct a novel human protein complex network by integrating HPRD protein interaction network and CORUM protein complexes. The result shows that a genome-wide disease gene prioritization for multi-factorial diseases can be obtained using such a human protein complex network. Using our method, the top ranking candidate disease genes that are found to be closely associated with protein complex can potentially be used to guide the prediction of disease-related protein complexes.

It should be acknowledged that the proposed RWPCN algorithm can be improved further. As RWPCN relies on the human protein complex interaction network, the coverage of the protein complex data can affect the performance of prediction. Since the current protein complex data is by no means complete, predicted human protein complexes with high quality could be taken into consideration. Combining the predicted and experimentally validated complex data into the prioritization process (e.g. using the method reviewed in [99]), can increase the power of prediction as long as the quality of the complex data is ensured. RWPCN also depends on the quality (i.e. reliability) of the PPI data which is considered in the current model. It is well-known that PPI data generated with high-throughput methods can be of inferior quality. One possible improvement is to assign weights to protein-protein interactions using diverse biological evidences (e.g. protein sequences, domain, motif, topological properties of PPI network [100], protein localization, molecular function, biological process and gene expression profiles, etc) to improve the reliability of the PPI data that we use for disease gene prioritization. We are currently exploring these and other approaches to further improve our RWPCN algorithm for discovering gene-phenotype associations.

# Chapter 4.

# Positive Unlabeled Learning for disease gene identification

Machine learning methods can be applied to discover new disease genes based on the known ones. Existing machine learning methods typically use the known disease genes as the positive training set $P$ and the unknown genes as the negative training set $N$ (non-disease gene set does not exist) to build classifiers to identify new disease genes from the unknown genes. However, such kind of classifiers is actually built from a noisy negative set $N$ as there can be unknown disease genes in $N$ itself. As a result, the classifiers do not perform as well as they could be.

Instead of treating the unknown genes as negative examples in $N$, we treat them as an unlabeled set $U$. We design a novel Positive-Unlabeled (PU) learning algorithm PUDI (PU learning for Disease gene Identification) to build a classifier using $P$ and $U$. We first partition $U$ into four sets, namely, reliable negative set $RN$, likely positive set $LP$, likely negative set $LN$, and weak negative set $WN$. The Weighted Support Vector Machines are then used to build a multi-level classifier based on the four training sets and positive training set $P$ to identify disease genes. Our experimental results demonstrate that our proposed PUDI algorithm outperformed the existing methods significantly.

The proposed PUDI algorithm is able to identify disease genes more accurately by treating the unknown data more appropriately as unlabeled set *U* instead of negative set *N*. Given that many machine learning problems in biomedical research do involve positive and unlabeled data instead of negative data, it is possible that the machine learning methods for these problems can be further improved by adopting PU learning methods, as we have done here for disease gene identification.

## 4.1   Introduction

Recent studies have revealed that genes associated with similar disorders have been shown to demonstrate higher probabilities of similar gene expression profiling [47], high functional similarities [26] and physical interactions between their gene products [1] [31]. As such, those unknown genes that share similar gene expression profiles with the confirmed disease genes, have high functional similarities with disease genes and interact with disease gene products are likely to be disease genes as well. Xu *et al.* [34] employed the K-nearest neighbor (KNN) classifier to predict disease genes based on the topological features in PPI networks, such as proteins' degree, the percentage of disease genes in proteins' neighborhood, etc. Smalter et al. [35] applied support vector machines (SVMs) classifier using PPI topological features, sequence-derived features, evolutionary age features, etc. Radivojac et al. [101] first built three individual SVM classifiers using three types of features, i.e. PPI network, protein sequence and protein functional information, respectively. It then built a final classifier by combining the predictions from three individual classifiers for candidate gene prediction.

The above works employed machine learning methods to build a binary classifier by using the confirmed disease genes as positive training set $P$ and some unknown genes as negative training set $N$. However, since the negative set $N$ will contain unconfirmed disease genes (false negatives), which confuses the machine learning techniques for building accurate classifiers. As such, the classifiers built based on the positive set $P$ and noisy negative set $N$ do not perform as well as they could in identifying new disease genes.

To address this issue, we design a novel positive-unlabeled (PU) learning algorithm PUDI (PU learning for disease gene identification) to build a more accurate classifier based on $P$ and $U$ [37] [38] [39]. First, we use a comprehensive combination of biological process, molecular function, cellular component, protein domain and PPI data to represent the genes into feature vectors. We design a novel feature selection method to reduce the dimensionality of the feature vectors. Then, we partition $U$ into four label sets, namely, reliable negative set, likely positive set, likely negative set, and weak negative set, based on their likelihoods being positive/negative class. Finally, we build multi-level weighted SVMs using these four sets together with positive set $P$ for identifying disease genes.

To the best of our knowledge, PUDI is the first to design a novel multi-level PU learning algorithm for building a classifier for disease gene identification. We have compared PUDI with three state-of-the-art techniques, namely, Smalter's method [35], Xu's method [34] and ProDiGe [36] method. Our experimental results showed that PUDI outperforms the existing methods significantly for predicting *general* disease genes and for identifying disease genes in eight *specific* disease classes, such as cardiovascular diseases, endocrine diseases, psychiatric diseases, metabolic diseases and cancer, etc.

## 4.2   Method

In section 4.2.1, we introduce a method to characterize genes into feature vectors using different biological features. In section 4.2.2, we propose a novel feature selection method to choose distinguishing features for better classification. Finally, we describe our proposed positive unlabeled learning procedure in section 4.2.3. The system schema and data flow of PUDI are shown in Figures 4.1 and 4.2 respectively.

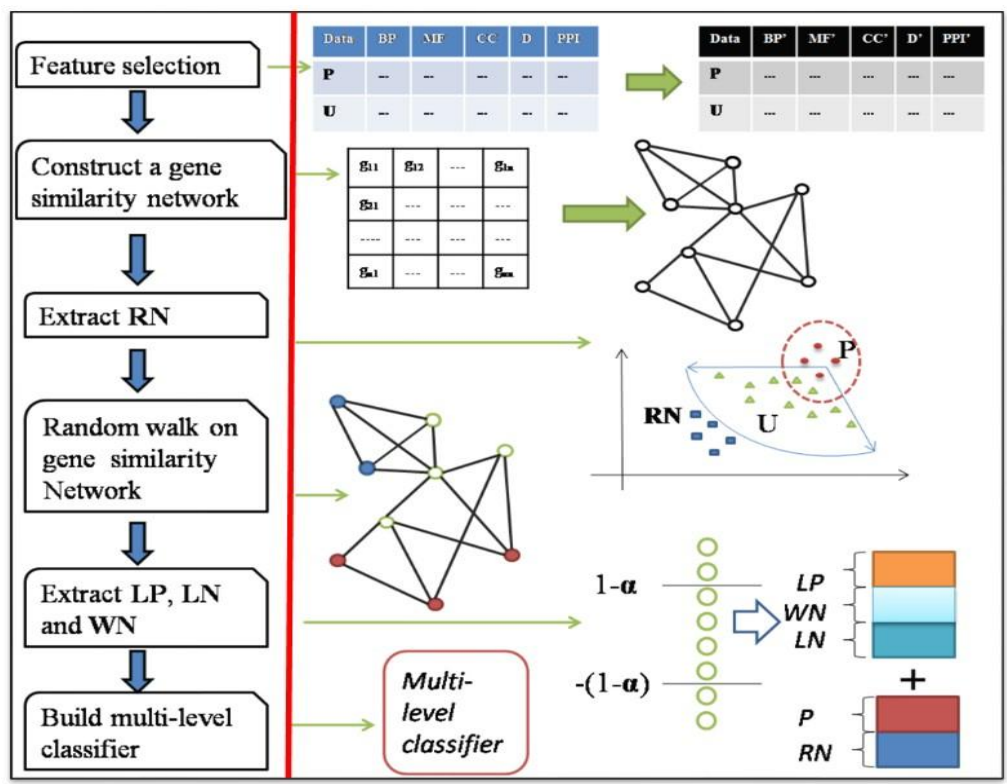

**Figure 4.1**: The Schema of PUDI algorithm

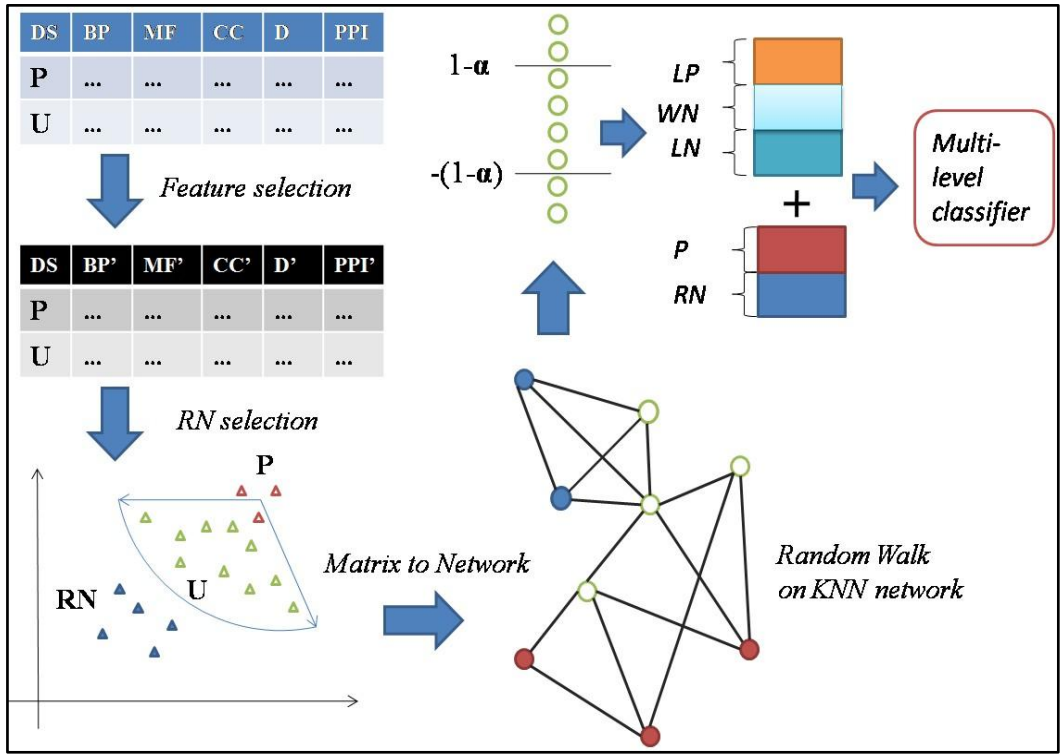

**Figure 4.2**: The data follow of PUDI algorithm

## 4.2.1　Gene characterization

Our approach is to characterize genes (or corresponding gene products) using a comprehensive range of biological information. The information includes protein domains (*D*), molecular functions (*MF*), biological processes (*BP*), cellular components (*CC*), as well as the genes' corresponding topological properties in the protein interaction networks (*PPI*). In other words, each gene $g_i$ is represented as a vector $Vg_i$ which consists of a *domain* component $Dg_i$, a *molecular function* component $MFg_i$, a *biological process* component $BPg_i$, a *cellular component* component $CCg_i$ and a *protein interaction* component $PPIg_i$, i.e. $Vg_i = (Dg_i, MFg_i, BPg_i, CCg_i, PPIg_i)$. We describe each of these components in details below.

Protein domains are evolutionarily conserved modules of amino acid sub-sequence postulated that as nature's functional "building blocks" for constructing the vast array of different proteins. Protein domains are thus regarded as essential units for such biological functions as the participation in transcriptional activities and other intermolecular interactions. Databases, such as the Protein families (Pfam) database and others, have been compiled to comprise comprehensive information about domains (http://www.sanger.ac.uk/Software/Pfam) [102]. In this study, we only used Pfam-A, a collection of manually curated and functionally assigned domains, instead of Pfam-B, which is computationally derived collection of domains (and hence less accurate), to ensure accuracy in our predictions. The *domain* component $Dg_i$ of the given gene $g_i$ is represented as $Dg_i = (d_{i1}, d_{i2}, \ldots, d_{i|\text{Pfam-A}|})$ where $d_{ij}$ ($1 \leq j \leq |\text{Pfam-A}|$) is equal to 1 if $g_i$'s gene product contains the corresponding domain in Pfam-A; 0 otherwise.

For the *molecular function* component $MFg_i$, *biological process* component $BPg_i$, and *cellular component* component $CCg_i$, we use the Gene Ontology (GO, *http://www.geneontology.org/*) database, which provides a common vocabulary that can be used to describe the biological processes (*BP*), molecular functions (*MF*) and cellular components (*CC*) for the genes [50].

Let $SMF = \{MF_1, MF_2, \ldots, MF_{|SMF|}\}$, $SBP = \{BP_1, BP_2, \ldots, BP_{|SBP|}\}$ and $SCC = \{CC_1, CC_2, \ldots, CC_{|SCC|}\}$ represent the set of *MF, BP* and *CC* in GO respectively. Then $MFg_i = (mf_{i1}, mf_{i2}, \ldots, mf_{i|SMF|})$, $BPg_i = (bp_{i1}, bp_{i2}, \ldots, bp_{i|SBP|})$, $CCg_i = (cc_{i1}, cc_{i2}, \ldots,$ $cc_{i|SCC|}$). Let us take $MFg_i$ as an example (similar for $BPg_i$, $CCg_i$) to show how to compute each element $mf_{ij}$ ($1 \le j \le |SMF|$). Note that each $g_i$ can be annotated by many GO terms at different levels in GO's DAG structure (Direct Acyclic Graphs). For example, the gene ADH4 is annotated by molecular function term set {0004022, 004024, 0004174, 0046872, 0008270, 0004023} in the GO database. Assume that $g_i$ has the following molecular functions $FUNg_i$ ={$fun_1$, $fun_2$, …, $fun_k$}, $mf_{ij}$ can be computed as follows:

$$mf_{ij} = max_{f_l \in FUN_{g_i}} sim\_go(fun_l, MF_j), 1 \le l \le k, \qquad (22)$$

where $sim\_go(fun_l, MF_j)$ is the GO term similarity between two functions $fun_l$ and $MF_j$. Since the GO terms of BP, MF and CC are organized into DAG structure, we use the computational method proposed in [103] to compute the similarity between two GO terms A and B. Let the GO term A be represented as $DAG_A = (T_A, E_A)$, where $T_A$ includes term A and all of its ancestor GO terms in the DAG graph, and $E_A$ is the set of edges (semantic relations) connecting the GO terms in $T_A$. For a term $t$ in $DAG_A = (T_A, E_A)$, its S-value related to term A, $S_A(t)$, is defined as:

$$\begin{cases} S_A(t) = 1 \quad t = A \\ S_A(t) = max\{w_e * S_A(t^{'}) \big| t^{'} \epsilon \ children \ of \ (t)\} \quad t \ne A \end{cases} \qquad (23)$$

where $w_e$ is the weight for edge $e \in E_A$ linking term $t$ with its child term $t^{'}$. The weights $w_e$ for two types of edges *"is a"* and *"part of"* are assigned as 0.8 and 0.6 respectively, as recommend in [103].

Given $DAG_A = (T_A, E_A)$ and $DAG_B = (T_B, E_B)$ for GO terms *A* and *B* respectively,

the similarity between $A$ and $B$, $sim(A, B)$, is defined as:

$$sim\_go(A, B) = \sum_{t \in T_A \cap T_B}(S_A(t) + S_B(t))/(SV(A) + SV(B)) \qquad (24)$$

where $SV(A) = \sum_{t \in T_A} S_A(t)$.

For the *protein interaction* component $PPIg_i$, we exploit a protein interaction network $G_{PPI} = (V_{PPI} \; E_{PPI})$ where $V_{PPI}$ represents the set of the interacting proteins and $E_{PPI}$ denotes all the detected pairwise interactions between proteins in $V_{PPI}$. We use four topological features from $G_{PPI}$ [34] for gene $g_i$ as $PPIg_i = (degree_i, 1N_i, 2N_i, Cluster_i)$. $degree_i = |N_i| = |\{u | u \in V_{PPI}, (g_i, u) \subseteq E_{PPI}\}|$ where $N_i$ is the set of $g_i$'s direct neighbors in $G_{PPI}$ and degree of $g_i$ is the cardinality of $N_i$. $1N_i$ represents the proportion of disease genes in $N_i$ which is defined as $1N_i = |\{u | u \in N_i \cap P\}|/degree_i$. Similarly, $2N_i$ represents the proportion of disease genes in $g_i$'s larger neighborhood (with radius 2, i.e. including $g_i$'s direct neighbors and indirect neighbors). $Cluster_i$ is the clustering coefficient which measures the degree to which $g_i$'s direct neighbors in $G_{PPI}$ tend to cluster together [104].

## 4.2.2   Feature Selection

We have represented each gene $g_i$ using a comprehensive list of biological features. In this section, we propose a novel feature selection method to choose subsets of features that are useful for distinguishing disease genes from non-disease genes.

For each feature $f$ in BP, MF, CC and D, we compute its affinity frequency in the positive set P *af*(*f*, *P*) and the unlabeled set U *af*(*f*, *U*):

$$af(f, P) = \sum_{g_i \in P} asso(g_i, f) \tag{25}$$

$$af(f, U) = \sum_{g_i \in U} asso(g_i, f) \tag{26}$$

where $asso(g_i, f)$ is the association score between a gene $g_i$ in P (or U) and the feature $f$. If ($f \in$ BP∪MF∪CC), then

$$asso(g_i, f) = max_{go_j \in GO(g_i)} sim\_go(go_j, f), 1 \leq j \leq |GO(g_i)| \tag{27}$$

In other words, we compute the association score using the maximal GO term similarity between feature $f$ and each of the $gi$'s GO terms. In the case of $f \in$ D, $asso(g_i, f) = 1$ if $f \in D(g_i)$ (or feature f belongs to gene $g_i$'s domain set); 0 otherwise. We evaluate each feature $f$ by its discrimination ability score:

$$da(f) = \left(af(f, P) + af(f, U)\right) * log\left(\frac{|P|}{af(f,P)} + \frac{|U|}{af(f,U)}\right) \tag{28}$$

Our objective is to choose those distinguishing features that either frequently occurred in the disease gene set $P$ but seldom occurred in unlabeled gene set $U$ (assuming large portion of unknown genes are still negatives), or frequently occurred in $U$ but seldom occurred in $P$. In this way, we choose the features which can help us to distinguish disease genes from non-disease genes. Let us see how equation 28 helps us do that. We can see from the equation that given a feature $f$, if its affinity frequency in $P$ $af(f, P)$ is large while its frequency in $U$ $af(f, U)$ is small, or the frequency in $U$ $af(f, U)$ is large while the frequency in $P$ $af(f, P)$ is small, then the value of $da(f)$ will be large since both factors $log(|P|/af(f, P) + |U|/af(f, U))$ and $af(f, P) + af(f, U)$ are large. When $af(f, P)$ and $af(f, U)$ are both large, then the value of $log(|P|/af(f,P) + |U|/af(f,U))$ will be small, hence, $da(f)$ will be relatively small. Similarly, when $af(f, P)$ and $af(f, U)$ are both small, the value of $af(f,P) + af(f,U)$ will be small, and $da(f)$ will also be relatively small.

With a reduced feature set formed by equation 28, we are able to speed up the computation for building a classification model, as well as avoid potential model over-fitting. Table 4.1 and 4.2 list some examples of highly-ranked GO and domain features, indicating the features selected are indeed associated with various diseases.

**Table 4.1**: The distinguishing features for BP, MF and CC

| GO term | GO | Definition | Disease Gene Number |
|---|---|---|---|
| GO:0007165 | BP | signal transduction; signaling cascade | 389 |
| GO:0050896 | BP | Response to stimulus; Physiological response to stimulus | 172 |
| GO:0007166 | BP | Cell surface receptor signaling pathway; Cell surface receptor linked signaling pathway; | 89 |
| GO:0035556 | BP | Intracellular signal transduction; Intracellular signaling cascade | 64 |
| GO:0000166 | MF | Nucleotide binding | 538 |
| GO:0008134 | MF | Transcription factor binding | 101 |
| GO:0019899 | MF | Enzyme binding | 89 |
| GO:0016020 | CC | Membrane | 782 |
| GO:0005634 | CC | Nucleus; cell nucleus | 1146 |

**Table 4.2**: The distinguishing features for domain D

| Domain | Disease name | Disease gene |
|---|---|---|
| | HYPERTROPHIC NEUROPATHY OF DEJERINE-SOTTAS: 145900; NEUROPATHY,CONGENITAL HYPOMYELINATING: 605253; CHARCOT-MARIE-TOOTH DISEASE, DEMYELINATING, | EGR2 |
| | TYPE 1D; CMT1D: 607678 | |
| | NEUTROPENIA, NONIMMUNE CHRONIC IDIOPATHIC, OF ADULTS: 607847 | GFI1 |
| | PALLISTER-HALL SYNDROME; PHS:146510; POLYDACTYLY, POSTAXIAL, TYPE A1:174200; POLYDACTYLY, PREAXIAL IV:174700; GREIG CEPHALOPOLYSYNDACTYLY SYNDROME; GCPS: 175700; HYPOTHALAMIC HAMARTOMAS CONGENITAL HYPOTHALAMIC HAMARTOMA SYNDROME, INCLUDED; CHHS, INCLUDED: 241800 | GLI3 |
| | GASTRIC CANCER:137215;PROSTATE CANCER:176807 | KLF6 |
| | SALIVARY GLAND ADENOMA, PLEOMORPHIC": 181030 | PLAG1 |
| **PF00096** | DIABETES MELLITUS, TRANSIENT NEONATAL, 1: 601410 | PLAGL1 |
| | TOWNES-BROCKS SYNDROME; TBS:107480 | SALL1 |
| | IVIC SYNDROME:147750; DUANE-RADIAL RAY SYNDROME; DRRS:607323 | SALL4 |
| | PIEBALD TRAIT; PBT:172800; WAARDENBURG SYNDROME, TYPE IID:608890 | SNAI2 |
| | ASTHMA, SUSCEPTIBILITY TO:600807 OBESITY LEANNESS, INCLUDED:601665 | ADRB2 |
| | OBESITY LEANNESS, INCLUDED:601665 | ADRB3 |
| | HYPERTENSION, ESSENTIAL:145500 RENAL TUBULAR DYSGENESIS;   RTD:267430 | AGTR1 |
| | Neuroepithelioma:612219 | EWSR1 |
| | AMYOTROPHIC LATERAL SCLEROSIS 6:608030 | FUS |
| | DIABETES MELLITUS, NONINSULIN-DEPENDENT; NIDDM:125853 | IGF2BP2 |
| | OCULOPHARYNGEAL MUSCULAR DYSTROPHY; OPMD:164300 | PABPN1 |
| | OBESITY LEANNESS, INCLUDED:601665 | PPARGC1B |
| **PF00076** | OSLER-RENDU-WEBER SYNDROME 2; ORW2:600376 | ACVRL1 |
| | BREAST CANCER:114480 COLORECTAL CANCER; CRC:114500 :167000 PROTEUS SYNDROME:176920 SCHIZOPHRENIA; SCZD:181500 | AKT1 |
| | DIABETES MELLITUS, NONINSULIN-DEPENDENT; NIDDM:125853 HYPOGLYCEMIA, NEONATAL, SIMULATING FOETOPATHIA DIABETICA:240900 | AKT2 |
| | "PERSISTENT MULLERIAN DUCT SYNDROME, TYPES I AND II; PMDS":261550 | AMHR2 |

### 4.2.3 PU learning to identify the disease genes from *U*

With the above feature representation and feature selection methods, we are now ready to build a classifier using the given confirmed disease gene set *P* and unlabeled gene set *U*. We call our proposed algorithm PUDI -- PU learning for Disease gene Identification. Given that we do not have any negative genes, the first step is to extract a set of reliable negative genes *RN* from *U* by computing the similarities of the unlabeled genes in *U* with the positive genes in *P*, based on the idea that those genes in *U* that are very dissimilar to the genes in *P* are likely to be reliable negatives [37].

The detailed algorithm is given in Algorithm 4.1. We initialize the reliable negative set RN as an empty set, and represent each gene $g_i$ in *P* and *U* as a vector $Vg_i$ using the feature representation method discussed in Section 4.2.1 and the feature selection method presented in Section 4.2.2. We build a "positive representative vector" (*pr*) by summing up the genes in *P* and normalizing it (Line 3). Lines 4-6 compute the average distance of each gene $g_i$ in *U* from *pr* using the Euclidean distance, $dist(pr, Vg_i)$ [105]. For each gene $g_i$ in *U*, if its Euclidean distance $dist(pr, Vg_i) > Ave\_dist$, we regard it as a reliable negative example and store it in *RN* (lines 7-9); since it is very far away from the positive examples, it is thus safe for us to treat it as a negative example.

Algorithm 4.1 Selection of Reliable Negative samples *RN* from Unlabeled set *U*

Input:   Set *P* and set *U* // training positive data and negative data vectors

Output:    RN: // output Reliable Negative

1. $RN = \varnothing$;

2. Represent each gene $g_i$ in $P$ and $U$ as a vector $Vg_i$;

3. $pr = \sum_{i=1}^{|P|} V_{g_i} / |P|$;

4. $Ave\_dist = 0$;

5. **For** each $g_i \in U$   **do**

6.    $Ave\_dist += dist(pr, Vg_i)/|U|$;

7. **For** each $g_i \in U$   **do**

8.    **If** $(dist(pr, Vg_i) > Ave\_dist)$

9.       $RN = RN \cup \{g_i\}$

At this point, we have a positive set $P$, a reliable negative set $RN$ and a refined unlabeled set $U$-$RN$, so we can build a classifier using $P$ and $RN$ with any supervised learning method. However, the reliable negatives in $RN$ may still be far away from the desired boundary between the actual positive and negative data. To build a robust classifier, an important next step in our PUDI algorithm is to further extract the likely positive examples $LP$ and the likely negative examples $LN$ from genes in the $U$-$RN$ which are near the positive and negative classification boundary.

To do so, we construct a gene similarity network $G_{SIM} = (V_{SIM}, E_{SIM})$, in which a vertex $v$ in vertex set $V_{SIM}$ represents a gene in $P \cup U$ and an edge $(g_i, g_j)$ in edge set $E_{SIM}$ represents a connection between two distinct genes $g_i$ and $g_j$. To construct $G_{SIM}$, we define the pairwise similarity matrix $W_{ij}$ between any two genes $g_i$ and $g_j$ as follows:

$$W_{ij} = 1 - \frac{dist(g_i, g_j) - min_{k \in [1, |P \cup U|]} \, dist(g_i, g_k)}{max_{k \in [1, |P \cup U|]} \, dist(g_i, g_j) - min_{k \in [1, |P \cup U|]} \, dist(g_i, g_k)} \qquad (29)$$

A high value in $W_{ij}$ indicates that the two genes $g_i$ and $g_j$ share the similar biological evidence and thus likely belong to same category (disease or non-disease). For each gene $g_i \in V_{SIM}$, we connect it with another gene if their similarities are among top Q most similar ones to gene $g_i$. This is to ensure that we keep only those robust connections in the network. With the resulting gene similarity network $G_{SIM} = (V_{SIM}, E_{SIM})$, we can then perform a random walk with restart algorithm to detect the likely positives and likely negatives, as follows:

*Step 1. Initialize the prior probabilities of positives and reliable negatives.* Let $P_0$ and $N_0$ denote the prior probability vector of the positives and reliable negatives, respectively. In $P_0$ the prior probabilities of positive examples in *P* are assigned an equal probability +1 (with the sum of the probabilities equal to |P|). In $N_0$, the prior probabilities of the reliable negative examples in *RN* are assigned as -|P|/|RN| (so the sum of the probabilities equals to -|P|). This guarantees fair allocation of prior probabilities from the two sets of labeled data. We represent the overall prior probability vector for the training data as $G_0 = (P_0, U_0, N_0)^T$, where $\sum P_0 = \sum N_0$. The prior probabilities in $U_0$ are assigned 0 and we will decide their posterior probabilities in step 2.

*Step 2. Propagate the label information influence from G0 to the genes of U-RN in the network.* After initialing the prior probabilities for positive examples and reliable negative examples as above, we score all the remaining unlabeled genes in the network by propagation. We propose to do flow propagation for this and adopt the Random Network algorithm [80] to our network $G_{SIM}$. The prior influence flows of labeled genes are distributed to their neighbors, which continue to spread the influence flows to other nodes iteratively. Formally, let $G_0$ be the initial probability vector, $G_r$, the probability vector at step $r$, can be calculated as follows:

$$G_r = (1 - \alpha)W_{ij}G_{r-1} + \alpha G_0, (r \geq 2) \tag{30}$$

where $G_1 = G_0$ and $W_{ij} = D^{-1}W_{ij}$. Here $D$ is the diagonal matrix with $D_{ii} = \sum_k W_{ik}$. The parameter $\alpha$ provides a probabilistic weighting of the prior information returning back to initial genes at every step. In this work, we set parameter $\alpha$ to 0.8, as recommend in [22]. At the end of the iterations, the prior information held by every vertex/gene in the network will reach a steady state as proven by [80]. This is determined by the probability difference between $G_r$ and $G_{r-1}$, represented as $Dif = |G_r - G_{r-1}|$ (measured by $L1$ norm). When $Dif \leq 10^{-6}$ [25] we consider that a steady stage has been reached and terminated the iterative process.

*Step 3. Label the likely positives and likely negatives.* According to the posterior probabilities of $U_0$, we further partition the remaining unlabeled data *U-RN* data set into three parts: likely positives (*LP*), likely negative (*LN*) and weak negative (*WN*) using the following criteria:

$$Likely\_Label(g_i) = \begin{cases} LP & G_r(g_i) > 1 - \alpha \\ LN & G_r(g_i) < -(1 - \alpha) \\ WN & otherwise \end{cases} \tag{31}$$

We can now build a classifier using the given positive set *P*, and four extracted sets from *U*, namely, the reliable negative set *RN*, the likely positive set *LP*, the likely negative set *LN*, and the weak negative set *WN*. To take into account of the inherently different levels of trustworthiness of labels in *P*, *RN*, *LP*, *LN* and *WN*, we use a multi-level examples learning technique, Weighted Support Vector Machines [106] [107], to build a classifier. The objective function of Weighted Support Vector Machine can be defined as [108]:

$$minimize: \frac{1}{2}\|w\|^2 + c'_+ \sum_{i \in P} \xi_i + c''_+ \sum_{i \in LP} \xi_i + c'_- \sum_{i \in RN} \xi_i \qquad (32)$$

$$+ c''_- \sum_{i \in LN} \xi_i + c'''_- \sum_{i \in WN} \xi_i$$

Subject to:    $y_i(W^T x_i + b) \geq 1 - \xi_i \ (i = 1,2, \dots, n)$

where $\xi_i$ is a slack variable which allows the misclassification of some training examples, and $c'_+$, $c''_+$, $c'_-$, $c''_-$ and $c'''_-$ represent the penalty factors for SVM to penalize the wrongly classified examples in *P*, *RN*, *LP*, *LN* and *WN* respectively. In particular, $c'_+ > c''_+$ since we are more confident with positive set *P* than the likely positive set *LP*. Correspondingly, we give a larger penalty if examples from *P* are classified as negative class than if examples from *LP* are classified as negative class. Similarly, condition $c'_- > c''_- > c'''_-$ holds since we are more confident with *RN* than *LN*, and we are also more confident with *LN* than *WN*. We used ten-fold cross validation to decide the values for these penalty factors.

## 4.3 Result

In this section, we present our experimental results on the comparisons of our proposed PUDI method with state-of-the-art techniques on general disease genes prediction, feature selection, parameter sensitivity analysis, specific disease gene prediction, and novel disease gene prediction.

### 4.3.1 Experimental data, settings and evaluation metrics

**Experimental data.** We downloaded the latest versions of disease gene data from GENECARD [109] and OMIM [81]. GENECARD and OMIM were then combined into our disease gene benchmark. There are 5405 known disease genes spanning 2751 disease phenotypes after combining GENECARD data together with OMIM. Gene Ontology, consisting of three sub-ontology MF, BP and CC are downloaded from GO (*http://www.geneontology.org/*). Protein domains were obtained from http://www.sanger.ac.uk/Software/Pfam [102]. Human PPI data were downloaded from the HPRD [110] and OPHID [111]. The combined PPI dataset contained 143939 PPIs involving a total of 13035 human proteins.

**Experimental settings.** We chose the known disease genes with at least two-thirds non-zero features as our positive training set P. Here, $|P|$=3849 since not all the genes possess the MF, BP, CC, D and PPI features in the current data sources. We used ~16k genes from Ensembl [112] as the unknown gene set from which we randomly select the actual unlabeled set so that we have a balanced $|P| = |U|$, following the setting in [33] [34] [35].

We then performed feature selection and selected the top N scored features (the default value of N is 1000) for each of the four feature groups, i.e. BP, MF, CC, and D respectively. We executed ten-fold cross validation experiments to evaluate the performance of all the techniques on predicting general disease genes, and three-fold cross validation on predicting disease genes for specific disease groups. The average results are reported in Section 4.3.2.

**Evaluation metrics.** We use the F-measure [113] to evaluate the performance of our classification systems. The F-measure is the harmonic mean of precision ($p$) and recall ($r$), and it is defined as $F = 2 * p * r/(p+r)$. The F-measure reflects an average effect of both precision and recall. When either of them ($p$ or $r$) is small, the value will be small. Only when both of them are large, the F-measure will be large. This is suitable since having either too small a precision or too small a recall for disease gene prediction is unacceptable and would be reflected by a low F-measure.

## 4.3.2   Experimental Result

Firstly, we compared our proposed PUDI algorithm with three state-of-the-art techniques, namely, Smalter's method, Xu's method and ProDiGe method for predicting general disease genes, i.e. automatically classify an unknown gene into a disease gene or a non-disease gene. We employed 10-fold cross validation and all the four methods above use the same groups of training and test set for fair evaluation. As mentioned earlier, both Smalter's method and Xu's method directly treat U as negative set. ProDiGe uses its bagging method to choose random subsets

RS from U and aggregate all the individual classifiers built using P and different RS. Our PUDI method partitions U into 4 label sets and then builds a multi-level Weighted SVM classifier that takes the confidence levels of these label sets into consideration.

Table 4.3 shows that our proposed PUDI method is able to achieve 76.5% F-measure which is 14.2%, 15.1% and 2.0% better than Smalter's method, Xu's method (KNN with K=5) and ProDiGe method respectively. Particularly, compared with ProDiGe, our PUDI method achieves similar precision but 5.1% higher recall, indicating that our multi-level PUDI method can better handle the unlabeled data U for identifying the hidden disease genes in the test set. For Xu's method, we increased its K value from 1 to 21, but its F-measure only changes slightly, ranging from 61.2-61.5. The experimental results in Table 4.3 confirm the benefits of appropriately processing the unknown gene set U.

**Table 4.3**: Overall comparison among different techniques

| Techniques | Precision (*p*) | Recall (*r*) | F-measure (*F*) |
|------------|-----------------|--------------|-----------------|
| PUDI | 72.3% | 81.0% | 76.5% |
| ProDiGe | 72.4% | 75.9% | 74.5% |
| Smalter's method | 62.9% | 61.5% | 62.2% |
| Xu method (1) | 65.0% | 55.6% | 59.9% |
| Xu method (5) | 66.3% | 57.1% | 61.3% |

Recall that we chose those disease genes with at least two-thirds non-zero features since they can provide sufficient informative information for classifiers building. To further evaluate the generalization ability of PUDI, we constructed 10 new test sets which consist of all the 121 poorly annotated disease genes and 10 groups of randomly selected 121 unlabeled genes (both with less than two-thirds non-zero features). Interestingly, we observed that PUDI, in average, achieves 86.5% F-measure, indicating that PUDI classifier is robust enough to accurately identify those poorly annotated disease genes by automatically choosing those highly distinguishing biological features.

**Table 4.4**: Results of individual feature and combinations of features

| Category | Precision ($p$) | Recall ($r$) | F-measure ($F$) |
|---|---|---|---|
| BP | 63.4% | 81.3% | 71.3% |
| MF | 50.3% | 99.6% | 68.6% |
| CC | 54.5% | 93.5% | 67.8% |
| D | 56.2% | 86.5% | 68.1% |
| PPI | 55.1% | 88.2% | 67.8% |
| ALL-BP | 65.3% | 83.3% | 73.2% |
| ALL-MF | 66.0% | 84.7% | 74.2% |
| ALL-CC | 67.4% | 85.7% | 75.4% |
| ALL-D | 62.3% | 86.9% | 72.6% |
| ALL-PPI | 67.9% | 86.7% | 76.1% |

Secondly, we conducted an experiment to investigate the effectiveness of the individual feature category and their combinations, as shown in Table 4.4 (Rows 2-6 and 7-11 respectively). Among the five individual categories, using only the BP ontology achieves the highest F-measure (71.3%), higher than the other feature categories where they have higher recalls but much lower precisions. Further, we filtered out one category from the combined feature set each time. The results in Rows 7-11 showed that using a combined feature set without PPI category can gain better performance than those of other four kinds of combined feature groups. This is probably because we only have 4 PPI features, so removing them will only affect the classification performance slightly. Note the performance of using a combined feature set without protein domains leads to the worst performance, indicating protein domains, as proteins' evolutionarily conserved modules, are useful for identifying disease genes. The performance of using all the features (Table 4.3) is still the best, confirming that integrating all the available biological resources is very valuable for disease gene prediction task.

Thirdly, we perform a sensitivity study for all the three parameters used in the algorithm, i.e. parameter N (used in our feature selection method to control the number of features from MF, BP, CC and D), parameter Q (decides the number of neighbors used in our gene similarity network) and parameter α (used in Random Network to decide how much the influence flows returning back to initial nodes).

Recall that we have one parameter N in our feature selection method to control the number of features from MF, BP, CC and D. To study the effect of parameter N on the performance of our algorithm, we run our method with N from 500 to 2000 with step 500. The results are shown in Table 4.5. The performance is improved with increasing value of N from 500 to 1000, indicating that incorporating more features is helpful for classifying target disease genes. However, if we further include more features with low feature discrimination scores (say N=2000), noisy features will be included and eventually affect the performance of disease gene classification.

**Table 4.5**: Effect of parameter $N$ (in feature selection) to classification performance

| # Parameter $N$ | Precision ($p$) | Recall($r$) | F-measure ($F$) |
|---|---|---|---|
| 500 | 70.8% | 82.5% | 76.2% |
| 1000 | 72.4% | 81.0% | 76.5% |
| 1500 | 70.2% | 81.7% | 76.2% |
| 2000 | 69.9% | 82.0% | 75.5% |

To study the effect of the parameter Q, we run our algorithm with Q from 3 to 9 while fixing N = 1000. Results are shown in Table 4.6. The F-measure is slightly decreased with the value of Q from 5 to 9, indicating that incorporating more edges with relatively low similarities may introduce the noisy connections and thus affect the performance of disease gene identification. Nevertheless, the performance with parameter Q from 3 to 9 without very slight difference suggests that our algorithm is robust to the noisy gene connections and insensitive to the specific value of Q.

**Table 4.6**: Effect of parameter Q (in constructing gene similarity network) to classification performance

| Parameter $Q$ | Precision ($p$) | Recall($r$) | F-measure ($F$) |
|---|---|---|---|
| 3 | 71.9% | 81.3% | 76.3% |
| 4 | 72.2% | 81.0% | 76.3% |
| 5 | 72.4% | 81.0% | 76.5% |
| 6 | 72.5% | 80.7% | 76.4% |
| 7 | 72.0% | 80.8% | 76.2% |
| 8 | 72.3% | 80.3% | 76.1% |
| 9 | 72.6% | 80.1% | 76.2% |

Parameter α in random walk algorithm is used to control how much the influence flows returning back to initial nodes (Genes in *P* and *RN*) at each iteration of the algorithm. In addition, it is also used to be to judge unlabeled genes assigned to likely positive *LP* or likely negative *LN*. With a large α in random walk algorithm, the flows are likely to return to the seed nodes. Therefore the nodes near to seeds are likely to gain higher scores to be assigned to set *LP/LN*. On the contrary, with a small α in random walk algorithm, the flows are likely to flow out of the seed nodes and spread to nodes far away from seeds, therefore those nodes near to seeds are likely to gain relatively lower scores to be assigned to *weak negative set WN*.

When fixing parameters N = 1000, K = 5, we are able to obtain higher F-measure value with increasing value of α, as shown in Table 4.7. Biologically, this is reasonable since unlabeled genes which share various biological evidences with labeled ones more likely belong to same class, either disease genes or non-disease genes.

**Table 4.7**: Effect of parameter α (in random network propagation) to classification performance

| Parameter α | Precision (*p*) | Recall(*r*) | F-measure (*F*) |
|---|---|---|---|
| 0.6 | 66.5% | 82.3% | 73.5% |
| 0.7 | 70.4% | 82.5% | 76.0% |
| 0.8 | 72.4% | 81.0% | 76.5% |
| 0.9 | 73.0% | 79.7% | 76.2% |

These results showed that PUDI was insensitive to the specific values of N and Q. In addition, the best performance was obtained when α = 0.8 which coincided with the recommended value by [22].

Fourthly, we investigated the capability of our proposed algorithm to detect disease genes for specific disease classes/groups – this is much more practically useful than predict general disease genes, e.g. developing novel drugs to tackle disease genes associated with a specific disease for pharmaceutical industry. In this work, we chose all disease classes [31] which have at least 20 confirmed disease genes and we obtained 8 specific disease classes in total. Here we listed the results for 8 specific disease classes. Taking the two disease classes: cardiovascular and endocrine diseases as examples, we selected the disease genes containing the title

'cardiovascular' or 'endocrine' in the causative disease phenotype descriptions from GENECARD and OMIM. A total of 107 cardiovascular disease genes and 81 endocrine disease genes are collected respectively (both treated as positive set *P*). Then, 10 groups of unlabeled gene sets are randomly selected from all gene set as the 10 unlabeled sets *U* (*U* has the same size with P, i.e. |*P*|=|*U*|). Again, all the approaches are evaluated on the identical groups of test data. Given that we have relatively small number of disease genes, to avoid tiny partitions, we performed 3-fold cross validation for each of the 10 training groups and reported the average results in Table 4.8.

**Table 4.8**: The performance comparison of six disease classes

| Diseases | Number | Method | F-measure | AUC |
|---|---|---|---|---|
| Cancer | 210 | PUDI | **72.4%** | **0.806** |
| | | ProDiGe | 69.5% | 0.708 |
| | | Smalter's method | 66.6% | 0.778 |
| | | Xu's method (1) | 63.7% | ~ |
| Cardiovascular | 107 | PUDI | 80.4% | **0.845** |
| | | ProDiGe | 69.3% | 0.703 |
| | | Smalter's method | 70.6% | 0.723 |
| | | Xu's method (1) | 65.4% | ~ |
| Endocrine | 81 | PUDI | 79.2% | **0.801** |
| | | ProDiGe | 69.3% | 0.701 |
| | | Smalter's method | 66.5% | 0.733 |
| | | Xu's method (1) | 68.0% | ~ |
| Metabolic | 263 | PUDI | **82.4%** | **0.897** |
| | | ProDiGe | 69.3% | 0.668 |
| | | Smalter's method | 69.6% | 0.728 |
| | | Xu's method (1) | 71.4% | ~ |
| Neurological | 217 | PUDI | **76.3%** | **0.843** |
| | | ProDiGe | 68.1% | 0.646 |
| | | Smalter's method | 63.1% | 0.753 |
| | | Xu's method (1) | 63.0% | ~ |
| Nutritional | 22 | PUDI | **72.7%** | 0.754 |
| | | ProDiGe | 66.4% | 0.695 |

| | | Smalter's method | 69.4% | **0.769** |
|---|---|---|---|---|
| | | Xu's method (1) | 65.6% | ~ |
| Ophthalmological | 163 | PUDI | **74.9%** | **0.842** |
| | | ProDiGe | 66.6% | 0.647 |
| | | Smalter's method | 55.6% | 0.758 |
| | | Xu's method (1) | 58.8% | ~ |
| Psychiatric | 26 | PUDI | **69.2%** | **0.751** |
| | | ProDiGe | 65.5% | 0.734 |
| | | Smalter's method | 66.1% | 0.742 |
| | | Xu's method (1) | 55.7% | ~ |

Table 4.8 shows that our proposed PUDI algorithm is 9.8% and 9.9% better than the best results from Smalter's method, Xu's method and ProDiGe method for cardiovascular and endocrine diseases respectively. For Xu's method, we have also tried different K valued from 1 to 21. It achieved the best results 72.1% with K=17 for cardiovascular disease and 68.0% with K=1 for endocrine disease in terms of F-measure.

We observed ProDiGe performs 1.3% worse than Smalter's method for cardiovascular disease but 1.3-2.8% better than Xu's method and Smalter's method for endocrine diseases, showing that it cannot achieve consistently better results than other methods. As we mentioned earlier, since the subsets RS that are randomly selected from *U* may still contain unknown disease genes, it will affect the performance of individual classifiers built using *P* and *RS* as well as the final aggregated classifier. On the other hand, our proposed PUDI method partitions *U* into four label sets, so that the multi-level Weighted SVM classifier, can better exploit *U* as training sets by taking the varying confidence levels of the training sets into consideration.

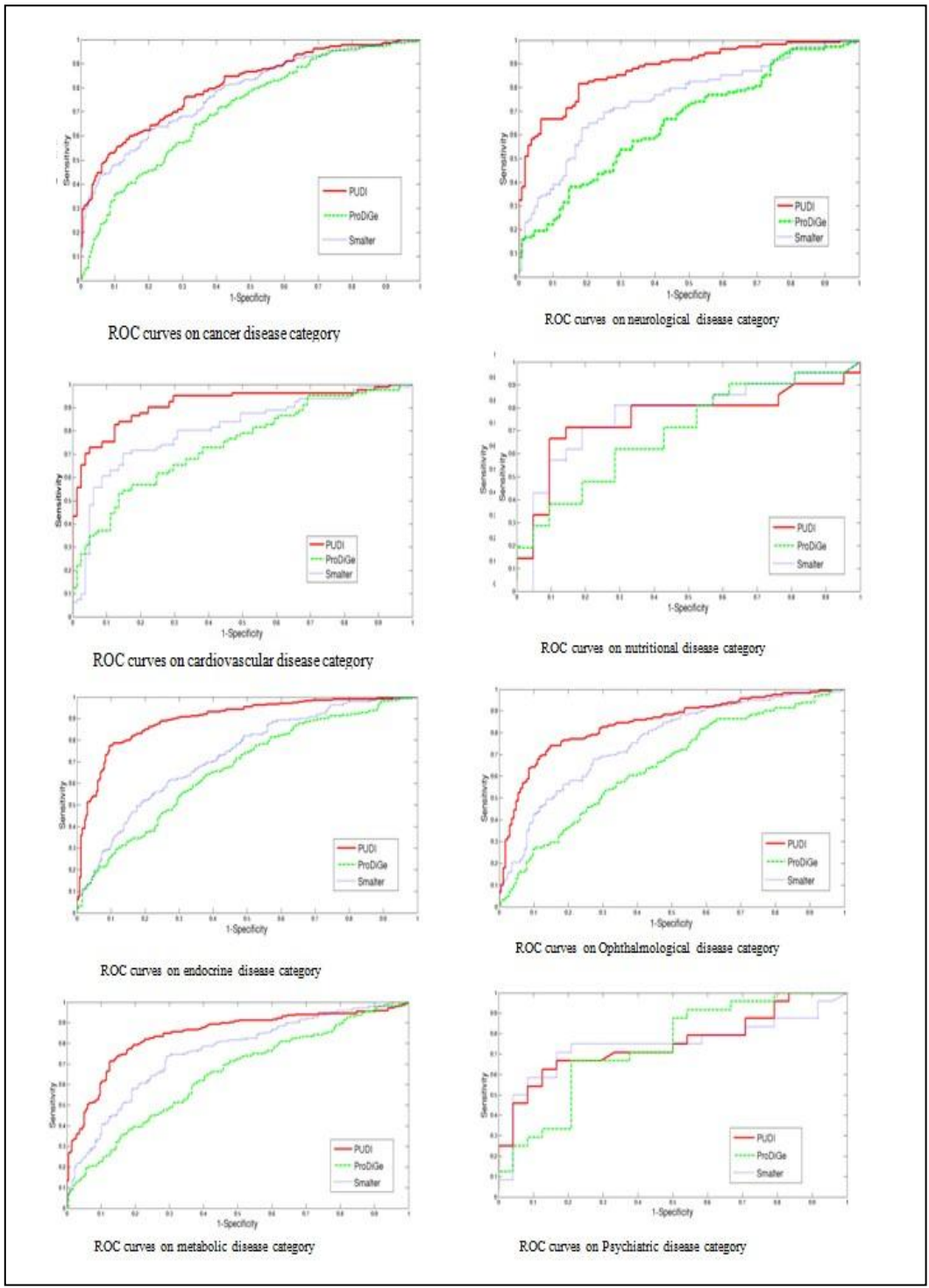

**Figure 4.3**: ROC curves on eight disease groups

ROC curve plots are drawn in Figure 4.3 and corresponding AUC from Table 4.8, indicating that PUDI outperform ProDiGe, and Smalter's method on most of eight disease groups. Since Xu's method did not provide score measures for ranking genes for ROC curves, we were not able to compare this method with others.

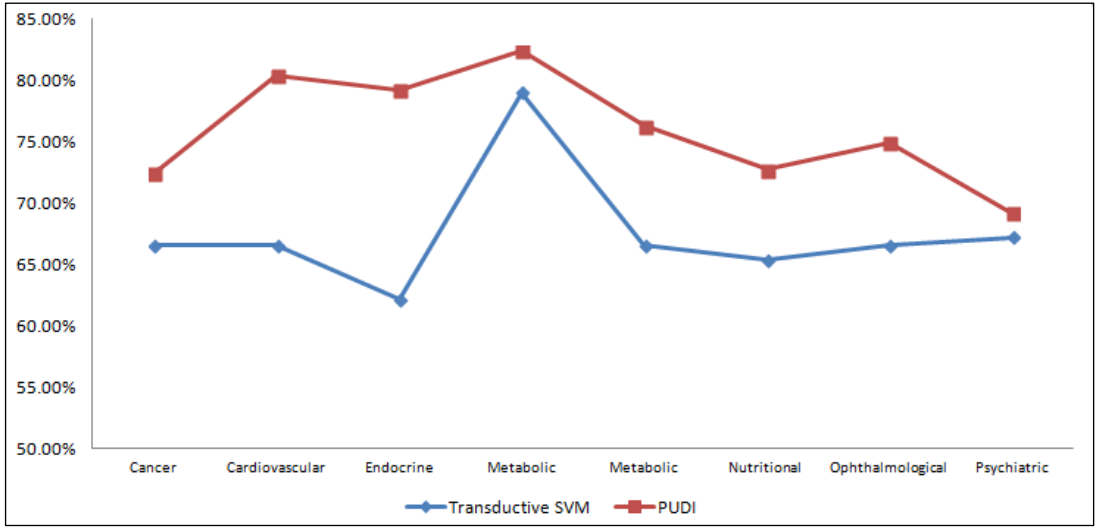

**Figure 4.4**: Comparison between PUDI and Transductive SVM

PUDI is a semi-supervised algorithm in which unlabeled data is exploited to improve the classification performance. To evaluate the efficiency of unlabeled data exploration, PUDI is compare with an existing semi-supervised learning technique, namely Transductive SVM on six disease groups. The comparison result in terms of F-measure in Figure 4.4 shows that PUDI consistently outperforms Transductive SVM, indicating that PUDI is effective to utilize unlabeled data for disease gene identification.

**Table 4.9**: Predicted novel disease genes using all confirmed genes

| Genes | Prob | Relevant Disease |
|-------|------|------------------|
| GP5 | 99.2% | Bernard-soulier syndrome |
| | | Gray platelet syndrome |
| | | Platelet disorder |
| | | Autoimmune thrombocytopenia |
| | | Coagulopathy |
| | | Thrombocytopenia |
| ALG13 | 97.9% | |
| ADPRHL1 | 96.7% | |
| PARVA | 96.6% | Tumors |
| | | Cancer |
| ODAM | 96.4% | |
| ANGPTL1 | 96.3% | Melanoma |
| | | Tumors |
| PTK7 | 96.1% | Panic |
| | | Panic attacks |
| | | Panic disorder |
| | | Premenstrual dysphoric disorder |
| | | Effects cardiovascular |
| | | Agoraphobia |
| | | Anxiety disorders |
| WSB1 | 95.7% | Neurobalstoma |
| AFF1 | 95.0% | Lymphoblastic leukemia acute |
| | | Acute leukemia |
| | | Leukemogenesis |
| | | Leukemia |
| | | Chromosomal aberrations |
| INHBB | 94.7% | Tumors |
| MAPK12 | 94.4% | Shock |
| PHLDA1 | 94.3% | Tumors |
| CABLES2 | 94.0% | |
| BDH2 | 94.0% | |
| CD97 | 94.0% | Thyroid carcinoma |
| | | Thyroid carcinoma anaplastic |
| | | Arthritis reactive |
| | | Colorectal tumors |
| | | Colorectal carcinoma |
| SLC29A4 | 93.9% | |
| FAIM | 93.8% | Leukemia, lymphocytic, Acute |
| EIF2AK2 | 93.8% | Virus infection |
| | | Vesicular stomatitis |
| | | Hepatitis c |
| | | Influenza |
| | | Herpes simplex |
| KRT20 | 93.7% | Carcinoma merkel cell |
| | | Carcinoma mucinous |
| | | Adenocarcinoma |
| ITGB1BP2 | 93.7% | Cardiac hypertrophy |
| | | Hypertrophy |

Fifth, we applied PUDI for uncovering novel disease genes. This is different from the evaluations above where we performed cross validations, i.e. we used part of the confirmed disease genes as the positive training set, and the remaining confirmed disease genes as positive test set. Here, we attempted to discover putative disease genes that are not presented in the current confirmed disease gene dataset. In other words, we will exploit all the confirmed disease genes to predict novel disease genes. As a case study, we applied our PUDI algorithm to discover novel disease genes for cardiovascular diseases. Our algorithm detected 10 unlabeled genes that were not in benchmark/confirmed disease gene dataset. We then performed literature search to check if any of these putative disease genes predicted is indeed associated to cardiovascular diseases. We found that four of the predicted disease genes, namely, ATF4, MBNL1, NCKAP1 and CXCL14, have been reported to be related to cardiovascular diseases. For ATF4, it has been verified to play an important role in cardiovascular diseases using reverse transcription/real-time polymerase chain reaction and western blotting [114]. For MBNL1, it exhibited a regionally restricted pattern of expression in canal region endocardium and ventricular myocardium during endocardia cushion development in chicken [115]. Also, mutations of NCKAP1 showed specific morphogenetic defects: these mouse failed to close the neural tube, also failed to form a single tube (cardia bifida), and showed delayed migration of endoderm and mesoderm [116]. In addition, for CXCL14, it enhanced the insulin-induced tyrosine phosphorylation of insulin receptors and insulin receptor substrate-1, suggesting that CXCL14 played a causal role in high-fat diet-induced obesity, which was frequently associated with hypertension (one type of cardiovascular diseases) [117].

We perform our proposed PUDI algorithm on endocrine disease and find that out of 11 predicted disease genes, three novel genes associated with endocrine diseases: EPHB6, CAMK2D, HEC6. Methylation-specific polymerase chain reaction (MSP) of EPHB6 is associated with breast cancer that is an endocrine-related cancer. In fact, studying the EPHB6 MSP is helpful for the prognosis and/or diagnosis of breast cancer [118]. Calmodulin and calmodulin-dependent protein kinase II (CaMKII) plays important rules in neuroendocrine cell. In Lu *et al.* [119], CaMKII negatively contributes to the regulation of parathyroid hormone (PTH) secretion via a pathway. Finally, HEC6 has medical implication in metastatic neuroendocrine prostate cancer, breast cancer and metastatic colon carcinoma.

Furthermore, we performed our PUDI algorithm using all the confirmed disease genes as positive training set P (not focus on 1 specific disease). We predicted 1110 novel disease genes and we selected the top 20 genes based on their SVM probabilities (we transformed the outputs from SVM into probabilities). Based on the literature search, the results in Table 4.9 show that 14 out of 20 (70%) predicted disease genes are indeed associated with one or more diseases.

Then, we will discuss the time complexity of various computational methods for disease gene prediction and then show the actual time spent by each individual methods. We compare the time complexity of the new approach PUDI with three existing methods, namely ProDiGe, Smalter's method and Xu's method.

PUDI, ProDiGe and Smalter's method are all SVM-based approaches and the training time complexity of SVM is $O(N^2)$ where $N$ is number of training samples. For PUDI, it needs three additional steps: (i) to extract $RN$ (with time complexity $O(N)$), (ii) to construct a gene similarity matrix and a gene similarity network (with time complexity $O(N^2)$), and (iii) to run a random walk algorithm to extract $LN$, $LP$ and $WN$. According to [120], Step (iii) has time complexity $O(w*N^2)$ in which w is number of iterations to converge. However, since w is typically very small (in our experiments w=20) compared to N, $O(w*N^2)$ can be reduced to $O(N^2)$. As such, the overall time complexity of PUDI is still $O(N^2)$. Similarly, the additional steps in ProDiGe and Smalter's methods do not increase their time complexity as well, so they still end up with an overall time complexity of $O(N^2)$. Although Xu's method is based on KNN algorithm, which classifies each target gene based on its similarities to all the other genes in the training set, the complexity of KNN algorithm is also $O(N^2)$. In summary, all the tools have exactly the same time complexity.

Next, we compare the actual running times using different tools across different disease groups. All the experiments were performed on the same machine with 1.83GHz CPU and 1GB (987MHz) memory.

**Table 4.10**: Speed comparisons using three algorithms

| Disease Group | NO. of Samples | Approaches | Time (s) |
|---|---|---|---|
| | 210 | PUDI | 15.410 |
| Cardiovascular disease | 210 | ProDiGe | 13.328 |
| | 210 | Smalter's method | 14.953 |
| Metabolic disease | 526 | PUDI | 45.578 |
| | 526 | ProDiGe | 34.000 |
| | 526 | Smalter's method | 41.015 |
| Neurological disease | 434 | PUDI | 40.486 |
| | 434 | ProDiGe | 32.538 |
| | 434 | Smalter's method | 34.75 |
| Ophthalmological disease | 214 | PUDI | 17.316 |
| | 214 | ProDiGe | 14.11 |
| | 214 | Smalter's method | 16.078 |
| Cancer disease | 350 | PUDI | 29.075 |
| | 350 | ProDiGe | 22.391 |
| | 350 | Smalter's method | 24.116 |

Table 4.10 shows the actual running time for the three SVM-based methods. First, we observe that PUDI did spend 10%-20% more time than ProDiGe and Smalter's method (see also Figure 4.5), as it needed to perform a number of steps before building the multi-level SVM classifier. However, we also notice that the additional time spent was quite small, i.e. only a few more seconds for all the disease groups, and with that our PUDI was able to achieve at least 10%-20% improvement in terms of F-measure than the other existing methods on most of the specific disease gene groups. Furthermore, once the final classifiers are built, the efficiency of prediction procedures is more or less same among all these methods. As such, our PUDI method which can provide more accurate prediction is certainly preferable.

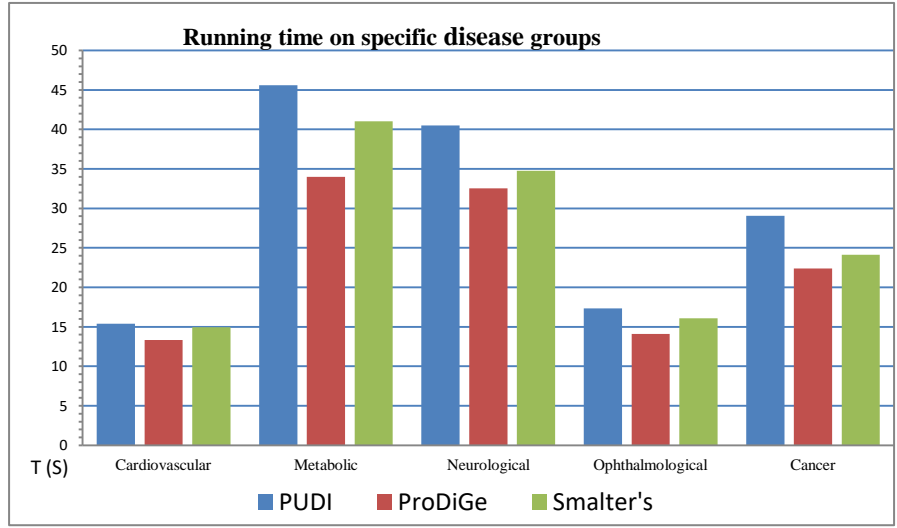

**Figure 4.5**: Running time using different algorithms

**Parameter Setting in Weighted SVM**

For Multi-level SVM, we set its penalty factors in following way: let SVM penalty factors $c = c_-^{'''}$, and $w_+^{'} = c_+^{'}/c_-^{'''}$, $w_+^{''} = c_+^{''}/c_-^{'''}$, $w_-^{'} = c_-^{'}/c_-^{'''}$, $w_-^{''} = c_-^{''}/c_-^{'''}$, then we can get an optimizing goal function using formula 32: $w_+^{'}$, $w_+^{''}$, $w_-^{'}$, and $w_-^{''}$ are used as weights for training sets $P$, $LP$, $RN$, and $LP$ respectively. The weight for $WN$ equals one in equation 32. Let $\Psi$ denote the weight vector as ($w_+^{'}$, $w_+^{''}$, $w_-^{'}$, $w_-^{''}$), we vary $c$ and $\Psi$ to obtain the empirical best parameter through 10 fold cross validation on whole disease gene set (3 fold cross validation on particular disease gene set), i.e. try different parameter values using the 9 fold training data and compute the classifiers' performance using the remaining 1 fold validation set. The parameter values with the best average results will be set the final parameters values.

In particular, we vary $c$ with $2^{-8}$, $2^{-7}$, $2^{-6}$, ..., $2^{7}$, $2^{8}$, and vary $\Psi$ in following ways: $\Psi$ is initialized by (1, 1, 1, 1), and then vary each $w$ of $\Psi$ by turns while keeping other $w$ ($w \in \Psi$) stable. Firstly, we vary $w_+^{'}$ from $2^{-6}$, $2^{-5}$, $2^{-4}$, …,$2^5$, and we obtain the empirical optimal value $2^b$ for $w_+^{'}$. Then, we vary $w_+^{''}$ with $2^{-6}$, $2^{-5}$, …,$2^b$, which guarantee the weight of $LP$ is lower than that of $P$. After turning parameters $w_+^{'}$ and $w_+^{''}$ for positive weights, we vary $w_-^{'}$ and $w_-^{''}$ respectively following the same step as $w_+^{'}$ and $w_+^{''}$ in the range of $2^{-6}$, $2^{-5}$, …,$2^b$. We discover that it is good enough to tune $w_-^{'}$ and $w_-^{''}$ in this range from our multiple experimental trials. For each parameter in weight vector, the maximal turning time is ($b$+6). The turning times for our weight vector are no more than 4*($b$+6). In summary, we set the values for $w_+^{'}$, $w_+^{''}$, $w_-^{'}$, and $w_-^{''}$ in turn so that they can achieve best average performances using cross validation experiments.

We used the criteria of Weight SVM parameter tuning procedure in [108]. In our experiments on general disease gene identification, we found that we could obtain the best performance when parameter $C$ was around 256, $w_+^{'}$ from 1.1 to 1.9, $w_+^{''}$ and $w_-^{''}$ from 1 to 1.1, and $w_-^{'}$ from 1.1 to 1.2 (note we can run a number of times cross-validation to get the average values). For example, one best performance for general disease gene identification was achieved when $C$ = 256, $w_+^{'}$ =1.5, $w_-^{'}$ =1.2, $w_+^{''}$ =1 and $w_-^{''}$ =1.1. We show the actual procedure for parameter tuning below:

**Algorithm 4.2** The procedure of parameter tuning in PUDI

1. Initialize ($w_+^{'}$ ,$w_+^{''}$ ,$w_-^{'}$ ,$w_-^{''}$ ) by (1,1,1,1);

2. Vary $C$ with $2^{-8}$, $2^{-7}$, $2^{-6}$, ..., $2^{7}$, and $2^{8}$ to get best result using cross-validation.

3. Vary $w_{+}^{'}$ from $2^{-6}$, $2^{-5}$, $2^{-4}$, ..., $2^{5}$ to obtain optimal value $2^{b}$ for $w_{+}^{'}$;

4. Vary $w_{+}^{''}$ with $2^{-6}$, $2^{-5}$, ..., $2^{b}$, to obtain value $2^{b'}$, $b' < b$;

5. Vary $w_{-}^{'}$ and $w_{-}^{''}$ respectively following the same step (3 and 4) as $w_{+}^{'}$ and $w_{+}^{''}$.

## 4.4   Summary

To identify disease genes, traditional machine learning methods typically build a binary classification model using confirmed disease genes as positive set $P$ and unknown genes as negative set $N$. The negative set $N$ is noisy because the unknown gene set $U$ contains some unknown disease genes. As such, the classifiers built do not perform as well as they could have.

In this work, we have proposed a novel PU learning approach PUDI for disease gene prediction. We introduced a new feature selection method to identify the discriminating features and performed a further partitioning of the unlabeled set $U$ into multiple training sets for a more refined treatment of $U$ to build the final classifier. We found that PUDI could better model the classification problem for disease gene prediction as it achieved significantly better results than the state-of-the-art methods. Given that many machine learning problems in biomedical research do involve positive and unlabeled data instead of negative data, we believe that the performance of machine learning methods for these problems can possibly be further improved by adopting a PU learning approach [121] [36], as we have done here for disease gene identification. For future work, we will consider to integrate more biological resources [122], such as gene expression data etc. In addition, we may explore more complicated machine learning methods to better model the positive and unlabeled data distributions.

# Chapter 5.

# Ensemble based Positive Unlabeled Learning for Disease Gene Identification

Identifying the association between human genetic diseases and their causative genes has significant impact to healthcare. With the rapid development of biomedical research, increasing numbers of genes have been confirmed as causative genes to diseases. Machine learning methods can be applied to discover new disease causative genes based on their genetic associations to those confirmed disease causative genes. Particularly, positive unlabeled learning (PU learning) methods have been recently proposed to build a classification model where the causative genes are treated as positive training set $P$ and unknown genes are treated as unlabeled set $U$ (instead of negative set $N$) as unknown genes contain undiscovered disease causative genes.

In this chapter, we investigate how to integrate multiple biological sources, including phenotype similarity, gene ontology, protein domain, gene expression, and protein interactions, to extract potential positive and negative sets with corresponding confidence scores from unlabeled set $U$, for building a number of PU learning classifiers. In addition, we have also designed a novel ensemble-based PU learning method EPU to integrate multiple PU learning classifiers for more accurate and robust disease gene prediction. We observe that EPU has achieved significant better results compared with the state-of-the-art methods as well as individual PU learning classifiers across six disease groups. Through integrating the outputs of several PU learning classifiers, we are able to minimize the potential bias and risk of individual predictions, so that the expected errors by our ensemble approach can be expected to be largely reduced.

The proposed EPU method is effective to integrate multiple biological data sources and numerous computational classifiers for disease gene prediction. Given that more reliable biological data sources and powerful computational classifiers will be available in the future, we can expect that our EPU method can be further improved by integrating these additional high-quality biological sources and computational methods.

## 5.1   Introductory

Identification of interaction between phenotype and its causative genes is a crucial part of healthcare. In recent years, a large number of biological data sources are available by high throughput experiments. This provides an invaluable resource for developing machine learning methods to identify novel disease genes on various types of biological datasets.

Recent studies have revealed that genes associated with similar disorders have been shown to demonstrate higher probabilities of similar gene expression profiling [123], high functional similarities [124] [125] and physical interactions between their gene products [126] [127]. In addition, with disease phenotype similarity data, genes associated with same/similar disease phenotypes are likely to share similar biological functions. Given a phonotype, we can infer its potential disease genes from those disease genes associated with other similar phenotypes [128]. From DNA sequence, proteins involved in hereditary diseases tend to be long, with more homologs with distant species, but fewer paralogs within human genome [129]. Furthermore, disease genes associated with similar disease phenotypes are likely to attach together to be a functional modules, such as protein complexes, pathways [130].

A number of methods above have been proposed to prioritize candidate genes based on different kinds of biological data, such as gene sequence data, gene expression profile, evolutionary features, functional annotation data and PPI dataset. Adie et al. [131] employed a decision tree algorithm based on a variety of genomic and evolutionary features, such as coding sequence length, evolutionary conservation, presence, closeness of paralogs in the human genome, etc. In addition to sequence and evolutionary information, topological information on PPI network has been demonstrated to be useful for disease gene prediction. Smalter et al. [35] applied support vector machines (SVM) classifier using PPI topological features, sequence-derived features, evolutionary age features, etc. Radivojac et al. [101] first built three individual SVM classifiers using three types of features, i.e. PPI network, protein sequence and protein functional information, respectively. It then built a final classifier by combining the predictions from three individual classifiers for candidate gene prediction.

The research work mentioned above employed machine learning methods to build a binary classifier where the confirmed disease genes are used as positive training set $P$ and unknown genes are used as negative training set $N$. However, since the negative set $N$ contains unconfirmed disease genes (false negatives), these machine learning techniques do not perform well. Recently, positive unlabeled learning (PU learning) methods have been proposed to build a classification model where unknown genes are treated as unlabeled set $U$ (instead of negative set $N$) as unknown genes contain undiscovered disease causative genes. For example, Mordelet et al. proposed a bagging method ProDiGe for disease gene prediction. This method iteratively chooses random subsets (RS) from $U$ and trains multiple classifiers using bias SVM to discriminate $P$ from each subset RS. It then aggregates all the classifiers to generate the final classifier [132]. As RS could contain less noise (unknown disease genes) than original set $U$, it performs better than standard binary classification models which directly use $U$ as negative training data. More recently Yang et al. designed a novel multi-level PU learning algorithm PUDI to build a classifier for disease gene identification where unlabeled set $U$ are partitioned into multiple positive and negative sets with confidence scores which can be used to enhance classifier building [133] [134].

In this chapter, we design a novel ensemble learning framework, called EPU (Ensemble Positive Unlabeled learning) for disease gene identification. We first extract multiple positive and negative samples from unlabeled set $U$ through performing random network with restart algorithm on three networks, namely protein interaction network, gene expression similarity network, and GO similarity network. Then, we build three independent PU learning models that utilize these extracted positive and negative samples as training data with different confidence scores. Finally, we design a novel ensemble strategy EPU via minimizing the overall error rate and giving different weights to different PU learning models. We have compared EPU with multiple state-of-the-art techniques, namely, multi-level example based learning [108], Smalter's method [35], Xu's method [135] and ProDiGe method [132]. The experimental results show that EPU outperforms the existing methods significantly for identifying disease genes on 6 disease groups. In addition, our proposed EPU algorithm also achieves better results compared to three individual PU learning classifiers, demonstrating that proposed ensemble-based approach is able to effectively utilize each of PU learning methods. We also conduct a case study to show how our proposed EPU algorithm can discover novel disease genes for endocrine and cancer diseases.

## 5.2   Material and Method

In this section, we begin with the description of the experimental data that we have used and briefly introduce how to build protein interaction network, gene expression similarity network, GO similarity network [25] [122] [136]. Then, in Section 5.2.1, we will present our proposed EPU algorithm, including how to learn multiple classification models, which learns an accurate classification model from the given positive set $P$ and unlabeled data $U$.

### 5.2.1 Experimental data and gene network modeling

In this section, we have exploited the following biological data, including human protein interaction data, gene expression data, gene ontology, and phenotype-gene association data.

*Human protein interaction data* (PPI) is downloaded from the Human Protein Reference Database (HPRD) [137] and Online Predicted Human Interaction Database (OPHID) [138]. The combined PPI dataset contains 143,939 PPIs among a total of 13,035 human proteins. We build a protein interaction network $G_{PPI} = (V_{PPI}, E_{PPI})$ where $V_{PPI}$ represents the set of vertices (proteins) and $E_{PPI}$ denotes all edges (detected pairwise interactions between proteins). $G_{PPI}$ can be represented as its matrix format, i.e. $W_{PPI}=[w_{ij}]$ where $w_{ij}=1$ if corresponding protein pairs $(V_i, V_j) \in E_{PPI}$; 0 otherwise.

*Gene expression data* is obtained from RNASeq data which is made publicly available in the EBI ArrayExpress, by the Illumina Human BodyMap 2.0 project, obtained from: http://www.ncbi.nlm.nih.gov/geo/query/acc.cgi?acc=GSE30611. The dataset comprises Fastq reads from the paired-end sequencing of cells from 16 human tissue types, including colon, heart, kidney, white blood cells and so on, using the Illumina HiSeq next generation sequencing platform. This dataset represents the expression values of 17,652 human genes on 16 human tissue types. Suppose gene $g_i$ and $g_j$ are represented into their profile vectors ($e_{i1}, e_{i2},\ldots, e_{is}$) and ($e_{j1}, e_{j2},\ldots, e_{js}$) respectively where $e_{ik}$ ($k$=1, 2, …, $s$) denotes the expression value of gene $i$ from $k$-th tissue. Pearson correlation coefficient is employed to measure the similarity between $g_i$ and $g_j$:

$$sim_{GE}(e_i, e_j) = \frac{\sum_{k=1}^{s}(e_{ik}-\bar{e}_i)(e_{jk}-\bar{e}_j)}{\sqrt{\sum_{k=1}^{s}(e_{ik}-\bar{e}_i)^2}\sqrt{\sum_{i=1}^{k}(e_{jk}-\bar{e}_j)^2}} \qquad (33)$$

where $\bar{e}_i = \frac{1}{s}\sum_{i=1}^{s}e_{ik}$, $\bar{e}_j = \frac{1}{s}\sum_{j=1}^{s}e_{jk}$.

We build a gene expression similarity network $G_{GE} = (V_{GE}, E_{GE})$, where $V_{GE}$ represents a set of genes occurred in the gene expression data and $E_{GE}$ represents a set of edges between the genes in $V_{GE}$. Particularly, for each gene $g_i$, we have a link between $g_i$ and $g_j$ if their similarity $sim(e_i, e_j)$ is among the top 5 out of all the similarities between $g_i$ and other genes, which filter those low similar pairs and potential noise in gene expression data.

*Gene Ontology* (GO, http://www.geneontology.org/) is a set of controlled vocabulary to annotate the attribution of genes and gene products [139]. Gene Ontology provides three sub-ontologies, namely, biological process (BP), molecular function (MF) and cellular components (CC) [139]. For each gene, we represent it into a feature vector where features include all the three sub-onotolgies, i.e. {*MF_1*,...,*MF_{|SMF|}*, *BP_1*,...,*BP_{|SBP|}*, *CC_1*,...,*CC_{|SCC|}*}. We then build GO similarity network $G_{GO} = (V_{GO}, E_{GO})$, where $V_{GO}$ is the gene set annotated in GO dataset and $E_{GO}$ is a set of edges between the genes in $V_{GO}$. Similarly to the gene expression similarity network, for each gene, we keep those top 5 edges which have highest similarities. $G_{GO}$ can be represented as its matrix format, i.e. $W_{GO}=[w_{ij}]$ where

$$w_{ij} = 1 - \frac{Dis(\boldsymbol{g_i}, \boldsymbol{g_j}) - min_{g_k \in V_{GO}} Dis(\boldsymbol{g_i}, \boldsymbol{g_k})}{max_{g_k \in V_{GO}} Dis(\boldsymbol{g_i}, \boldsymbol{g_k}) - min_{g_k \in V_{GO}} Dis(\boldsymbol{g_i}, \boldsymbol{g_k})} \qquad (34)$$

where $Dis(g_i, g_j)$ denotes Euclidean distance between two gene vectors $g_i$ and $g_j$. Note that $0 \leq w_{ij} \leq 1$.

*Phenotype-gene association data*: 4260 phenotype-gene association data spanning 2659 known disease genes and 3200 disease phenotypes, is obtained from the latest version of OMIM (http://omim.org/) [140]. Goh et al. [126] further processed all the entries in OMIM database and categorized the 3200 disease phenotypes into 22 disease groups/classes, i.e. Cancer, Metabolic, Neurological, Endocrine, etc, through the physiological system affected. For example, Endocrine disease group has 62 OMIM phenotypes, including OMIM 241850 (Bamforth-Lazarus syndrome) and OMIM 304800 (Diabetes insipidus, nephrogenic) etc.

*Phenotype similarity network*: Disease phenotype similarity network [141], is defined as $G_{PH} = (V_{PH}, E_{PH})$, where $V_{PH}$ denotes the set of disease phenotypes and $E_{PH}$ denotes relevant phenotype pairs. Disease phenotypes in $V_{PH}$ are represented as feature vectors in which feature terms are Medical Subject Headings (MeSH) controlled vocabulary, and phenotype similarities in $E_{PH}$ are evaluated underline concept relevance and frequency of MeSH terms appeared in text description of OMIM documents. According to Vanunu, O. et al. [128], phenotype pairs with high similarities are regarded as relevant and stored in $E_{PH}$.

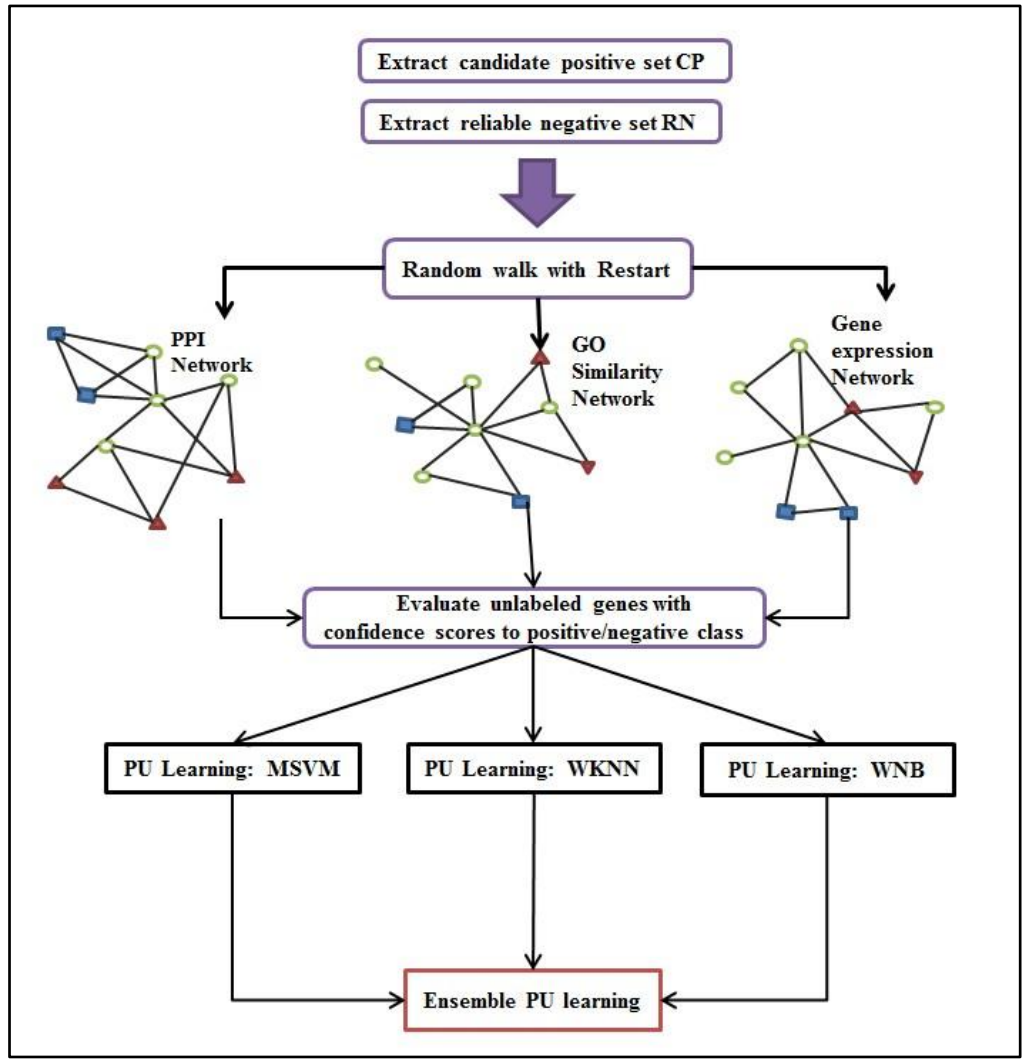

**Figure 5.1**: Overall schema of EPU learning algorithm

## 5.2.2. The proposed Technique EPU

The schema of EPU algorithm is present in Figure 5.1. EPU firstly selects candidate positives from positive genes and reliable negatives from unlabeled genes, then builds three gene similarity networks using PPI data, gene expression data and Gene Ontology data and applies random walk on three networks to weight confidences of unlabeled genes to positive/negative class. Secondly, we exploit these weighted genes to build three diverse classification models to predict "soft" labels for test genes. Finally, considering the prediction results from three classifiers, an ensemble learning approach is applied to make a final decision for test gene class.

Suppose all disease genes from OMIM are stored into a disease gene set *DIS*. Then, all the other genes without involving in *DIS* will be treated as unknown/unlabeled genes and stored into a set *UG* (contains 16, 570 genes) [142]. Note that each gene, both in *DIS* and *UG*, is represented as a feature vector, namely, $\vec{g} = \{g_1, \dots, g_m\}$ where $m$ is the total number of features from GO terms, protein domains and PPI topological features, following our previous work [133].

Now, we will elaborate how to predict novel disease genes given a particular disease or disorder. In particular, those confirmed disease genes for the given disorder group are treated as *positive set P* ($P \subset DIS$) and randomly selected unknown genes from *UG* are treated as *unlabeled set U* ($U \subset UG, |U| = |P|$), following the experimental settings in [131] [35] [135]. As we mentioned in introduction part, we will employ PU learning models for disease gene prediction.

**Step 1. Weighting unlabeled genes by integrating multiple biological evidences**

Given one specific disease class and its associated disease genes, we firstly prioritize candidate positives and reliable negatives based on their similarity to query disease class. Then we build three gene similarity networks using PPI, gene expression and GO as introduced in section 5.2.1, and perform a random walk with restart algorithm on three gene similarity networks to evaluate degree of unlabeled genes belonging to disease class or non-disease class.

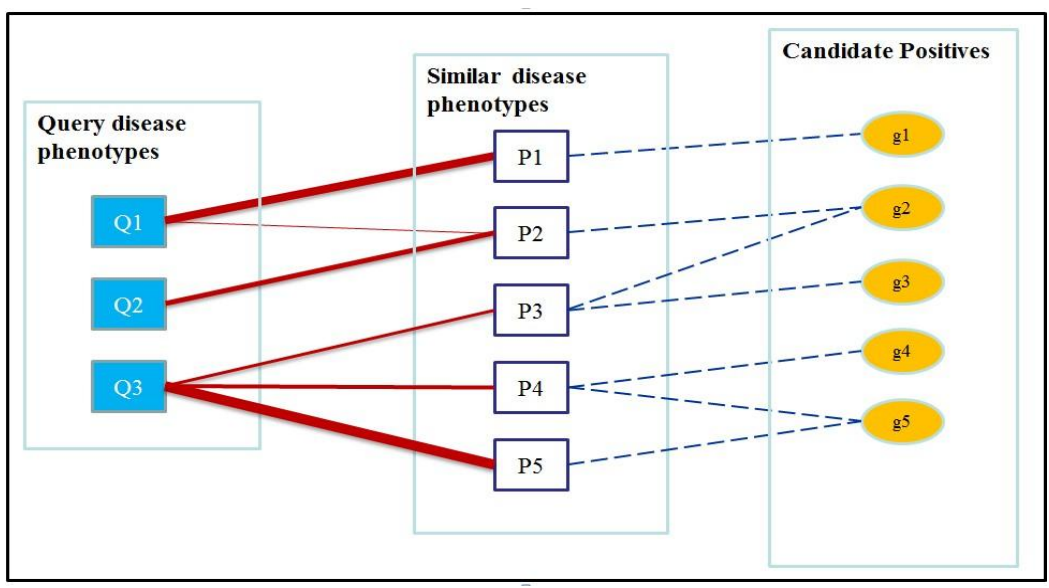

**Figure 5.2**: Procedure of extracting candidate positive set

## Step 2. Extracting candidate positives and reliable negatives

Typically, given positive set *P* is relatively small, we first want to find some positive candidate genes *CP* to complement original positive set *P*. Recent studies have revealed that similar phenotypes are often caused by functionally related disease genes [124] [126], suggesting we could find positive candidate genes *CP* through searching those genes associated similar/relevant phenotypes, by guilt-by-association principal. Particularly, given a disease group/class, we search its associated disease phenotypes, which serve as query phenotypes to uncover those similar disease phenotypes. Figure 5.2 shows the detailed procedure where Q1-Q3 are three disease phenotypes involving in current disease class (e.g. Cancer). We can find those similar phenotypes denoted by P1-P5 through *Phenotype similarity network* if they have a link, e.g. Q1 and P1 (thicker lines represent higher similarity). Known confirmed disease genes (denoted by g1-g5) associated with similar phenotypes P1-P5 will be regarded as candidate positive genes.

Now, we elaborate how to extract *reliable negative* gene set *RN*. Reliable negatives are those unlabeled genes are very different from positive set *P*. To extract *RN*, we build a "positive representative vector" (*pr*) by summing up gene vectors in *P* and normalizing it, following the work [133]. Then we compute average *Euclidean distance* [143] of every unlabeled gene $g_i$ in *U* from *pr*. Finally, we regard an unlabeled gene $g_i$ as a member of *RN* if its distance from *pr* is longer than the average distance (of all the genes in *U*) from *pr*, formalized as:

$$RN = \{g_i | dis(pr, g_i) > \overline{D}\} \tag{35}$$

where $dis(pr, g_i)$ is the Euclidean distance between gene $g_i$ and positive representative vector *pr*. Here we compute an average distance $\overline{D}$ of all the unlabeled gene in *U* from *pr* as: $\overline{D} = \frac{1}{|U|} \sum_{i=1}^{|U|} dis(pr, g_i)$.

## Step 3. Weighting unlabeled genes by performing label propagation on multiple networks

At this point, we have the given positive set *P*, a candidate positive set *CP*, a reliable negative set *RN* and a remaining unlabeled set $U' = U - RN$. In order to build a good classification model, we need to extract those examples/genes with reliable labels which are near the decision boundary between the actual positive and negative class. In this chapter, Random Walk with Restart algorithm [144] is adapted to perform flow propagation which spreads the label information from *P*,

*CP* and *RN* to those unlabeled genes in $U'$ on the three networks we have constructed, namely a PPI network $G_{PPI}$, a GO similarity network $G_{GO}$ and a gene expression similarity network $G_{GE}$.

Formally, let $R_0$ be an initialization vector where primitive scores are assigned to all genes in three networks, which indicate genes' potential label information. Let $p_0$, $p_0'$ and $n_0$ denote the initial values for genes in *P*, *CP* and *RN* respectively. Particularly, all the genes $g_i \in P$ are given a score $p_0(g_i) = +1$, indicating their disease gene status. For each candidate positive gene $g_i \in CP$, a score $p_0'(g_i) = max_{\forall ph_i \in PH(g_i), \forall ph_j \in PH(P)} sim(ph_i, ph_j)$ is assigned to it (where $PH(g_i)$ denotes disease phenotypes caused by gene $g_i$, $PH(P)$ denotes disease phenotypes caused by disease set *P*), i.e. its maximal phenotypic similarity to the known disease genes in *P*. The higher the maximal phenotypic similarity to the known disease genes, the more reliable a gene in *CP* belongs to the disease/disorder class. On the other hand, for genes in reliable negative set *RN*, to balance total amount of flows between positive genes and negative genes, $n_0(g_i) = -(\sum_{g_i \in P} p_0(g_i) + \sum_{g_i \in CP} p_0'(g_i))/|RN|$. Note that remaining unlabeled genes in $U'$ are assigned a score 0 temporally and we will perform flow propagation and assign them the final scores.

For each of our three networks $G_{PPI}$, $G_{GE}$ and $G_{GO}$, the prior influence from seed nodes in *P*, *CP* and *RN*, are first distributed and pumped to their direct neighbors, which continue to spread the influence flows to other nodes iteratively across the whole network. Given $R_0$ be the initial score vector (step 0), $R_t$, the score vector at step $t$, can be calculated as follows:

$$R_t = (1 - \alpha)WR_{t-1} + \alpha R_0, (t \geq 2) \qquad (36)$$

where $R_1 = R_0$ and $W = D^{-1}W$ is a normalized format of matrix $W$, $W \in \{W_{PPI}, W_{GO}, W_{GE}\}$. Here $D$ is the diagonal matrix and $D_{ii} = \sum_k W_{ik}$. $\alpha$ represents the percentage of flow back to original seed nodes in $P$, $CP$ and $RN$ during each iteration of propagation. Default value of $\alpha$ is set as 0.7, following the papers in [25] [133] [130].

Eventually, the information flow will converge to steady state [144]. In our case, the Random Walk with Restart algorithm will stop its iterative process when difference between two steps $R_t$ and $R_{t-1}$ is less than $10^{-6}$ [25], measured by *L1 norm*. Finally, unlabeled gene scores, calculated from three gene networks respectively, are combined into one *integrated score*:

$$Int\_score(g) = \frac{1}{3}(R_t(g, W_{PPI}) + R_t(g, W_{GO}) + R_t(g, W_{GE})) \qquad (37)$$

where $R_t(g, W_{PPI})$, $R_t(g, W_{GO})$ and $R_t(g, W_{GE})$ are gene $g$ scores in PPI network, GO similarity network and gene expression similarity network respectively.

### 5.2.3   Ensemble positive unlabeled learning EPU

Given two classes $C = \{+, -\}$, where '+' denotes positive/disease class and '-' presents negative/non-disease class, we have built three classification models, including Support Vector Machine, K-Nearest Neighbor and Naïve Bayes classifier, to classify genes into positive and negative class.

***PU learning model 1: Multi-level Support Vector Machine* (MSVM)** Based on the integrated score $Int\_score(g)$, we further partition the unlabeled genes $g \in (U - RN)$ into three parts: likely positive set *LP* (genes get higher positive integrated scores), likely negative set *LN* (genes get lower negative integrated scores) and weak negative set *WN* (remaining genes) using the following criteria:

$$L(g) = \begin{cases} LP & Int\_score(g) > (1-\alpha) \\ LN & Int\_score(g) < -(1-\alpha) \\ WN & -(1-\alpha) \leq Int\_score(g) \leq (1-\alpha) \end{cases} \quad (38)$$

Finally, a multi-level classifier is built based on positive training set *P*, reliable negative set *RN*, and three newly generated sets *LP*, *LN*, and *WN*, via weighted support vector machine technique [106] [107], to take into account of the inherently different levels of trustworthiness of labels in the five gene set.

The objective function of Weighted SVM can be defined as [108]:

$$minimize: \frac{1}{2} \|w\|^2 + c'_+ \sum_{i \in P} \xi_i + c''_+ \sum_{i \in LP} \xi_i + c'_- \sum_{i \in RN} \xi_i \quad (39)$$

$$+ c''_- \sum_{i \in LN} \xi_i + c'''_- \sum_{i \in WN} \xi_i$$

Subject to: $\quad y_i(W^T x_i + b) \geq 1 - \xi_i \ (i = 1,2,\dots,n)$

Where the values of parameters $c'_+$, $c''_+$, $c'_-$, $c''_-$ and $c'''_-$ can be decided by using cross-validation techniques. Finally, we apply our MSVM model $P(Y = c_j | g_i, h = \text{MSVM})$ to classify test gene $g_i$ which indicates its probability with respect to class $c_j$ $(c_j \in C)$.

Note here we did not use candidate positive set *CP* directly. However, together with *P*, it has been used for choosing those likely positive genes in *LP* through propagating their influence across three networks.

***PU learning model 2: Weighted K-Nearest Neighbor (WKNN)***: KNN is an instance based learning method, which classifies a test unknown example/gene based on the class labels of its top *K* nearest training examples, i.e. majority class vote of its nearest *K* neighbors. The distance between the test gene and other training examples can be computed using some common distance metrics such as Euclidean distance. Given a test gene $g_i$ and its *k* nearest neighbor set $D_i$, we divide $D_i$ into positive and negative training subsets, namely $D_{i+} = \{g | int\_score(g) > 0, g \in D_i\}$ and $D_{i-} = \{g | int\_score(g) < 0, g \in D_i\}$ based on these neighbors' integrated scores. The conditional probability of the test gene $g_i$ with respect to disease (+) /non-disease class (-), is measured as

$$P(+|g_i, h = KNN) = \frac{\sum_{g_+ \in D_{i+}} |int\_score(g_+)|}{\sum_{g \in D_i} |int\_score(g)|} \;,$$

$$P(-|g_i, h = KNN) = \frac{\sum_{g_- \in D_{i-}} |int\_score(g_-)|}{\sum_{g \in D_i} |int\_score(g)|} \tag{40}$$

Note that the weighted KNN accumulated positive integrated scores and negative integrated scores and then the probability belonging to positive (or negative) class is propositional to their accumulated scores.

***PU learning model 3: Weighted Naïve Bayes (WNB)***: Given a test gene $g_i$,

according to Bayes' theorem, the probability that gene $g_i$ belongs to a class $c_j$ ($c_j \in C = \{+, -\}$) can be computed as:

$$P(Y = c_j | g_i, h = WNB) = \frac{P(g_i | Y = c_j) P(Y = c_j)}{P(g_i)} \qquad (41)$$

where the probability $P(g_i)$ is a constant for the positive and negative classes. The prior probabilities of positive and negative class are defined as 0.5, i.e. $P(Y = +) = P(Y = -) = 0.5$. Given a gene vector $\vec{g} = \{g_{f_1}, ..., g_{f_m}\}$, the conditional probability of feature $f_k$ associated with class $c_j$, denoted as $P(f_k | Y = c_j)$, is calculated as:

$$P(f_k | Y = c_j) = \frac{\sum_{g \in D_{c_j}} g(f_k) * int\_score(g)}{\sum_{k=1}^{m} \sum_{g \in D_{c_j}} g(f_k) * int\_score(g)} \qquad (42)$$

where $g(f_k)$ is value of feature $f_k$ in gene vector $\vec{g}$, $D_{c_j}$ is defined as either $D_+ = \{g \in D | int\_score(g) > 0\}$ or $D_- = \{g \in D | int\_score(g) < 0\}$, depending on $c_j$ is positive class + or negative class −.

Finally, assuming that the probabilities of features are independent given the class $c_j$, we obtain the Naïve Bayes classifier:

$$P(Y = c_j | g_i, h = WNB) = \frac{P(Y = c_j) \prod_{k=1}^{m} g_i(f_k) P(f_k | Y = c_j)}{\sum_{j=1}^{|C|} P(Y = c_j) \prod_{k=1}^{m} g_i(f_k) P(f_k | Y = c_r)} \qquad (43)$$

**Ensemble-based algorithm for integration of individual classifiers**

Note we can apply the three classification models constructed above to classify each unlabeled genes as disease or non-disease gene individually. In this section, in order to perform more robust classification, we design a novel ensemble learning model to integrate these models. The performance of our proposed ensemble model is evaluated via three fold cross validation. Particularly, we partition the genes in $P$ and $U$ into three folds where two folds are used for training set $D$ and the remaining one fold is used for test set $T$. We perform the experiments for three times and we report the average results in terms of F-measures as the evaluation metric.

Suppose $x_{ij} = P(Y = c|g_i, h_j)$ denotes probability value of gene $g_i$ with respect class $c$ predicted by individual classifier $j^{th}$. All genes in $D$ can be organized as the following matrix:

$$\begin{bmatrix} x_{11} & \cdots & x_{1k} \\ \vdots & \ddots & \vdots \\ x_{|D|1} & \cdots & x_{|D|k} \end{bmatrix} \tag{44}$$

where $k = 3$ is the number of individual classifiers and $|D|$ is the size of training set $D$.

Next, we train our ensemble model $o(\vec{x}_i)$, to integrate multiple classification models, which can be denoted as follows:

$$o(x_{i1}, \ldots, x_{ik}) = \text{sgn}(\vec{w} \cdot \vec{x}_i) = \begin{cases} 1, & if\ w_0 + w_1 x_{i1} + w_2 x_{i2} + \cdots + w_k x_{ik} > 0 \\ -1, & otherwise \end{cases} \tag{45}$$

where $\vec{w}$ is a weight vector that indicates the importance of individual models. The final output value "1" denotes disease/positive class and '-1' denotes non-disease/negative class.

Next, we elaborate how to learn the classifier weight $\vec{w}$ from training set $D$. We define $E(\vec{w})$ as training error of the hypothesis of our ensemble model:

$$E(\vec{w}) = \frac{1}{2}\sum_{i \in D}(y_i - o_i)^2 \tag{46}$$

where $y_i$ $(y_i \in \{-1, 1\})$ and $o_i$ $(o_i \in \{-1, 1\})$ are the actual class and predicted class by our ensemble model for training gene $g_i$ respectively. $E(\vec{w})$ is a linear square error function that evaluates the difference between $y_i$ and $o_i$. We minimize $E(\vec{w})$ to guarantee the classification output $o$ with minimal error rate and calculate the weight vector $\vec{w}$.

Here, Gradient decent is applied to search the probable weight vectors in error surface. The gradient of $E$ for $\vec{w}$, denoted as $\nabla E(\vec{w}) = \left[\frac{\partial E}{\partial w_0}, \frac{\partial E}{\partial w_1}, ..., \frac{\partial E}{\partial w_k}\right]$, is the derivative of $E$ with respect to each component of the vector $\vec{w}$. From above equation, we could get each component of $\nabla E(\vec{w})$ as follows:

$$\frac{\partial E}{\partial w_j} = \frac{\partial}{\partial w_j}\frac{1}{2}\sum_{i \in D}(y_i - o_i)^2 \tag{47}$$

$$= \frac{1}{2}\sum_{i \in D}\frac{\partial}{\partial w_j}(y_i - o_i)^2$$

$$= \sum_{i \in D}(y_i - o_i)\frac{\partial}{\partial w_j}(y_i - \vec{w}\cdot\vec{x}_i)$$

$$\frac{\partial E}{\partial w_j} = \sum_{i \in D}(y_i - o_i)(-x_{ij})$$

The training rule of gradient descent is to guarantee $\vec{w}$ is changed in direction that moves to steepest descent along the error surface: $\vec{w} \leftarrow \vec{w} + \Delta\vec{w}$, where $\Delta\vec{w} = -\eta\nabla E(\vec{w})$. $\eta$ is a small positive constant, called learning rate, to determine the step size in gradient decent exploration. We set $\eta = 0.001$, following previous work [145]. The negative gradient $-\nabla E(\vec{w})$ gives the direction of steepest decrease.

According to equations above, we update the gradient descent rule, as:

$$\Delta w_j = -\eta \frac{\partial E}{\partial w_j} = \eta \sum_{i \in D}(y_i - o_i)\, x_{ij} \qquad (48)$$

The overall ensemble learning method is described in Algorithm 5.1: we first pick up an initial random weight vector for $\vec{w}$. The ensemble model is applied to all training genes and compute $\Delta w_j$ for each weight of individual classifiers according to equation ($\Delta w_j$) above. Each weight is then updated by adding $\Delta w_j$. This process is repeated until $\vec{w}$ converges. When $\eta$ is a large number, the search exploration might overstep the minimum point in the error surface rather than settling into it. Therefore, the value of $\eta$ is supposed to be gradually reduced as the number of gradient descent grows.

Algorithm 5.1 Ensemble based Positive Unlabeled learning algorithm

1.  Initialize each element in weight vector $\vec{w}$ with a small random real number;

2.  Do following operations until $\vec{w}$ converges:

3.      Initialize each $\Delta w_i = 0$;

4.      For each decision vector $<\vec{x}_i, y>$ in training samples, Do:

5.          compute output $o(\vec{x}_i)$;

6.          For each linear unit weight $w_j$, Do:

7.              $\Delta w_j = \Delta w_j + \eta(y - o)x_j$

8.      For each $w_j$ in $\vec{w}$, Do:

9.          $w_j = w_j + \Delta w_j$

10.     Record the weight vector as optimal vector;

11. Predict the class of test samples using $o(x_{i1}, \ldots, x_{ik}) = \text{sgn}(\vec{w} \cdot \vec{x}_i)$

## 5.3  Experimental Results

In this section, we begin with introduction about experimental setting and evaluation metrics. Then we present the experimental results, including the comparisons of our proposed EPU algorithm with four state-of-the-art techniques for disease gene prediction, including PUDI method [133], Smalter's method [35], Xu's method [135] and ProDiGe [132]. In addition, we also compare with three individual component classifiers, namely, MSVM, WKNN and WNB. Finally, we also compare EPU with existing ensemble model and perform *novel* disease gene prediction.

### 5.3.1  Experimental setting

From the 22 specific disease classes [126], we choose six largest disease classes with at least 50 confirmed disease genes, which allows us to build classification models as well as for evaluation purpose. The table 5.1 lists number of disease genes for six disease/disorder classes, including cardiovascular disease, endocrine disease, cancer disease, metabolic disease, neurological disease, and ophthalmological disease. Given a particular disease, its disease genes are treated as positive set $P$, and the unlabeled set $U$ is randomly selected from all the unknown genes, with a balanced set $|P| = |U|$, following the setting in [131] [35] [135]. To avoid bias sampling, 10 groups of unlabeled set $U$ are randomly selected. Note all approaches build classification models and evaluate performance on the identical groups of training and test data.

**Table 5.1**: Number of disease genes associated six specific diseases

| Disease category | cardiovascular | Endocrine | metabolic | neurological | ophthalmological | cancer |
|---|---|---|---|---|---|---|
| No. of gene samples | 107 | 81 | 263 | 217 | 163 | 210 |

## 5.3.2   Evaluation metrics

In this study, we adopt the precision, recall and F-measure to measure the performance of the classification model on six specific disease classes. The F-measure is the harmonic mean of precision (denoted as $p$) and recall (denoted as $r$), defined as

$$F = 2 \times \frac{p \times r}{p+r} \qquad (49)$$

Therefore, F-measure indicates an average effect between precision and recall, and F-measure is large only when both of $p$ and $r$ are good, is small when either of them is poor. This is suitable to our objective to accurately predict disease genes in each disease class. Having either too small a precision or too small a recall is unacceptable, and reflecting a low F-measure.

Note that we will compute the F-measure for all six disease gene classes and report the average F-measure base on 10 groups of training sets of each disease class.

### 5.3.3 Experimental result

**Compare our EPU ensemble learning algorithm over state-of-the-art techniques**

Firstly, we compared our ensemble-based algorithm with four state-of-the-art techniques, namely, PUDI method [133], Smalter's method [35], Xu's method [135] and ProDiGe [132] for specific disease/disorder group gene classification.

Table 5.2 shows that our proposed EPU, in average, is able to achieve 6.5%, 15.1%, 16.2% and 16.4% better than PUDI, ProDiGe, Smalter's method, Xu's method in terms of F-measure. Particularly, when we compare with PUDI, a recently proposed method, our EPU can achieve much better precision and consistently better recall. These results indicate that our EPU method can effectively extract hidden positive and negative data from unlabeled data, which in turn boost the performance of our EPU algorithm.

**Table 5.2**: Overall comparison among different state-of-the-art techniques

| Disease group | Techniques | Precision ($p$) | Recall ($r$) | F-measure ($F$) |
|---|---|---|---|---|
| Cardiovascular | PUDI | 0.820 | 0.803 | 0.804 |
| | ProDiGe | 0.543 | 0.963 | 0.693 |
| | Smalter's method | 0.754 | 0.676 | 0.706 |
| | Xu's method | 0.721 | 0.600 | 0.654 |
| | EPU | 0.852 | 0.810 | **0.841** |
| Endocrine | PUDI | 0.836 | 0.753 | 0.792 |
| | ProDiGe | 0.573 | 0.877 | 0.693 |
| | Smalter's method | 0.764 | 0.588 | 0.665 |
| | Xu's method | 0.754 | 0.620 | 0.680 |
| | EPU | 0.881 | 0.877 | **0.879** |
| Neurological | PUDI | 0.703 | 0.801 | 0.749 |
| | ProDiGe | 0.631 | 0.740 | 0.681 |
| | Smalter's method | 0.606 | 0.659 | 0.631 |
| | Xu's method | 0.597 | 0.667 | 0.630 |
| | EPU | 0.782 | 0.804 | **0.786** |
| metabolic | PUDI | 0.801 | 0.848 | 0.824 |
| | ProDiGe | 0.587 | 0.845 | 0.693 |
| | Smalter's method | 0.591 | 0.847 | 0.696 |
| | Xu's method | 0.656 | 0.783 | 0.714 |
| | EPU | 0.833 | 0.939 | **0.909** |
| ophthalmological | PUDI | 0.716 | 0.785 | 0.749 |
| | ProDiGe | 0.583 | 0.777 | 0.666 |
| | Smalter's method | 0.567 | 0.778 | 0.655 |
| | Xu's method | 0.642 | 0.713 | 0.674 |
| | EPU | 0.893 | 0.810 | **0.847** |
| cancer | PUDI | 0.763 | 0.800 | 0.780 |
| | ProDiGe | 0.711 | 0.798 | 0.753 |
| | Smalter's method | 0.738 | 0.790 | 0.763 |
| | Xu's method | 0.710 | 0.797 | 0.751 |
| | EPU | 0.812 | 0.845 | **0.826** |
| Average performance | PUDI | 0.773 | 0.798 | 0.783 |
| | ProDiGe | 0.605 | 0.833 | 0.697 |
| | Smalter's method | 0.670 | 0.723 | 0.686 |
| | Xu's method | 0.680 | 0.697 | 0.684 |
| | EPU | 0.842 | 0.848 | **0.848** |

**Compare our EPU method with individual component classifiers**

In this subsection, we compared the performance among 4 techniques, including 3 individual component classifiers MSVM, WNB, WKNN and our proposed EPU. As shown in Table 5.3, in average, MSVM achieves the highest F-measure (81.3%), much higher than WNB (69.5%) and WKNN (68.7%). This is not surprising as MSVM can take multiple positive and negative sets with different confidence scores into consideration for building its classification model. Furthermore, SVM has performed significantly better than NB and KNN in many real-world applications.

In addition, comparing our ensemble learning method EPU with the three individual component classifiers, we observe that EPU is able to achieve 84.8% in terms of F-measure, which is 3.5%, 15.3% and 16.1% better than MSVM, WNB, WKNN respectively. This is because our proposed EPU can effectively integrate multiple classification models and minimize the overall error rate of our final ensemble classifier via dynamically assigning different weights to different classification models.

**Table 5.3**: Overall comparison to single-expert classifiers

| Disease group | Techniques | Precision ($p$) | Recall ($r$) | F-measure ($F$) |
|---|---|---|---|---|
| Cardiovascular | MSVM | 0.743 | 0.876 | 0.804 |
| | WNB | 0.573 | 0.725 | 0.639 |
| | WKNN(3) | 0.601 | 0.686 | 0.640 |
| | EPU | 0.852 | 0.810 | **0.841** |
| Endocrine | MSVM | 0.834 | 0.852 | 0.842 |
| | WNB | 0.613 | 0.704 | 0.653 |
| | WKNN(3) | 0.645 | 0.531 | 0.579 |
| | EPU | 0.881 | 0.877 | **0.879** |
| Neurological | MSVM | 0.693 | 0.837 | 0.758 |
| | WNB | 0.611 | 0.744 | 0.670 |
| | WKNN(3) | 0.623 | 0.671 | 0.646 |
| | EPU | 0.782 | 0.804 | **0.786** |
| Metabolic | MSVM | 0.840 | 0.913 | 0.874 |
| | WNB | 0.688 | 0.799 | 0.739 |
| | WKNN(3) | 0.766 | 0.788 | 0.776 |
| | EPU | 0.833 | 0.939 | **0.909** |
| ophthalmological | MSVM | 0.784 | 0.861 | 0.819 |
| | WNB | 0.612 | 0.787 | 0.688 |
| | WKNN(3) | 0.673 | 0.722 | 0.696 |
| | EPU | 0.893 | 0.810 | **0.847** |

| Cancer | MSVM | 0.734 | 0.839 | 0.783 |
|--------|------|-------|-------|-------|
| | WNB | 0.725 | 0.851 | 0.783 |
| | WKNN(3) | 0.764 | 0.810 | 0.786 |
| | EPU | 0.812 | 0.845 | **0.826** |
| Average performance | MSVM | 0.786 | 0.863 | 0.813 |
| | WNB | 0.637 | 0.768 | 0.695 |
| | WKNN(3) | 0.679 | 0.701 | 0.687 |
| | EPU | 0.842 | 0.848 | **0.848** |

**Comparing EPU with existing ensemble learning approach**

We compare our EPU with two ensemble baselines, one is to adopt a uniform combination of the three models trained individually, the other applies a weighted combination based on accuracy of component models. Table 5.4 performs evaluation of three ensemble approaches on six disease groups, and EPU consistently outperforms other ensemble methods significantly, which indicates either uniform or weighted combination is unable to balance component classifiers with proper weights. Uniform combination (UComb) has the worst performance due to equally weighting all components for any disease group evaluations. On the other hand, weighted combination (WComb) roughly equates single classifier scenario with that in ensemble classifiers. Unlike above two approaches, EPU uses Gradient decent to optimize the weights of each component classifiers under each disease group, which specifies the weights corresponding to different disease groups.

**Table 5.4** Number of disease genes associated with six disease classes

| Disease group | Techniques | Precision ($p$) | Recall ($r$) | F-measure ($F$) |
|---------------|------------|-----------------|--------------|------------------|
| Cardiovascular | Wcomb | 73.7% | 87.3% | 80.0% |
| | Ucomb | 56.3% | 80.0% | 66.0% |
| | EPU | 85.2% | 81.0% | **84.1%** |
| Endocrine | Wcomb | 86.1% | 84.0% | 85.0% |
| | Ucomb | 65.5% | 73.3% | 67.9% |
| | EPU | 88.1% | 87.7% | **87.9%** |
| Neurological | Wcomb | 69.6% | 83.5% | 75.9% |
| | Ucomb | 65.3% | 74.7% | 70.0% |
| | EPU | 78.2% | 80.4% | **78.6%** |
| Metabolic | Wcomb | 86.6% | 92.5% | 89.5% |
| | Ucomb | 68.4% | 89.1% | 77.4% |
| | EPU | 83.3% | 93.9% | **90.9%** |
| Ophthalmological | Wcomb | 76.9% | 87.3% | 81.8% |
| | Ucomb | 59.4% | 78.7% | 67.7% |
| | EPU | 89.3% | 81.0% | **84.7%** |
| Cancer | Wcomb | 78.7% | 80.3% | 79.5% |
| | Ucomb | 69.7% | 93.7% | 79.9% |
| | EPU | 81.2% | 84.5% | **82.6%** |
| Average performance | Wcomb | 78.5% | 86.0% | 81.8% |
| | Ucomb | 64.1% | 81.6% | 71.5% |
| | EPU | 84.2% | 84.8% | **84.8%** |

**Sensitivity analysis of parameters in EPU algorithm**

We perform a sensitivity study for the parameter used in the algorithm. Parameter $\eta$ is a step size, sometimes called the learning rate in machine learning.

To study the effect of the parameter $\eta$, we run our algorithm with $\eta$ from 0.001 to 0.03 in the scales of 0.005. The performance of the algorithm is measured on cardiovascular disease using three fold cross validation. Results are shown in Table 5.5. The F-measure is relevant steady with the value of $\eta$ from 0.0005 to 0.001, indicating that step size is small enough to move the area of minimum points in hypothesis space, would be helpful to find the optimal weights on error rate space. However, if we further increase the step size of learning rate, the algorithm exploration might cross the optimal (minimum) value point in hypothesis space, instead of staying that point, the searched large weight vector eventually affects the performance of our ensemble algorithm. Nevertheless, the un-stable result in Table 4 with $\eta$ from 0.005 to 0.03 suggests that our algorithm is robust and steady when $\eta$ becomes small.

**Table 5.5**: Effect of parameter $\eta$ to classification performance of cardiovascular disease

| Parameter $\eta$ | Precision | Recall | F-measure |
|---|---|---|---|
| **0.0005** | 0.852 | 0.810 | 0.840 |
| **0.001** | 0.844 | 0.819 | 0.841 |
| **0.005** | 0.895 | 0.762 | 0.819 |
| **0.01** | 0.888 | 0.762 | 0.813 |
| **0.015** | 0.894 | 0.743 | 0.804 |
| **0.025** | 0.867 | 0.810 | 0.832 |
| **0.03** | 0.909 | 0.743 | 0.809 |

**Sensitivity study of noisy data and data coverage in biological networks**

We conduct experiments to analyze how three biological networks affect disease gene prediction model. Table 5.6 studies the effect of the parameter *k* that decides the number of neighbors of each gene in PPI network, GO similarity network and gene expression similarity network. Through tuning number of neighbor interactions in three biological networks, EPU prediction performance is affected by the coverage and noisy of three biological networks. We ran EPU with k from 1 to 9 with $\eta$ = 0.001. On several disease groups like Neurological, Metabolic and Ophthalmological, the performance of EPU algorithm did improve with increasing value of *k* from 1 to 5, indicating that incorporating more informative similar neighbors is helpful for prioritizing disease genes. However, if we further include more neighbors (e.g. when *k* > 8) with low genetic similarities, noisy and un-meaningful neighbors will be included and eventually affects the performance of disease gene prediction. For example, the results in Cardiovascular and Neurological disease groups showed that the performance with $k = 9$ has worsened. Nevertheless, the performance of EPU algorithm with $k$ in wide range was consistently better than that of PUDI and ProDiGe, suggesting that EPU is insensitive to the specific value of $k$.

**Table 5.6**: Sensitive analysis on biological network noise to disease gene prediction

| Disease group | KNN ($k$) | F-measure ($F$) |
|---|---|---|
| Cardiovascular | 1 | 84.1% |
| | 2 | 83.8% |
| | 3 | 82.3% |
| | 4 | 82.9% |
| | 5 | 82.3% |
| | 6 | 82.0% |
| | 7 | 82.6% |
| | 8 | 82.6% |
| | 9 | 82.6% |
| Endocrine | 1 | 87.1% |
| | 2 | 85.2% |
| | 3 | 87.3% |
| | 4 | 87.1% |
| | 5 | 87.1% |
| | 6 | 87.9% |
| | 7 | 87.9% |
| | 8 | 87.9% |
| | 9 | 87.9% |
| Neurological | 1 | 75.0% |
| | 2 | 75.2% |
| | 3 | 75.1% |
| | 4 | 75.2% |
| | 5 | 78.0% |
| | 6 | 75.7% |
| | 7 | 75.9% |
| | 8 | 76.6% |
| | 9 | 76.0% |
| Metabolic | 1 | 90/1% |
| | 2 | 89.4% |
| | 3 | 89.8% |
| | 4 | 90.5% |
| | 5 | 90.9% |
| | 6 | 90.5% |
| | 7 | 90.5% |
| | 8 | 90.1% |
| | 9 | 89.9% |
| | 1 | 83.0% |
| | 2 | 83.0% |
| | 3 | 84.0% |
| | 4 | 84.0% |
| Ophthalmological | 5 | 83.6% |
| | 6 | 83.6% |
| | 7 | 84.0% |
| | 8 | 84.0% |
| | 9 | 83.2% |
| | 1 | 81.8% |
| | 2 | 81.4% |
| | 3 | 80.8% |
| | 4 | 82.2% |
| Cancer | 5 | 81.2% |
| | 6 | 81.7% |
| | 7 | 82.2% |
| | 8 | 82.2% |
| | 9 | 82.4% |

**Predicting novel disease genes for disease groups**

Given a particular disease class, the set of confirmed disease genes are obtained from OMIM and GENECARD. Using all these disease genes as positive training set, we perform experiments by applying our proposed EPU algorithm to prioritize novel disease genes from all the unlabeled gene set. We have chosen two important disease groups, namely, metabolic and cancer, as a case study.

We first applied our EPU algorithm to discover novel disease genes for metabolic diseases. 12 unlabeled genes are detected to be associated with target disease in our algorithm. We then search literature to check whether any of these predicted disease genes are really related to metabolic. We found that two predicted genes, namely, RHEB and DOK5, have been reported to be associated with metabolic diseases. Particularly, Rheb, a GTP-binding protein, is inactivated to protect cardiomyocyte during energy deprivation via activation of autophagy. Therefore, RHEB is a key regulator of autophagy during myocardial ischemia, which has implications in patients with obesity and metabolic syndrome [146]. Tabassum et al. had identified that DOK5 is a novel candidate disease genes associated with type 2 diabetes, a metabolic disorder due to obesity [147]. From the samples in North Indian, the variants of DOK5 might lead to modulation of type 2 diabetes susceptibility.

For cancer disease gene prioritization, 32 unlabeled genes are predicted as candidate disease genes by our EPU model. Seven of them, SIGLEC7, PRDX4, PRDX5, HNRNPL, SRPK1, ABCB10 and PHF10 are reported to be associated with cancer diseases. Table 5.7 lists these candidate disease genes and related literature evidence to support their association to cancer.

For suspicious disease genes without literature evidence support, seven genes, PMM1, SRCIN1, ISY1, KDM4A, CIR1, PPP2R5A and NOL3, are similar to/interacted with confirmed cancer disease genes in terms of gene ontology, gene expression and protein-protein interaction. From GO similarity network, PMM1 is one of top 5 nearest neighbors of cancer disease gene PPM1D and SCRIN1 is one of neighbors of disease gene CTNNB1. In GE similarity network, ISY1 is linked to disease gene P2RX7, KDM4A and CIR1 are interacted with disease genes

CTNNB1 and MSH2 respectively, indicating that three suspicious genes are higly correlated with cancer disease genes in terms of gene expression. From PPI network, PPP2R5A is directly interacted with two disease genes, BCL2 and TP53, and NOL3 is linking to two disease genes, BAX and CASP8. Besides the biological networks above, other biological knowledge can also be useful to provide insightful information to infer association between genes and phenotypes, such as gene expression and pathway.

**Table 5.7**: Cancer-related genes predicted by EPU

| Gene ID | Supported literatures |
|---------|----------------------|
| SUGLEC7 | Ito A. et al. (2001) Binding specificity of siglec7 to disialogangliosides of renal cell carcinoma: possible role of disialogangliosides in tumor progression. FEBS Lett. |
| PRDX4 | Lee S.U. et al. (2008) Involvement of peroxiredoxin IV in the 16alpha-hydroxyestrone-induced proliferation of human MCF-7 breast cancer cells. Cell Biol Int 32(4): 401-5 |
| | Park H.J. et al. (2008) Proteomic profiling of endothelial cells in human lung cancer. J Proteome Res 7(3):1138-50. |
| PRDX5 | Enqman L. et al. (2003) Thioredoxin reductase and cancer cell growth inhibition by organotellurium compounds that could be selectively incorporated into tumor cells. Bioorg Med Chem 11(23): 5091-100. |
| | McNaughton M., et al. (2004) Cyclodextrin-derived diorganyl tellurides as glutathione peroxidase mimics and inhibitors of thioredoxin reductase and cancer cell growth. J Med Chem 47(1): 233-9. |
| | Enqman L., et al. (2000) Water-soluble organotellurium compounds inhibit thioredoxin reductase and the growth of human cancer cells. Anticancer Drug Des. 15(5): 323-30. |
| HNRNPL | Goehe, R.W., et al. (2010) hnRNPL regulates the tumorigenic capacity of lung cancer xenografts in mice via caspase-9 pre-mRNA processing. J. Clin. Inves. 120(11): 3923. |
| | Hope N.R., et al. (2011) The expression profile of RNA-binding proteins in primary and metastatic colorectal cancer: relationship of heterogeneous nuclear ribonucleoproteins with prognosis. Hum Pathol. 42(3): 393-402. |
| SRPK1 | Hayes, G.M., et al. (2007) Serine-arginine protein kinase 1 overexpression is associated with tumorigenic imbalance in mitogen-activated protein kinase pathways in breast, colonic, and pancreatic carcinomas. Cancer Res. 67(5): 2972-80. |
| ABCB10 | Tang, L., et al. (2009) Exclusion of ABCB8 and ABCB10 as cancer candidate genes in acute myeloid leukemiaLetter to the Editor. Leukemia 23: 1000-2. |
| PHF10 | Wet M., et al. (2010) Preparation of PHF10 antibody and analysis of PHF10 expression gastric cancer tissues. Journal of Xiao Bao Yu Fen Zi Mian Yi Xue 26(9): 874-6. |

| | Li C., et al. (2012) MicroRNA-409-3p regulates cell proliferation and apoptosis by targeting PHF10 in gastric cancer. Cancer Lett 320(2): 187-97. |

## 5.4   Summary

In this work, we design a novel ensemble learning method EPU, to classify disease genes for different disease groups. Firstly, we extract multiple positive and negative samples from unlabeled set $U$ through performing random network with restart algorithm on three networks, namely protein interaction network, gene expression similarity network, and GO similarity network. Secondly, we build three PU learning models independently to utilize these extracted positive and negative samples as training data with different confidence scores. Finally, we design a novel ensemble strategy EPU to integrate multiple PU learning models which can minimize the overall error rate and give reasonable weights to different PU learning models. Experimental results illustrate the effectiveness of our proposed methods. Our proposed EPU method performs much better than the existing state-of-the-art techniques for disease gene prediction.

For further work, we will explore if there are other biological data sources are useful for disease gene prediction. In addition, we will integrate additional PU learning methods for further enhancing our EPU methods.

# Chapter 6.

# Conclusions and Future Directions

In this chapter, we firstly envision an overview of our research contributions presented in the entire Ph.D thesis, and then provide the possible directions for future work.

## 6.1    Conclusion and Discussion

In our research work, we focus on the protein complex network model, Positive-Unlabeled Learning and ensemble learning for disease gene prioritization and classification. In order to discover the functional modules in the protein complex network, we propose a three-layer heterogeneous network and investigate its capability for novel disease gene prediction. To discover reliable and efficient approaches for disease gene classification, we have applied the PU learning based framework PUDI and ensemble-based model EPU.

In label propagation approach, we built a novel protein complex network by fitting HPRD protein interaction network and CORUM protein complexes. Our experimental results showed that disease genes associated with complex diseases can be prioritized using such a human protein complex network. We have verified disease gene prediction ability of the RWPCN through extensive experiments. Our RWPCN outperformed other existing approaches on both whole genome evaluation and *ab* initio evaluation and consistently better over different coverage of phenotype interactome. As RWPCN relies on the human protein complex interaction network, the coverage of the protein complex data could affect prediction performance. Due to limited number of experimental validated protein complex data, predicted human protein complex with high quality and functional modules (such as pathway) could be taken into consideration. One possible improvement is to weight predicted protein complexes using protein localization, molecular function and biological process. Protein members within same or similar biological features are more likely to form protein complexes to perform biological functions.

Next, a novel PU learning approach PUDI was proposed for disease gene prediction. Traditional machine learning methods typically build a binary classification model using confirmed disease genes as the positive set $P$ and unknown genes as the negative set $N$, which may suffer from false negative samples, . Due to imbalanced data characteristics that only a few disease genes are identified from thousands of unlabeled genes on whole genome, binary classifier would lose efficacy. To address this issue, we propose PUDI for disease gene prediction. Assumed genes associated identical disease groups are more likely to form functional modules on biological interaction networks, PUDI first learned representative knowledge of confirmed disease genes, and then applied a semi-supervised learning algorithm on biological networks to prioritize candidate positives and reliable negatives from unlabeled genes. Labeling unlabeled genes with different weights, PUDI is more confident to build a reliable and accurate classifier for disease gene identification. Given that many machine learning problems in biomedical research do involve positive and unlabeled data instead of negative data, we believe that the performance of machine learning methods for these problems can potentially be further improved by adopting a PU learning approach [121] [36], as we have done here for disease gene identification. For future work, we will consider to integrate more biological resources. Such as gene expression data, etc. In addition, we may explore efficient PU learning methods to model positive and unlabeled data distribution and address imbalance data issue.

Finally we propose an ensemble-based framework, namely EPU, to prioritize disease genes associated with six disease classes. By using multiple biological data sources, EPU is less susceptible to potential bias, incompleteness and noise in single data source. By employing an ensemble approach for prediction, EPU also minimizes the inherent limitations of single learning models. However, ensemble learning model works when component data sources/ learning classifiers are independent, unrelated and complementary. Since correlated components are likely to make similar decisions, ensemble framework with correlated components tends to build a bias classifier. One possible improvement of our work is to reduce highly correlated learning models from the ensemble learning framework and retain complementary and reliable learning models. For future work, we will consider other methods to search for global optimal points on error spaces more efficiently, like the Markov chain Monte Carlo (MCMC) method. On the other side, we may explore better machine learning methods to combine multiple data sources with different data distributions.

## 6.2 Future Directions

Here we provide several possible directions for future study in the area of disease gene prediction.

### 6.2.1 Integration of more biological interactions to improve protein complex interaction network

In this thesis, we propose a three-layer network model (RWPCN) for disease genes prediction. However, the protein complex data is extracted from CORUM, which contains only part of the existing human protein complexes [78]. We did not consider the high-quality protein complexes obtained from the computational approaches. Another problem is that we measure the strength of protein complex interactions using only PPI data, which has a high false-positive rate. Integration of more biological interactions might increase the quality of protein complex interactions. A possible direction for modeling a reliable protein complex network is to extend protein complexes using reliable protein complexes detected by computational approaches [99]. Another improvement is to evaluate protein complex interaction using diverse biological evidences (e.g. biological process, gene expression profiles [148] and metabolic reactions [149]).

### 6.2.2 Phenotype Entities Similarity Calculation

The phenotype similarity could also be improved. In this thesis, the similarity between two phenotype entities is calculated based on the text description in OMIM [81]. To calculate the similarity of phenotypes, they used OMIM Mesh terms instead of a controlled vocabulary [74]. Recently, the Human Phenotype Ontology (HPO) has been proposed [53]. HPO provides a standard vocabulary to describe phenotypic abnormalities in human diseases, such as *atrial septal defect*. However, different phenotype data sources have different predictive power for different diseases due to the data characteristics and quality. It is necessary to balance the phenotype similarities, according to the prediction performance on different disease classes. With the availability of well annotated phenotype data and accurate similarity measurements, a better quality phenotype network could be obtained to improve the prediction ability of our method.

### 6.2.3   Improving the network propagation based method (RWPCN) using machine learning based classification approaches

We want to explore how to integrate the network propagation based method (RWPCN) with other machine learning based classification methods. Given the different characteristics of two methods, we could minimize the potential bias and risk of each individual method and thus further improve the prediction accuracy.

Classification methods, such as SVM and Neuron network, can assign weights to test examples, which represent a kind of similarity to positive or negative class, just like the scores assigned by Random Walk with Restart to nodes based on topological similarity to initial labeled nodes on the networks. Both continuous outputs provided by the machine learning and network propagation approaches are interpreted as the degree of support given to that class (such as disease class and non-disease class). Classifier fusion [150] is a method that provides a strategy for combining the outputs generated by individual approaches. In this method, classifier outputs are normalized to the [0, 1] interval, allowing formation of an ensemble through algebraic combination rules, such as majority voting, maximum/minimum/sum/product [151] [152] [153] and kernel method [154] [155].

### 6.2.4   Prioritization of loci using GWAS data

Recently, many scientific teams have been examining the genomes of thousands of people in an attempt to find mutations presented only in individuals with certain traits. Interactions amongst genome loci associated with diseases have been largely mapped from data generated through forward genetic approaches, such as recombination hotspot [156] [157] [158] [159] [160], genome-wide linkage [161] or genome-wide association studies (GWAS) [162] [163]. Such methods leverage naturally occurring mutations in the genome to pinpoint loci that have associations, ideally causal associations, with a trait of interest [164].

## 6.3   Final Remarks

Identifying disease genes from the human genome is a crucial but challenging task in the area of bioinformatics research and medical health. In wet-lab experiments, disease genes are identified using mutation analysis, which is very expensive and labor intensive. In this thesis, we proposed novel computational approaches to prioritize and identify disease genes. The experimental results show that our work is more robust and accurate than other state-of-the-the-art techniques for disease gene identification.

27(3): 122-126, 2002.

# Author's Publications